\documentclass[12pt,english]{article}
\usepackage{mathptmx}
\usepackage[T1]{fontenc}
\usepackage[latin9]{inputenc}
\usepackage{geometry}
\usepackage{float}
\usepackage{amsmath}
\usepackage{graphicx}
\usepackage{setspace}
\usepackage{esint}
\usepackage[numbers]{natbib}
\makeatletter

\providecommand{\tabularnewline}{\\}

\numberwithin{equation}{section}

\usepackage{hyperref}
\date{}

\makeatother

\usepackage{babel}
\begin{document}

\title{Density Forecasts and the Leverage Effect: Some Evidence from Observation
and Parameter-Driven Volatility Models}

\author{Leopoldo Catania$^{a}$, Nima Nonejad$^{b}$\thanks{b: Department of Mathematical Sciences, Aalborg University. Fredrik
Bajers Vej, 7G, 9220, Aalborg East, Denmark. E-mail: \texttt{nimanonejad@gmail.com}.
a: Department of Economics and Finance, University of Rome ``Tor
Vergata''. Via Columbia, 2, 00133, Rome, Italy. E-mail: \texttt{leopoldo.catania@uniroma2.it}.
A previous version of this paper was circulated under the title ``Density
Forecasting Comparison of Volatility Models''.}}
\maketitle
\begin{abstract}
\noindent The leverage effect refers to the well-established relationship
between returns and volatility. When returns fall, volatility increases.
We examine the role of the leverage effect with regards to generating
density forecasts of equity returns using well-known observation and
parameter-driven volatility models. These models differ in their assumptions
regarding: The parametric specification, the evolution of the conditional
volatility process and how the leverage effect is accounted for. The
ability of a model to generate accurate density forecasts when the
leverage effect is incorporated or not as well as a comparison between
different model-types is carried out using a large number of financial
time-series. We find that, models with the leverage effect generally
generate more accurate density forecasts compared to their no-leverage
counterparts. Moreover, we also find that our choice with regards
to how to model the leverage effect and the conditional log-volatility
process is important in generating accurate density forecasts.\\


\noindent (JEL: C11, C22, C51, C53) \newpage{}
\end{abstract}

\section{Introduction \label{sec: Introduction}}

How to model volatility is a very important issue in financial econometrics.
The behavior of the volatility process has important implications
in derivative pricing and portfolio optimization. It also receives
a great deal of concern from policy makers, central banks and market
participants because it can be used as a proxy to measure risk. Changes
in bonds, commodity, exchange rates and stock prices raise important
questions regarding the stability of financial markets and the impact
of price variations on the economy. For instance, for oil dependent
nations, unexpected changes in crude oil volatility can imply huge
losses (gains) and thus lower revenues (higher revenues) with drastic
negative (positive) consequences on the economy. For these reasons,
over the years a large number of volatility models have been developed,
see Poon and Granger (2003) for a review.

However, although volatility plays a very big role in financial econometrics,
there is an inherent problem with using it: Volatility is latent and
cannot be directly observed. Moreover, there is no unique or universally
accepted way to define it. By far, the most popular approach to model
volatility is the Generalized Autoregressive Conditional Heteroscedasticity
(GARCH) framework, introduced Engle (1982) and Bollerslev (1986).
GARCH belongs to the class of observation-driven models, as defined
in Cox (1981)\footnote{In an observation-driven volatility model, conditional volatility
is updated using deterministic functions of lagged observations and
conditional volatilities. In a parameter-driven volatility model,
conditional volatility is modeled as a dynamic processes with idiosyncratic
innovations.}. Within the GARCH framework, conditional volatility is a deterministic
function of lagged observations and conditional volatilities. This
simple specification is able to capture important stylized facts of
financial time-series such as heavy tails in log-returns, mean reversion
and volatility clustering. Moreover, the likelihood function for GARCH
is available in closed form via the prediction error decomposition.
This in turn leads to simple and fast estimation procedures through
maximum likelihood and partly explains the popularity of GARCH models
in applied econometrics. Recently, Hansen and Lunde (2005) compare
over three hundred GARCH-type models in terms of their ability to
describe the conditional variance. Out-of-sample comparison finds
no evidence that a simple GARCH(1,1) is outperformed by more sophisticated
models in the context of exchange rates. On the other hand, the GARCH(1,1)
model is clearly inferior to models that can accommodate the leverage
effect in the context of equity data.

Recently, Creal et al. (2013) and Harvey (2013) propose a new class
of observation-driven models referred to as Generalized Autoregressive
Score (GAS), or, equivalently, Dynamic Conditional Score (DCS) models.
Similar to GARCH, estimation of GAS models is straightforward using
maximum likelihood techniques. However, contrary to GARCH, the mechanism
to update the parameters occurs through the scaled score of the conditional
distribution for the observable variables. Creal et al. (2013) demonstrate
that this approach provides a unified and consistent framework for
introducing time-variation in the model parameters for a wide class
of nonlinear models, which also encompasses the GARCH framework. Harvey
(2013) and Koopman et al. (2016) argue that GAS provides the same
degree of generality as for nonlinear non-Gaussian state space models,
which compared to GAS are generally harder to estimate. Using Monte
Carlo simulations, Koopman et al. (2016) show that when the data generating
process is a nonlinear state-space model, the predictive accuracy
(in terms of point forecasts) of a (misspecified) GAS models is similar
to that of a (correctly specified) nonlinear state-space model. For
nine time-varying parameters models, the loss in mean squared error
from GAS instead of the correct state-space model is less than $1\%$
most of the times and never higher than $2.5\%$, see Koopman et al.
(2016) for more details.

An alternative to GARCH and GAS models is the stochastic volatility
(SV) model introduced in Taylor (1986), which represents our example
of a parameter-driven volatility model in this paper. In this framework,
conditional log-volatility is modeled as an unobserved process with
idiosyncratic innovations. Typically, we assume that conditional log-volatility
follows an autoregression of order one, AR(1), which is a discrete
time approach to the diffusion process used in the option pricing
literature, see Hull and White (1987). Generally, SV has proven to
be more attractive than GARCH-type models. For example, Jacquier et
al. (1994) find that compared to GARCH, SV yields a better and more
robust description of the autocorrelation pattern of the squared returns.
Kim et al. (1998) show that the basic SV model provides a better in-sample
fit than GARCH. However, a closed form likelihood function is not
available for the SV model. Therefore, in most cases, we must resort
to simulation techniques to estimate it, see Kim et al. (1998), Koopman
and Uspensky (2002), Flury and Shephard (2011) and Koopman et al.
(2016) for more details.

Ever since Black (1976) and Christie (1982), the relationship between
returns and changes in volatility has been of great interest. Variations
in prices and their relationship with volatility can imply huge losses
or gains to investors involved in financial markets. Leverage also
plays an important role at micro level such as the asset volatility
of a firm, see for example, Choi and Richardson (2016). The usual
claim dating back to Crane (1959) states that returns and changes
in volatility are negatively correlated. The theoretical background
behind this relationship has been developed by financial economists
based on the well-known Modigliani-Miller framework, see Black and
Scholes (1973) . The theory suggests that, a fall in equity value
increases the debt to equity ratio (leverage) and consequently the
riskiness of a firm, which in turn translates into a higher volatility
level. Nowadays, it is generally well-known that this theory does
not apply to the real word, see Figlewski and Wang (2000). However,
the negative relation between returns and volatility remains a well-established
stylized fact of financial time-series, see McNeil et al. (2015).

Observation and parameter-driven volatility models incorporate the
leverage effect differently. For the former, the leverage effect is
incorporated directly in the conditional volatility equation. Negative
and positive return innovations impact the conditional volatility
asymmetrically. For the later, the leverage effect is expressed through
a correlation coefficient between return and log-volatility innovations.
Obviously, an important purpose of observation and parameter-driven
models is to generate out-of-sample forecasts. Given the different
nature of these models, it is very interesting to assess the relative
merits of these approaches. Generally, most research focuses on conditional
volatility forecasts, see for example, Hansen and Lunde (2005) and
Koopman et al. (2016). However, given the interest among practitioners
to obtain a complete description of the conditional return distribution
(return density) a comparison of density forecasts among different
volatility models therefore seems in order. Furthermore, as mentioned
in Koopman et al. (2016), comparative analyses of different volatility
models often exclude parameter-driven models because compared to observation-driven
models they are computationally more brutal.

To our knowledge, too little is known about the ability of well-known
observation and parameter-driven model's ability to generate accurate
density forecasts. Therefore, a large-scale analysis using a large
number of different time-series seems in order. The aim of this paper
is to address this point. Particularly, following Koopman et al. (2016),
we consider four volatility models (three observation and one parameter-driven)
each with and without the leverage effect (eight models in total)
and perform an out-of-sample density forecast comparison using more
than four hundred financial time-series. First, we consider a very
long time-series of daily (weekly) Dow Jones returns from $1902$
to $2016$. Then, we focus on individual return-series from the S\&P
$500$ index, covering sectors such as: Energy, financials, telecommunications
and utilities from $2004$ to $2014$. This way, we can (a): Determine
to which extent accounting for the leverage effect enables a model
to generate accurate density forecasts, (b): Which model-type performs
best at a given forecast horizon, and (c): How do results change across
forecast horizons and returns.

The remaining of this paper is as follows: Model framework is discussed
in Section \ref{sec: Econometric Framework}. Data and models are
introduced in Section \ref{sec: Data and Models}. Results are discussed
in Section \ref{sec: Empirical Results} and the last section concludes.
The supplementary material accompanying the paper contains additional
results.

\section{Econometric Framework \label{sec: Econometric Framework}}

Let $y_{1},...,y_{T}$ denote a $T\times1$ sequence of returns. In
this paper, we are interested in modeling the volatility of the conditional
distribution of returns given all available information. Our general
notation assumes that the observed return at time $t$, $y_{t}$,
is generated from $y_{t}=\varLambda\left(h_{t}\right)\varepsilon_{t}$,
where $h_{t}$ is the conditional log-volatility at time $t$, $\varLambda\left(\cdot\right)$
is a nonlinear link function and $\varepsilon_{t}$ is a white-noise
independent of $\varLambda\left(h_{t}\right)$. We then forecast the
$h$ $\left(h>0\right)$ step ahead conditional density of $y_{t}$,
$p\left(y_{t+h}\mid\mathcal{F}_{t},\theta\right)$, where $\mathcal{F}_{t}$
denotes the information set at time $t$ and $\theta$ is the vector
of the model parameters that govern $h_{t}$. Model comparison is
performed using the weighted Continuous Ranked Probability Score (wCRPS)
criterion of Gneiting and Ranjan (2011). wCRPS also is used to asses
a model's ability to predict specific parts of the conditional distribution
such as the center and tails.

\subsection{Observation-driven models \label{sub: Observation-driven models}}

In observation-driven models, $h_{t}$ depends on its owned lagged
values and lagged values of $y_{t}$ in a deterministic way. In this
paper, our simplest specification is the t-EGARCH(\emph{p},\emph{q})
model introduced in Nelson (1991). We set $p=1$ and $q=1$\footnote{We find that $p=1$ and $q=1$ generally works well and increasing
$p$ or $q$ does not add any significant improvements in terms of
generating density forecasts. The same conclusion holds for Beta-t-EGARCH
and SPEGARCH. }. Thus, our t-EGARCH(1,1) is given as
\begin{eqnarray}
y_{t} & = & \exp\left(h_{t}/2\right)\varepsilon_{t},\quad\varepsilon_{t}\sim St\left(v\right)\label{eq:2.1}\\
h_{t+1} & = & \omega+\alpha\left(\left|\varepsilon_{t}\right|-E\left[\left|\varepsilon_{t}\right|\right]\right)+\gamma\varepsilon_{t}+\beta h_{t},\label{eq:2.2}
\end{eqnarray}
where $St\left(v\right)$ stands for the Student's t-distribution
with $v>2$ degrees of freedom. The choice of the Student's t-distribution
in (\ref{eq:2.1}) is due to the fact that financial returns often
exhibit excess kurtosis (i.e. kurtosis$>3$), which is not properly
accounted for by the Gaussian distribution, see for example, Conero
et al. (2004) and McNeil et al. (2015). In (\ref{eq:2.2}), $\omega$
is the level of conditional log-volatility, $\alpha$ and $\beta$
determine the impact of past observations and conditional volatilities
on $h_{t}$ and $\gamma$ controls the leverage effect. The term,
$E\left[\left|\varepsilon_{t}\right|\right]$, is the mean of the
folded Student's t-distribution, see Psarakis and Panaretoes (1990).
The inclusion of this term implies that $\varepsilon_{t}-E\left[\left|\varepsilon_{t}\right|\right]$
is a Martingale difference sequence with respect to $\mathcal{F}_{t-1}$.
Thus, the unconditional log-volatility level, $\overline{h}$, is
given as $\omega/\left(1-\beta\right)$. Moreover, $h_{1}$ is treated
as an additional parameter to be estimated along with $\omega$, $\alpha$,
$\gamma$, $\beta$ and $v$. Due to the exponential link function
in (\ref{eq:2.1}), conditional volatility is always positive and
the only constraint imposed during the estimation procedure to ensure
stationarity of $h_{t}$ is $\left|\beta\right|<1$, see also Nelson
(1991). When $\varepsilon_{t}>0$, then $\alpha+\gamma$ determines
the response to past observations. When $\varepsilon_{t}<0$, then
the magnitude of the response is $\alpha-\gamma$. Evidently, when
$\gamma<0$, we have the leverage effect and thus decreases in returns
increases volatility.

Recently, Harvey (2013) introduces the Beta-t-EGARCH(\emph{p},\emph{q})
model, which belongs to the class of Score-Driven models, see also
Creal et al. (2013). Similar to (\ref{eq:2.1})-(\ref{eq:2.2}), we
set $p=1$ and $q=1$. Following the same notation as Harvey (2013),
our Beta-t-EGARCH(1,1) model is given as
\begin{eqnarray}
y_{t} & = & \exp\left(h_{t}\right)\varepsilon_{t},\quad\varepsilon_{t}\sim St\left(v\right)\label{eq:2.3}\\
h_{t+1} & = & \omega+\alpha u_{t}+\textrm{sgn}\left(-\varepsilon_{t}\right)\gamma\left(u_{t}+1\right)+\beta h_{t},\label{eq:2.4}
\end{eqnarray}
where in (\ref{eq:2.4}), $\textrm{sgn}\left(x\right)$ returns the
sign of the variable $x$ and $u_{t}$ is the score of the distribution
of $y_{t}$ with respect to $h_{t}$ given as $u_{t}=\left(\left(v+1\right)\varepsilon_{t}^{2}/\left(\left(v-2\right)+\varepsilon_{t}^{2}\right)\right)-1$.
As argued in Harvey (2013), (\ref{eq:2.3})-(\ref{eq:2.4}) has the
nice property of being robust to extreme observations compared to
the simpler t-EGARCH model. The robustness properties of Beta-t-EGARCH
with respect to t-EGARCH can be easily seen by comparing the response
of $h_{t}$ to $\varepsilon_{t}$. Indeed, taking apart the leverage
effect controlled by $\gamma$, the response of $h_{t}$ for t-EGARCH
is piece-wise linear in $\varepsilon_{t}$, while for Beta-t-EGARCH
it is a smooth function bounded by $v$. The inclusion of the leverage
effect for the Beta-t-EGARCH model is more intuitive than for t-EGARCH.
Indeed, since $u_{t}+1$ is always positive, if $\varepsilon_{t}<0$,
then the volatility level at time $t+1$ is increased by an amount
$\gamma\left(u_{t}+1\right)$ if $\gamma>0$. Since $u_{t}$, is a
Martingale difference with respect to $\mathcal{F}_{t-1}$, the unconditional
mean of $h_{t}$ is given by $\omega/\left(1-\beta\right)$. Similar
to t-EGARCH, $h_{1}$ is estimated along with the model parameters,
$\theta$.

t-EGARCH and Beta-t-EGARCH both assume the same parametric specification
for $\varepsilon_{t}$, namely, $\varepsilon_{t}\sim St\left(v\right)$.
In order to investigate the role of the leverage effect in a semiparametric
framework, we also consider the semiparametric EGARCH model, which
we label as SPEGARCH(1,1). Particularly, we follow Pascual et al.
(2006) and assume that
\begin{eqnarray}
y_{t} & = & \exp\left(h_{t}/2\right)\varepsilon_{t},\quad\varepsilon_{t}\overset{iid}{\sim}\left(0,1\right)\label{eq:2.5}\\
h_{t+1} & = & \omega+\alpha\left|\varepsilon_{t}\right|+\gamma\varepsilon_{t}+\beta h_{t},\label{eq:2.6}
\end{eqnarray}
where, $\varepsilon_{t}\overset{iid}{\sim}\left(0,1\right)$ means
that $\varepsilon_{t}$ is an iid sequence of white-noise shocks with
mean $0$ and variance $1$. Similar to t-EGARCH, the only constraint
imposed during the maximization is $\left|\beta\right|<1$. Note that,
contrary to (\ref{eq:2.1})-(\ref{eq:2.2}), (\ref{eq:2.6}) does
not include $E\left[\left|\varepsilon_{t}\right|\right]$ due to the
obvious lack of parametrical assumption of $\varepsilon_{t}$ in (\ref{eq:2.5}),
see also Straumann and Mikosh (2006) for a similar approach.

\subsection{Parameter-driven models \label{sub: Parameter driven models}}

In the context of parameter-driven models, we consider the stochastic
volatility (SV) model, see Kim et al. (1998), Malik and Pitt (2011)
and Flury and Shephard (2011). The SV model is given as
\begin{eqnarray}
y_{t} & = & \exp\left(h_{t}/2\right)\varepsilon_{t},\quad\varepsilon_{t}\sim N\left(0,1\right)\label{eq:2.7}\\
h_{t+1} & = & \mu+\phi h_{t}+\sigma\eta_{t}\quad\eta_{t}\sim N\left(0,1\right),\label{eq:2.8}
\end{eqnarray}
where $\mu$ is the level of conditional volatility, $\phi$ and $\sigma$,
denote the persistence and the conditional volatility of volatility,
respectively. We follow Kim et al. (1998) and impose that $\left|\phi\right|<1$,
with the initial condition, $h_{1}\sim N\left(\mu,\sigma^{2}/\left(1-\phi^{2}\right)\right)$.
The usual way to incorporate leverage in (\ref{eq:2.7})-(\ref{eq:2.8})
is to assume correlation between $\varepsilon_{t}$ and $\eta_{t}$,
i.e. $E\left[\varepsilon_{t}\eta_{t}\right]=\rho$ and $\left|\rho\right|<1$.
Thus, a negative shock at time $t$ increases volatility at time $t+1$.
Moreover, (\ref{eq:2.7})-(\ref{eq:2.8}) differs from models in Section
\ref{sub: Observation-driven models} in important aspects. First,
we have two sources of innovations, namely $\varepsilon_{t}$ and
$\eta_{t}$. Second, the leverage effect is expressed in terms of
correlation between these disturbances. Third, the one step ahead
predictive density of (\ref{eq:2.7})-(\ref{eq:2.8}) is the continuous
mixture distribution
\begin{eqnarray}
p\left(y_{t}\mid\mathcal{F}_{t-1},\theta\right) & = & \int_{-\infty}^{\infty}p\left(y_{t}\mid h_{t},\mathcal{F}_{t-1},\theta\right)p\left(h_{t}\mid\mathcal{F}_{t-1},\theta\right)dh_{t}.\label{eq:2.9}
\end{eqnarray}
Contrary to observation-driven models, (\ref{eq:2.9}) takes into
account the complete density structure of past observations. Moreover,
in the context of (\ref{eq:2.7})-(\ref{eq:2.8}), we have to distinguish
between the measurement density, $p\left(y_{t}\mid h_{t},\mathcal{F}_{t-1},\theta\right)$,
which is obviously Gaussian and the predictive density, $p\left(y_{t}\mid\mathcal{F}_{t-1},\theta\right)$.
The reason for this is that contrary to $p\left(y_{t}\mid h_{t},\mathcal{F}_{t-1},\theta\right)$,
(\ref{eq:2.9}) is leptokurtic, see Carnero et al. (2004) and Koopman
et al. (2016). In other words, similar to t-EGARCH (Beta-t-EGARCH),
(\ref{eq:2.9}) has fat tails even though $\varepsilon_{t}$ and $\eta_{t}$
in (\ref{eq:2.7})-(\ref{eq:2.8}) follow a $N\left(0,1\right)$.
However, (\ref{eq:2.9}) is not available in closed form and therefore
we must resort to simulation, see Section \ref{sub: Models}.

\section{Data and Models \label{sec: Data and Models}}

Below, we outline our data-sets and also briefly discuss how we operationalize
the different models. All the computations are performed in \texttt{R}
(\texttt{R} Core Team, 2016). In order to reduce computation time,
we code the majority of the routines in \texttt{C++} using the \texttt{armadillo}
library of Sanderson (2010) and made available in \texttt{R} using
the \texttt{Rcpp} and \texttt{RcppArmadillo} packages of Eddelbuettel
et al. (2016a) and Eddelbuettel et al. (2016b). Parallel computations
relied on the \texttt{OpenMP API} (OpenMP, 2008). Numerical optimizations
of the likelihood functions for observation-driven models was performed
using the General Nonlinear Augmented Lagrange Multiplier optimization
method of Ye (1988) available in the \texttt{R }package \texttt{Rsolnp}
of Ghalanos and Theussl (2016).

\subsection{Data \label{sub: Data}}

We divide our analysis into several parts. First, we use a very long
time-series of daily Dow Jones (DJ) returns from $3$ March $1902$
to $15$ April of $2016$, for a total of $30921$ daily observations.
Our sample contains a very large number of historical events such
as world wars (we remove the missing observations during WWI and WWII),
economic depressions, oil and financial crises such as the Arab\textendash Israeli
Wars and the Great Recession of $2008$. We also consider shorter
datasets using data from $5$ January, $2004$ to $31$ December,
$2014$ consisting of $2768$ observations for a cross sectional dimension
of $432$ firms from the S\&P $500$ index\footnote{As of January $2016$, there are $504$ firms in the S\&P $500$ index
according to the composition available from Datastream. We then remove
the series where no data is available at January $2007$ leaving us
with a total of $432$ firms.}. The index composition is that of $1$ January $2016$. We use the
Global Industry Classification Standard (GICS) to classify groups
of firms in $11$ economic sectors, see Table \ref{Table 5}. The
number of firms for each sector oscillates between $20$ and $60$,
apart from telecommunication services which displays only $5$ firms
and consumer discretionary which displays $70$ firms. Overall, our
sample is heterogeneous with respect both to the level of capitalization
and the use of debt, see Table \ref{Table 5}. We convert the price
series into the logarithmic percentage return series, using $100\times\ln\left(P_{t}/P_{t-1}\right)$,
where $P_{t}$ $\left(P_{t-1}\right)$ is the price at time $t$ $\left(t-1\right)$.

\subsection{Models \label{sub: Models}}

For models in Section \ref{sec: Econometric Framework}, we consider
both a version with leverage and one version without, which we use
the acronym ``NL'' indicating no-leverage to distinguish the model
with leverage from the model without leverage, see Table \ref{Table 1}
for more details. Estimation of t-EGARCH and Beta-t-EGARCH (semiparametric
EGARCH) models is routinely performed by (quasi) Maximum Likelihood,
ML, (QML). The particle marginal Metropolis-Hastings (PMMH) sampler
of Flury and Shephard (2011) is a computationally and efficient technique
for obtaining an unbiased estimate of the likelihood function of nonlinear
non-Gaussian state-space models. In this paper, we implement the same
particle filter algorithm as Flury and Shephard (2011). Compared to
other techniques such as maximum simulated likelihood, MSL, PMMH has
several nice properties: First, the relative variance of the estimated
likelihood only increases linearly in time, which makes it work even
for large time-series. Second, contrary to MSL, where the number of
particles, $M$, plays an important role in determining the asymptotic
properties of $\theta$, the only concern with regards to $M$ in
our context is to choose $M$ to ensure that the likelihood estimate
is not too jittery, such that the chain does not get stuck at a particular
value, see Figure 3 of Flury and Shephard (2011). In this paper, we
find that $M=100$ works very well providing us with reasonable MH
acceptance ratios.

One final question is whether estimation output from ML based observation-driven
and Bayesian PMMH based parameter-driven models can be compared. We
refer the reader to Hoogerheide et al. (2012), where it is shown that
density predictions delivered by models estimated in a classical or
Bayesian framework are indeed comparable.

\section{Empirical Results \label{sec: Empirical Results}}

We use the models in Table \ref{Table 1} to obtain and evaluate $h=1$,
$5$ and $20$ days ahead forecasts. Each forecast is based on a re-estimation
of the underlying model using a rolling window of $1000$ observations,
which corresponds to roughly $4$ years of data. In other words, at
each step, as a new observation arrives the model is re-estimated
and a density forecast $h$ periods ahead is computed using the recursive
method of forecasting, see Marcellino et al. (2006). For DJ data,
the out-of-sample period consists of $29921$ observations running
from $7$ July $1905$ until $15$ April of $2016$.

As previously, mentioned, the context of S\&P $500$ equity returns
(given that we consider $432$ return-series) in order to reduce the
computational burden, we re-estimate the models every $40$ days instead
of each day. In other words, the parameters are fixed within the $40$
days window, and only the data are updated. The out-of-sample period
consists of $1768$ observations running from $24$ December $2007$
until $31$ December, $2014$. The hypothesis of equal predictive
ability between different forecasts is tested using the procedure
of Diebold and Mariano (1995).

\subsection{Forecast evaluation methodology \label{sub: Forecast evaluation methodology}}

We evaluate density forecasts from our models based on the weighted
Continuous Ranked Probability Score (wCRPS), introduced in Gneiting
and Ranjan (2011). wCRPS circumvents some of the drawbacks of the
usually employed log-score (the logarithm of the predictive density),
as log-score does not reward values from the predictive density that
are close but not equal to the actual realized value, see Gneiting
and Raftery (2007) for more details. Log-score it is also very sensitive
to outliers, which is a common feature of financial data. Furthermore,
Gneiting and Ranjan (2011) show that it is invalid to use weighted
log-scores to emphasize certain areas of the distribution.

The wCRPS for a model $i$ measures the average absolute distance
between the empirical cumulative distribution function (CDF) of $y_{t+h}$,
which is simply a step function in $y_{t+h}$, and the predicted CDF
that is associated with model $i$\textquoteright s predictive density.
Furthermore, the comparison between the empirical and the predicted
CDF can also be weighed by a function that emphasizes particular regions
of interest, for example the center or the tails of the predictive
density. We define wCRPS for model $i$ at at time $t+h$ conditional
on information at time $t$ as
\begin{eqnarray}
wCRPS_{t+h\mid t}^{i} & = & \int_{-\infty}^{\infty}w\left(z\right)\left(\hat{F}_{t+h\mid t}^{i}\left(z\right)-I_{\left(y_{t+h}<z\right)}\right)^{2}dz,\label{eq:4.1}
\end{eqnarray}
where $w\left(z\right)$ is the weight function and $\hat{F}_{t+h\mid t}^{i}\left(z\right)$
is the $h$-step ahead cumulative density function of model $i$,
evaluated at $z$. The simplest case is $w\left(z\right)=1$, which
we refer to as ``uniform''. As its name suggests, when $w\left(z\right)=1$,
we put the same amount of weight on each region of the predictive
density, see also Gneiting and Ranjan (2011). Besides, $w\left(z\right)=1$,
we also consider different alternatives formulations of $w\left(z\right)$,
see Table \ref{Table 2}. This way, we are also able to focus on the
different parts of the distribution and better understand where the
eventual improvements of one model over another comes from. However,
(\ref{eq:4.1}) is not available in closed form. Therefore, we use
the approximation
\begin{eqnarray}
wCRPS_{t+h\mid t}^{i} & \approx & \frac{y_{u}-y_{l}}{K-1}\sum_{k=1}^{K}w\left(y_{k}\right)\left(\hat{F}_{t+h\mid t}^{i}\left(y_{k}\right)-I_{\left(y_{t+h<}y_{k}\right)}\right)^{2},\label{eq:4.2}
\end{eqnarray}
where
\begin{eqnarray}
y_{k} & = & y_{l}+k\frac{y_{u}-y_{l}}{K}.\label{eq:4.3}
\end{eqnarray}
In (\ref{eq:4.2}), $y_{u}$ and $y_{l}$ are the upper and lower
values, which defines the range of integration. The accuracy of the
approximation can be increased to any desired level by $K$. In this
paper, we set $y_{l}=-100$, $y_{u}=100$ and $K=1000$, which work
well for daily returns in percentage points. The model with lower
average wCRPS is always preferred. In other words, if the ratio of
wCRPS for model $i$ over $j$ is greater then one, model $j$ is
preferred to model $i$ and vice versa.

\subsection{Dow Jones \label{sub: Dow Jones}}

We begin the analysis with a pairwise model comparison, where we compare
the leverage model with the model without the leverage effect. We
report the ratios of wCRPS for the model without the leverage effect
over the version with the leverage effect in Table \ref{Table 3}.
For instance, the column ``t-EGARCH'' reports the average wCRPS
of t-EGARCH over t-EGARCH-NL for various choices of $w\left(z\right)$,
see Table \ref{Table 2} at $h=1$, $5$ and $20$. Obviously, when
$w\left(z\right)=1$, which we label as uniform, $w\left(z\right)$
weights equally across the conditional distribution. In this case,
for each model-type, the version which accounts for the leverage effect
is able to generate statistically significant more accurate density
forecasts than the version without the leverage effect. At $h=1$,
on average, we obtain reductions in wCRPS around $3\%$ to $5\%$.
At $h=5$ and $h=20$, incorporating the leverage effect does not
result in any major changes for Beta-t-EGARCH and SV. On the contrary,
Beta-t-EGARCH-NL significantly outperforms Beta-t-EGARCH at $h=20$.
Conversely, t-EGARCH and SPEGARCH dominate their no leverage counterpart
by more than $10\%$ at $h=20$.

Next, we experiment with different $w\left(z\right)$, see Table \ref{Table 1}
fore more details. We observe that models with the leverage effect
are particularly able to predict the tails of the conditional distribution
better than the models without the leverage effect. This is very evident
at $h=1$. Conversely, we obtain less improvements from the center.
This is particularly notable for SV at $h=1$ as compared to SV-NL,
where we obtain improvements of around $8\%$ at the tails compared
to $5\%$ when $w\left(z\right)=1$ or $3\%$ when we focus on the
center of the conditional distribution. At $h=5$ and $h=20$, we
continue to observe similar trends, however, the improvements are
of smaller magnitudes. For Beta-t-EGARCH and SV, the trend is reversed.
Similar to when $w\left(z\right)=1$, t-EGARCH and SPEGARCH are able
to predict the tails and the center of the conditional return distribution
significantly better than t-EGARCH-NL and SPEGARCH-NL as we increase
$h$.

The connection between the business cycle and relative forecast performance
may shed light on the sources of the predictive power from the model
with the leverage effect. Let $\triangle wCRPS$  denote the difference
between the wCRPS for the model without the leverage effect and the
wCRPS for the model which accounts for the leverage effect. Each panel
in Figure \ref{Figure 1} reports the cumulative $\triangle wCRPS$
over the out-of-sample period at $h=1$. Periods when the line slopes
upward represent periods in which the model with the leverage effect
outperforms the model without the leverage effect, while downward-sloping
segments indicate the opposite. The blue vertical bars indicate business
cycle peaks, i.e., the point at which an economic expansion transitions
to a recession based on NBER business cycle dating. Intuitively, one
would expect these plots to trend steadily downward during tranquil
periods as additional estimation error associated with the more heavily
parameterized leverage model increases the wCRPS relative to the model
without the leverage effect. The opposite is of course expected during
high volatility periods as neglecting the leverage effect ought to
increase the wCRPS for the model without the leverage effect.

Figure \ref{Figure 1} shows that the gains from the models with the
leverage effect are typically concentrated near the highest point
between the end of an economic expansion and the start of a contraction
(peak) and the period marking the end of a period of declining economic
activity and the transition to expansion (trough). The former is very
evident during the Great Recession of $2008$, where we obtain very
notable gains in favor of the model with the leverage effect. The
latter is very evident, with regards to the period after the recession
in the early $1980$s and during the mid $2000$s. The models with
leverage and without the leverage effect generate similar density
forecasts during tranquil periods. For instance, the lines are flat
during the Great Moderation before increasing towards the beginning
of $2000$s. Evidently, the additional error associated with estimating
the more complex leverage model in periods of economic turmoil does
not necessarily result in worse density forecasts. t-EGARCH and SPEGARCH
demonstrate similar patterns as Beta-t-EGARCH. We also find some differences
between models that account for the leverage effect and their relative
performance. Except for the Great Depression and early $2000$s till
the end of the sample Beta-t-EGARCH and Beta-t-EGARCH-NL generate
similar density forecasts. Thus, the better performance of Beta-t-EGARCH
in Table \ref{Table 3} is predominantly due to the model's ability
to generate more accurate density forecasts towards the end of the
sample. Beta-t-EGARCH's ability to predict the left tails of the conditional
distribution is also concentrated during the Great Recession. The
story is somewhat different for SV. Here, we observe large gains for
SV in the $1950$s and $1960$s.

Other interesting results come from comparison between different models.
In Table \ref{Table 4}, we report wCRPS using different $w\left(z\right)$
relative to t-EGARCH-NL. At $h=1$, all models except SV-NL and SPEGARCH-NL
generate statistically significant more accurate density forecasts
than the benchmark t-EGARCH-NL model. Beta-t-EGARCH is the top performer
followed by SV and t-EGARCH. Particularly, Beta-t-EGARCH is able to
predict the tails of the conditional distribution better than the
other models. At $h=5$, Beta-t-EGARCH, SV and Beta-t-EGARCH-NL are
able generate more accurate density forecasts than the remaining models.
SV-NL also outperforms t-EGARCH and SPEGARCH. However, for Beta-t-EGARCH
and SV, we observe less gains compared to Beta-t-EGARCH-NL and SV-NL.
Moreover, Beta-t-EGARCH-NL is able to generate very similar density
forecasts as SV at $h=5$ and does even better at $h=20$, especially
for the tails. Generally, compared to Beta-t-EGARCH and SV, t-EGARCH
and SPEGARCH are not able to generate more accurate density forecasts.
When $w\left(z\right)=1$, t-EGARCH and SPEGARCH outperform t-EGARCH-NL
by about $2\%$ at $h=1$, $4\%$ at $h=5$ and $13\%$ at $h=20$.

In Appendix \ref{sec: Supplementary results}, we repeat our forecasting
exercise, however, we focus on weekly DJ returns from $1921$ to $2016$
with $h=1$, $4$ and $12$, i.e., one week, one month and one semester,
see Table \ref{Table 14}. At $h=1$, pairwise model comparison shows
that models with the leverage effect generate more accurate density
forecasts than the models without the leverage effect. When $w\left(z\right)=1$,
Beta-t-EGARCH is the top performer, however, by only $1\%$ relative
to t-EGARCH, SPEGARCH and SV. t-EGARCH, Beta-t-EGARCH-NL, Beta-t-EGARCH
and SV are able generate statistically more accurate forecast than
t-EGARCH-NL in predicting the center and the left-tail of the conditional
distribution. At $h=4$, t-EGARCH-NL (SPEGARCH-NL) and t-EGARCH (SPEGARCH)
generate similar density forecasts, whereas Beta-t-EGARCH-NL and SV-NL
outperform Beta-t-EGARCH and SV, respectively. Thus, the leverage
model's ability to outperform its no-leverage counterpart seems to
be frequency-dependent. We observe the same pattern for these models
at $h=12$, whereas t-EGARCH and SPEGARCH generate more accurate density
forecasts than their non-leverage counterpart. At $h=4$ and $12$,
Beta-t-EGARCH-NL is the top performer, outperforming Beta-t-EGARCH
by about $2\%$. Finally, we also compute $\triangle wCRPS$ for the
specification without leverage relative to the specification with
the leverage effect for each model-type. Here, we obtain a slightly
different relationship between the leverage effect and the business
cycles. Indeed, we actually observe a downward sloping line during
the Great Moderation. Except for $2008$, upwards level-shifts in
$\triangle wCRPS$ are less apparent in periods of market distress
and recessions, see Figure \ref{Figure 2}.

\subsection{S\&P 500 \label{sub: S=000026P 500}}

As previously mentioned, we also consider shorter datasets from $2004$
to $2014$ using a cross sectional dimension of firms from the S\&P
$500$ index. We consider the same models as the previous section
and generate density forecasts at $h=1$, $h=5$ and $h=20$. The
out-of-sample period runs from $24$ December $2007$ until $31$
December $2014$. In Table \ref{Table 6}, we report the percentages
in which the model with the leverage effect generates more accurate
density forecasts than the model without the leverage effect for each
model-type. In the parentheses, we report the percentages where these
improvements are statically significant according to Diebold and Mariano
(1995). Evidently, for each model-type, the specification that accounts
for the leverage effect is generally able to generate more accurate
density forecasts than the model without the leverage effect. Similar
to Section \ref{sub: Dow Jones}, the improvements from considering
the leverage effect play a minor role with regards to Beta-t-EGARCH
and SV as we increase $h$, whereas the reverse is true for t-EGARCH
and SPEGARCH. We also experiment by changing $w\left(z\right)$ according
to Table \ref{Table 2}. Here, we observe very interesting results.
When $w\left(z\right)=\phi\left(z\right)$, $w\left(z\right)=1-\phi\left(z\right)/\phi\left(0\right)$
and $w\left(z\right)=\Phi\left(z\right)$, we generally obtain similar
results as when $w\left(z\right)=1$. However, when $w\left(z\right)=1-\Phi\left(z\right)$,
i.e. the left-tail, we find that the percentages are higher than the
remaining cases. This is primarily due to the fact that compared to
DJ, equity return series contain more frequent negative extreme observations.

Given that the models that account for leverage effect pairwise outperform
their non-leverage counterparts, we proceed to compare density forecasts
between the leverage models. In Tables \ref{Table 7} and \ref{Table 8},
we report the percentages of time where each leverage model is able
to generate more accurate density forecast than the other leverage
models when $w\left(z\right)=1$ and $w\left(z\right)=1-\Phi\left(z\right)$.
Results for when $w\left(z\right)=\phi\left(z\right)$, $w\left(z\right)=1-\phi\left(z\right)/\phi\left(0\right)$
and $w\left(z\right)=\Phi\left(z\right)$ are reported in Tables \ref{Table 15}
to \ref{Table 17} of Appendix \ref{sec: Supplementary results}.
Here, we find that Beta-t-EGARCH is the clear winner. Regardless of
$h$, Beta-t-EGARCH is able to generate more accurate density forecasts
than the other leverage models. When $w\left(z\right)=1-\Phi\left(z\right)$,
which is one of the interesting cases, results again confirm the superior
performance of Beta-t-EGARCH, see Table \ref{Table 8}. However, selecting
the second best model is more difficult. At $h=1$, t-EGARCH and SV
generate similar percentages. At the same time, they both tend to
outperform SPEGARCH. As we increase $h$, we find that SV tends to
outperform t-EGARCH and SPEGARCH. For instance, at $h=5$, SV outperforms
t-EGARCH and SPEGARCH in the majority of cases. At $h=20$, this percentage
increases to $76$ and $73$, respectively.

Next, we analyze the difference between the model that accounts for
the leverage effect and the model without the leverage effect by decomposing
equities according to sectors reported in Table \ref{Table 5}. In
Tables \ref{Table 9} and \ref{Table 10}, we report results for $w\left(z\right)=1$
and $w\left(z\right)=1-\Phi\left(z\right)$. Results for when $w\left(z\right)=\phi\left(z\right)$,
$w\left(z\right)=1-\phi\left(z\right)/\phi\left(0\right)$ and $w\left(z\right)=\Phi\left(z\right)$
are reported in Appendix \ref{sec: Supplementary results}. Overall,
we observe similar trends as Table \ref{Table 6} even when we consider
equities at sector-level. The models that account for the leverage
effect on average are able to generate more accurate density forecasts
than the models without the leverage effect. Beta-t-EGARCH and SV
generate more accurate density forecasts than Beta-t-EGARCH-NL and
SV-NL with very high percentages. However, the percentages decrease
as we increase $h$. For t-EGARCH and SPEGARCH, the reverse is the
case. We also observe some variations within sector and forecast horizons.
For example, the percentages in favor of Beta-t-EGARCH and SV are
generally lower in real estate and telecommunications, whereas the
percentages are higher in financials and healthcare. When $w\left(z\right)=1-\Phi\left(z\right)$,
we again observe that models that account for the leverage effect
are able to predict the left tail of the conditional return distribution
better than the model without the leverage effect.

In Tables \ref{Table 11} to \ref{Table 13}, we report the wCRPS
for twenty equities (the exact same equities as in Koopman et al.
(2016)) from the S\&P $500$ at $h=1$, $h=5$ and $h=20$ relative
to t-EGARCH. In order to save space, we report results for $w\left(z\right)=1$,
with similar results for the other cases. Generally, for each model,
the specification with the leverage effect tends to outperform the
one without the leverage effect. There are also cases, where the specification
without the leverage effect is able to produce more accurate density
forecasts than the model with the leverage effect, see for example,
Boeing and Caterpillar for SV at $h=1$. However, compared to t-EGARCH,
SPEGARCH and SV, Beta-t-EGARCH delivers the most consistent pattern,
i.e. Beta-t-EGARCH consistently generates more accurate density forecasts
than Beta-t-EGARCH-NL.

Similar to results reported in Section \ref{sub: Dow Jones}, other
interesting results are also obtained from comparison between model-types.
Here, we find that Beta-t-EGARCH is the top performer. At $h=1$,
Beta-t-EGARCH outperforms the other models that account for the leverage
effect by about $4\%$ to $10\%$. For instance, for Chevron Beta-t-EGARCH
outperform SV by $6\%$ and t-EGARCH by $4\%$. On the other hand,
the gains in favor of Beta-t-EGARCH compared to SV and t-EGARCH are
around $10\%$ for General Electric. At $h=1$, in most cases, the
second best model is Beta-t-EGARCH-NL. Indeed, among our twenty returns
series, Beta-t-EGARCH-NL outperforms t-EGARCH, SPEGARCH and SV models.
t-EGARCH is able to generate statistically significant more accurate
forecasts than t-EGARCH-NL, whereas SV and SV-NL generate similar
forecasts. At $h=5$ and $h=20$, we find that Beta-t-EGARCH still
is the top performer followed very closely by Beta-t-EGARCH-NL. At
the same time, the magnitude of the improvements over t-EGARCH and
SV increases. For instance, for Chevron Beta-t-EGARCH outperform SV
by $8\%$ and t-EGARCH by $6\%$ at $h=5$. For General Electric,
the gains are about $10\%$ and $14\%$ over SV and t-EGARCH at $h=5$.
We also find that, as we increase $h$, SV and SV-NL both are able
to generate statistically significant more accurate density forecasts
than t-EGARCH and t-EGARCH-NL. Finally, SPEGARCH tends to generate
similar forecasts as t-EGARCH.

\subsection{Summary \label{sub: Summary}}

So what do we learn from results reported in Sections \ref{sub: Dow Jones}
and \ref{sub: S=000026P 500}?
\begin{itemize}
\item For each model-type: (\ref{eq:2.1})-(\ref{eq:2.2}), (\ref{eq:2.3})-(\ref{eq:2.4}),
(\ref{eq:2.5})-(\ref{eq:2.6}) and (\ref{eq:2.7})-(\ref{eq:2.8}),
the specification with the leverage effect is able to generate more
accurate density forecasts than the specification without the leverage
effect. Beta-t-EGARCH and SV perform relatively better than their
no-leverage counterparts at $h=1$ and $h=5$ compared to $h=20$,
whereas we observe the opposite trend for t-EGARCH and SPEGARCH. Thus,
parametric specification and how we choose to incorporate the leverage
effect impacts out-of-sample performance at different forecast horizons.
Moreover, we find that the specification that accounts for the leverage
effect is able to predict the tails and in some cases also the center
of the conditional return distribution significantly better than the
no-leverage model. The later point is very evident when we consider
equities from the S\&P $500$ index, where for each model-type, the
specification that considers the leverage effect is able to predict
the left tail of the conditional distribution of returns considerably
better than the model without the leverage effect.
\item There is relationship between business cycles and the leverage effect.
Predictive gains from the model with the leverage effect are concentrated
near the highest point between the end of an economic expansion and
the start of a contraction (peak) and the period marking the end of
a period of declining economic activity and the transition to expansion
(trough). At the daily frequency, the models with leverage and without
the leverage effect generate similar density forecasts during tranquil
periods. Evidently, the additional estimation error associated with
estimating the more complex leverage model in tranquil periods does
not necessarily result in worse density forecasts, which means that
we can be less concerned about the bias-variance trade-off. When we
decrease the data frequency to weekly observations, adding the leverage
effect has a negative impact during tranquil periods. Furthermore,
the cumulative wCRPSs also reveal that the benefits of accounting
for the leverage effect decrease as we decrease data frequency. In
other words, the impact of accounting for the leverage effect depends
on the frequency of the data. Compared to daily frequency, in most
instances, the no-leverage model is able to predict the left tail
just as well as the model with the leverage effect.
\item Results indicate that Beta-t-EGARCH is the top performer. We also
find that in some cases, Beta-t-EGARCH-NL also outperforms t-EGARCH,
SPEGARCH and SV. In other words, besides accounting for the leverage
effect, how we choose to specify the parametric specification and
evolution of the conditional volatility process can play just as an
important role as the leverage effect with regards to generating accurate
density forecasts.
\item A practitioner is interested in being able to (i): Determine which
model generates the most accurate density forecasts, (ii): Perform
recursive model estimation in a rather parsimonious way, that is being
able to obtain parameter estimates for every out-of-sample observation,
while maintaining a reasonable computation time. Taking (i) and (ii)
into consideration, results indicate that Beta-t-EGARCH is the preferred
model.
\end{itemize}

\section{Conclusion \label{sub: Conclusion}}

\noindent In this paper, we examine the role of the leverage effect
with regards to generating accurate density forecasts of returns using
well-known observation and parameter-driven volatility models. These
models differ in their assumptions regarding: The parametric specification,
evolution of the conditional volatility process and how the leverage
effect is incorporated in the model.

Considering daily Dow Jones and more than four hundred equities from
the S\&P $500$ index, we find that models with the leverage effect
generally generate statistically significant more accurate density
forecasts compared to their no-leverage counterparts. Predictive gains
from the models with the leverage effect are concentrated near the
on onset of recessions and the period marking the end of a period
of declining economic activity and the transition to expansion. A
comparison between volatility models shows that Beta-t-EGARCH is the
top performer, regardless of forecast horizon. We also find that,
in some cases Beta-t-EGARCH-NL also outperforms t-EGARCH, SPEGARCH
and SV. In other words, besides accounting for the leverage effect,
how we choose to specify the parametric specification and evolution
of the conditional volatility process is also important with regards
to generate accurate density forecasts. Overall, we recommend Beta-t-EGARCH
as it performs best while at the same time maintaining a reasonable
computation time.

\newpage{}

\begin{center}
\begin{table}[H]
{\footnotesize{}\caption{\label{Table 1} Models and description}
}{\footnotesize \par}

\begin{centering}
\setlength\tabcolsep{3.8pt}{\footnotesize{}}%
\begin{tabular}{ll}
\hline
{\footnotesize{}Model} & {\footnotesize{}description}\tabularnewline
\hline
{\footnotesize{}t-EGARCH} & {\footnotesize{}t-EGARCH model. }\tabularnewline
{\footnotesize{}t-EGARCH-NL} & {\footnotesize{}t-EGARCH without the leverage effect, i.e. $\gamma=0$
in the estimation procedure.}\tabularnewline
{\footnotesize{}Beta-t-EGARCH} & {\footnotesize{}Beta-t-EGARCH model.}\tabularnewline
{\footnotesize{}Beta-t-EGARCH-NL} & {\footnotesize{}Beta-t-EGARCH without the leverage effect, i.e. $\gamma=0$
in the estimation procedure.}\tabularnewline
{\footnotesize{}SPEGARCH} & {\footnotesize{}semiparametric EGARCH model.}\tabularnewline
{\footnotesize{}SPEGARCH-NL } & {\footnotesize{}SPEGARCH without the leverage effect, i.e. $\gamma=0$
in the estimation procedure.}\tabularnewline
{\footnotesize{}SV} & {\footnotesize{}Stochastic volatility model.}\tabularnewline
{\footnotesize{}SV-NL} & {\footnotesize{}Stochastic volatility model without the leverage effect,
i.e. $\rho=0$ in the estimation procedure.}\tabularnewline
\hline
\end{tabular}
\par\end{centering}{\footnotesize \par}

\begin{centering}
\vspace*{0.15cm}
\par\end{centering}

{\footnotesize{}This table lists the model labels together with a
brief description of the models. The acronym ``NL'' denotes ``no
leverage''.}{\footnotesize \par}

\vspace{-0.45in}
\end{table}
\vspace{-0.20in}
\par\end{center}

\begin{center}
\begin{table}[H]
{\footnotesize{}\caption{\label{Table 2} Weight functions for weighted Continuous Ranked Probability
Score, wCRPS}
}{\footnotesize \par}

\begin{centering}
\setlength\tabcolsep{77.8pt}{\footnotesize{}}%
\begin{tabular}{ll}
\hline
{\footnotesize{}Emphasis} & {\footnotesize{}weight function}\tabularnewline
\hline
{\footnotesize{}Uniform} & {\footnotesize{}$w\left(z\right)=1$}\tabularnewline
{\footnotesize{}Center} & {\footnotesize{}$w\left(z\right)=\phi\left(z\right)$}\tabularnewline
{\footnotesize{}Tails} & {\footnotesize{}$w\left(z\right)=1-\phi\left(z\right)/\phi\left(0\right)$}\tabularnewline
{\footnotesize{}Right tail (tail-r)} & {\footnotesize{}$w\left(z\right)=\Phi\left(z\right)$}\tabularnewline
{\footnotesize{}Left tail (tail-l)} & {\footnotesize{}$w\left(z\right)=1-\Phi\left(z\right)$}\tabularnewline
\hline
\end{tabular}
\par\end{centering}{\footnotesize \par}

\begin{centering}
\vspace*{0.15cm}
\par\end{centering}

{\footnotesize{}This table reports the weight functions for wCRPS.
$\phi\left(z\right)$ and $\Phi\left(z\right)$ denote the pdf and
cdf of a $N\left(0,1\right)$ distribution.}{\footnotesize \par}

\vspace{1.15in}
\end{table}
\vspace{-0.20in}
\par\end{center}

\begin{center}
\begin{table}[H]
{\footnotesize{}\caption{\label{Table 3} Pairwise density forecast comparison using daily
Dow Jones returns}
}{\footnotesize \par}

\begin{centering}
\setlength\tabcolsep{23.3pt}{\footnotesize{}}%
\begin{tabular}{lllll}
\hline
{\footnotesize{}Model} & {\footnotesize{}t-EGARCH} & {\footnotesize{}Beta-t-EGARCH } & {\footnotesize{}SPEGARCH} & {\footnotesize{}SV}\tabularnewline
\hline
{\footnotesize{}$h=1$} &  &  &  & \tabularnewline
{\footnotesize{}Uniform } & {\footnotesize{}0.997$^{\left(a\right)}$ } & {\footnotesize{}0.997$^{\left(a\right)}$ } & {\footnotesize{}0.997$^{\left(a\right)}$ } & {\footnotesize{}0.995$^{\left(a\right)}$ }\tabularnewline
{\footnotesize{}Center } & {\footnotesize{}0.998$^{\left(a\right)}$ } & {\footnotesize{}0.998$^{\left(a\right)}$ } & {\footnotesize{}0.998$^{\left(a\right)}$ } & {\footnotesize{}0.997$^{\left(a\right)}$ }\tabularnewline
{\footnotesize{}Tails } & {\footnotesize{}0.994$^{\left(a\right)}$ } & {\footnotesize{}0.995$^{\left(a\right)}$ } & {\footnotesize{}0.994$^{\left(a\right)}$ } & {\footnotesize{}0.992$^{\left(a\right)}$ }\tabularnewline
{\footnotesize{}Tail-r } & {\footnotesize{}0.996$^{\left(a\right)}$ } & {\footnotesize{}0.996$^{\left(a\right)}$ } & {\footnotesize{}0.995$^{\left(a\right)}$ } & {\footnotesize{}0.994$^{\left(a\right)}$ }\tabularnewline
{\footnotesize{}Tail-l } & {\footnotesize{}0.998$^{\left(a\right)}$ } & {\footnotesize{}0.998$^{\left(a\right)}$ } & {\footnotesize{}0.998$^{\left(a\right)}$ } & {\footnotesize{}0.997$^{\left(a\right)}$ }\tabularnewline
{\footnotesize{}$h=5$} &  &  &  & \tabularnewline
{\footnotesize{}Uniform } & {\footnotesize{}0.996$^{\left(a\right)}$ } & {\footnotesize{}0.999$^{\left(b\right)}$ } & {\footnotesize{}0.995$^{\left(a\right)}$ } & {\footnotesize{}0.999$^{\left(a\right)}$ }\tabularnewline
{\footnotesize{}Center } & {\footnotesize{}0.997$^{\left(a\right)}$ } & {\footnotesize{}1.000} & {\footnotesize{}0.997$^{\left(a\right)}$ } & {\footnotesize{}0.999$^{\left(a\right)}$ }\tabularnewline
{\footnotesize{}Tails } & {\footnotesize{}0.993$^{\left(a\right)}$ } & {\footnotesize{}0.998$^{\left(a\right)}$ } & {\footnotesize{}0.992$^{\left(a\right)}$ } & {\footnotesize{}0.997$^{\left(a\right)}$ }\tabularnewline
{\footnotesize{}Tail-r } & {\footnotesize{}0.995$^{\left(a\right)}$ } & {\footnotesize{}0.999$^{\left(c\right)}$ } & {\footnotesize{}0.995$^{\left(a\right)}$ } & {\footnotesize{}0.998$^{\left(a\right)}$ }\tabularnewline
{\footnotesize{}Tail-l } & {\footnotesize{}0.996$^{\left(a\right)}$ } & {\footnotesize{}0.999$^{\left(c\right)}$ } & {\footnotesize{}0.995$^{\left(a\right)}$ } & {\footnotesize{}0.999$^{\left(a\right)}$ }\tabularnewline
{\footnotesize{}$h=20$} &  &  &  & \tabularnewline
{\footnotesize{}Uniform } & {\footnotesize{}0.986$^{\left(a\right)}$ } & {\footnotesize{}1.001$^{\left(a\right)}$ } & {\footnotesize{}0.985$^{\left(a\right)}$ } & {\footnotesize{}0.999$^{\left(a\right)}$ }\tabularnewline
{\footnotesize{}Center } & {\footnotesize{}0.987$^{\left(a\right)}$ } & {\footnotesize{}1.001$^{\left(a\right)}$ } & {\footnotesize{}0.986$^{\left(a\right)}$ } & {\footnotesize{}0.999$^{\left(a\right)}$ }\tabularnewline
{\footnotesize{}Tails } & {\footnotesize{}0.983$^{\left(a\right)}$ } & {\footnotesize{}1.001$^{\left(a\right)}$ } & {\footnotesize{}0.983$^{\left(a\right)}$ } & {\footnotesize{}0.999 }\tabularnewline
{\footnotesize{}Tail-r } & {\footnotesize{}0.986$^{\left(a\right)}$ } & {\footnotesize{}1.001$^{\left(a\right)}$ } & {\footnotesize{}0.986$^{\left(a\right)}$ } & {\footnotesize{}0.999$^{\left(b\right)}$ }\tabularnewline
{\footnotesize{}Tail-l } & {\footnotesize{}0.986$^{\left(a\right)}$ } & {\footnotesize{}1.001$^{\left(a\right)}$ } & {\footnotesize{}0.984$^{\left(a\right)}$ } & {\footnotesize{}1.000$^{\left(c\right)}$ }\tabularnewline
\hline
\end{tabular}
\par\end{centering}{\footnotesize \par}

\begin{centering}
\vspace*{0.15cm}
\par\end{centering}

{\footnotesize{}This table reports the average wCRPS for each model-type
according to different weights, $w\left(z\right)$, (see Table \ref{Table 2})
for the version that considers the leverage effect relative to the
version that does not consider the leverage effect.  For instance,
the column \textquotedblleft t-EGARCH\textquotedblright{} reports
the average wCRPS for t-EGARCH over t-EGARCH-NL. The apexes $a$,
$b$, and $c$ indicate rejection of the null-hypothesis of equal
predictive ability according to the Diebold and Mariano (1995) test
at $1\%$, $5\%$ and $10\%$, respectively. The out-of-sample period
consists of $29921$ observations from $7$ July $1905$ until $15$
April of $2016$.}{\footnotesize \par}

\vspace{-0.45in}
\end{table}
\vspace{-0.20in}
\par\end{center}

\begin{center}
\begin{table}[H]
{\footnotesize{}\caption{\label{Table 4} Density forecast comparison using daily Dow Jones
returns}
}{\footnotesize \par}

\begin{centering}
\setlength\tabcolsep{2.6pt}{\footnotesize{}}%
\begin{tabular}{llllllll}
\hline
{\footnotesize{}Model} & {\footnotesize{}t-EGARCH } & {\footnotesize{}Beta-t-EGARCH-NL} & {\footnotesize{}Beta-t-EGARCH } & {\footnotesize{}SPEGARCH-NL} & {\footnotesize{}SPEGARCH} & {\footnotesize{}SV-NL} & {\footnotesize{}SV }\tabularnewline
\hline
{\footnotesize{}$h=1$} &  &  &  &  &  &  & \tabularnewline
{\footnotesize{}Uniform } & {\footnotesize{}0.997$^{\left(a\right)}$ } & {\footnotesize{}0.998$^{\left(a\right)}$ } & {\footnotesize{}0.995$^{\left(a\right)}$ } & {\footnotesize{}1.001$^{\left(a\right)}$ } & {\footnotesize{}0.998$^{\left(a\right)}$ } & {\footnotesize{}1.001$^{\left(a\right)}$ } & {\footnotesize{}0.997$^{\left(a\right)}$ }\tabularnewline
{\footnotesize{}Center } & {\footnotesize{}0.998$^{\left(a\right)}$ } & {\footnotesize{}0.998$^{\left(a\right)}$ } & {\footnotesize{}0.996$^{\left(a\right)}$ } & {\footnotesize{}1.001$^{\left(b\right)}$ } & {\footnotesize{}0.999$^{\left(a\right)}$ } & {\footnotesize{}1.001$^{\left(a\right)}$ } & {\footnotesize{}0.998$^{\left(a\right)}$ }\tabularnewline
{\footnotesize{}Tails } & {\footnotesize{}0.994$^{\left(a\right)}$ } & {\footnotesize{}0.997$^{\left(a\right)}$ } & {\footnotesize{}0.991$^{\left(a\right)}$ } & {\footnotesize{}1.002$^{\left(b\right)}$ } & {\footnotesize{}0.996$^{\left(a\right)}$ } & {\footnotesize{}1.000 } & {\footnotesize{}0.993$^{\left(a\right)}$ }\tabularnewline
{\footnotesize{}Tail-r } & {\footnotesize{}0.996$^{\left(a\right)}$ } & {\footnotesize{}0.998$^{\left(a\right)}$ } & {\footnotesize{}0.994$^{\left(a\right)}$ } & {\footnotesize{}1.002$^{\left(a\right)}$ } & {\footnotesize{}0.997$^{\left(a\right)}$ } & {\footnotesize{}1.001$^{\left(a\right)}$ } & {\footnotesize{}0.996$^{\left(a\right)}$ }\tabularnewline
{\footnotesize{}Tail-l } & {\footnotesize{}0.998$^{\left(a\right)}$ } & {\footnotesize{}0.998$^{\left(a\right)}$ } & {\footnotesize{}0.995$^{\left(a\right)}$ } & {\footnotesize{}1.001} & {\footnotesize{}0.998$^{\left(c\right)}$ } & {\footnotesize{}1.001$^{\left(c\right)}$ } & {\footnotesize{}0.997$^{\left(a\right)}$ }\tabularnewline
{\footnotesize{}$h=5$} &  &  &  &  &  &  & \tabularnewline
{\footnotesize{}Uniform } & {\footnotesize{}0.996$^{\left(a\right)}$ } & {\footnotesize{}0.991$^{\left(a\right)}$ } & {\footnotesize{}0.990$^{\left(a\right)}$ } & {\footnotesize{}1.002$^{\left(a\right)}$ } & {\footnotesize{}0.997$^{\left(a\right)}$ } & {\footnotesize{}0.994$^{\left(a\right)}$ } & {\footnotesize{}0.992$^{\left(a\right)}$ }\tabularnewline
{\footnotesize{}Center } & {\footnotesize{}0.997$^{\left(a\right)}$ } & {\footnotesize{}0.994$^{\left(a\right)}$ } & {\footnotesize{}0.994$^{\left(a\right)}$ } & {\footnotesize{}1.001$^{\left(a\right)}$ } & {\footnotesize{}0.998$^{\left(a\right)}$ } & {\footnotesize{}0.996$^{\left(a\right)}$ } & {\footnotesize{}0.995$^{\left(a\right)}$ }\tabularnewline
{\footnotesize{}Tails } & {\footnotesize{}0.993$^{\left(a\right)}$ } & {\footnotesize{}0.985$^{\left(a\right)}$ } & {\footnotesize{}0.983$^{\left(a\right)}$ } & {\footnotesize{}1.002$^{\left(a\right)}$ } & {\footnotesize{}0.994$^{\left(a\right)}$ } & {\footnotesize{}0.988$^{\left(a\right)}$ } & {\footnotesize{}0.985$^{\left(a\right)}$ }\tabularnewline
{\footnotesize{}Tail-r } & {\footnotesize{}0.995$^{\left(a\right)}$ } & {\footnotesize{}0.991$^{\left(a\right)}$ } & {\footnotesize{}0.991$^{\left(a\right)}$ } & {\footnotesize{}1.001$^{\left(b\right)}$ } & {\footnotesize{}0.996$^{\left(a\right)}$ } & {\footnotesize{}0.993$^{\left(a\right)}$ } & {\footnotesize{}0.992$^{\left(a\right)}$ }\tabularnewline
{\footnotesize{}Tail-l } & {\footnotesize{}0.996$^{\left(a\right)}$ } & {\footnotesize{}0.991$^{\left(a\right)}$ } & {\footnotesize{}0.990$^{\left(a\right)}$ } & {\footnotesize{}1.002$^{\left(a\right)}$ } & {\footnotesize{}0.998$^{\left(a\right)}$ } & {\footnotesize{}0.994$^{\left(a\right)}$ } & {\footnotesize{}0.993$^{\left(a\right)}$ }\tabularnewline
{\footnotesize{}$h=20$} &  &  &  &  &  &  & \tabularnewline
{\footnotesize{}Uniform } & {\footnotesize{}0.986$^{\left(a\right)}$ } & {\footnotesize{}0.968$^{\left(a\right)}$ } & {\footnotesize{}0.969$^{\left(a\right)}$ } & {\footnotesize{}1.002$^{\left(a\right)}$ } & {\footnotesize{}0.987$^{\left(a\right)}$ } & {\footnotesize{}0.973$^{\left(a\right)}$ } & {\footnotesize{}0.973$^{\left(a\right)}$ }\tabularnewline
{\footnotesize{}Center } & {\footnotesize{}0.987$^{\left(a\right)}$ } & {\footnotesize{}0.973$^{\left(a\right)}$ } & {\footnotesize{}0.974$^{\left(a\right)}$ } & {\footnotesize{}1.002$^{\left(a\right)}$ } & {\footnotesize{}0.988$^{\left(a\right)}$ } & {\footnotesize{}0.977$^{\left(a\right)}$ } & {\footnotesize{}0.976$^{\left(a\right)}$ }\tabularnewline
{\footnotesize{}Tails } & {\footnotesize{}0.983$^{\left(a\right)}$ } & {\footnotesize{}0.958$^{\left(a\right)}$ } & {\footnotesize{}0.959$^{\left(a\right)}$ } & {\footnotesize{}1.002$^{\left(a\right)}$ } & {\footnotesize{}0.984$^{\left(a\right)}$ } & {\footnotesize{}0.965$^{\left(a\right)}$ } & {\footnotesize{}0.964$^{\left(a\right)}$ }\tabularnewline
{\footnotesize{}Tail-r } & {\footnotesize{}0.986$^{\left(a\right)}$ } & {\footnotesize{}0.968$^{\left(a\right)}$ } & {\footnotesize{}0.969$^{\left(a\right)}$ } & {\footnotesize{}1.000} & {\footnotesize{}0.986$^{\left(a\right)}$ } & {\footnotesize{}0.973$^{\left(a\right)}$ } & {\footnotesize{}0.972$^{\left(a\right)}$ }\tabularnewline
{\footnotesize{}Tail-l } & {\footnotesize{}0.986$^{\left(a\right)}$ } & {\footnotesize{}0.968$^{\left(a\right)}$ } & {\footnotesize{}0.969$^{\left(a\right)}$ } & {\footnotesize{}1.003$^{\left(a\right)}$ } & {\footnotesize{}0.988$^{\left(a\right)}$ } & {\footnotesize{}0.973$^{\left(a\right)}$ } & {\footnotesize{}0.973$^{\left(a\right)}$ }\tabularnewline
\hline
\end{tabular}
\par\end{centering}{\footnotesize \par}

\begin{centering}
\vspace*{0.15cm}
\par\end{centering}

{\footnotesize{}This table reports the average wCRPS using different
weights, $w\left(z\right)$, (see Table \ref{Table 2}) for models
relative to t-EGARCH-NL. The apexes $a$, $b$, and $c$ indicate
rejection of the null-hypothesis of equal predictive ability relative
to t-EGARCH-NL according to the Diebold and Mariano (1995) test at
$1\%$, $5\%$ and $10\%$, respectively. The out-of-sample period
consists of $29921$ observations from $7$ July $1905$ until $15$
April of $2016$.}{\footnotesize \par}

\vspace{-0.45in}
\end{table}
\vspace{-0.20in}
\par\end{center}

\begin{center}
\begin{table}[H]
{\footnotesize{}\caption{\label{Table 5} Information regarding sectors in the S\&P $500$
index}
}{\footnotesize \par}

\begin{centering}
\setlength\tabcolsep{15.4pt}{\footnotesize{}}%
\begin{tabular}{lllll}
\hline
{\footnotesize{}Sector} & {\footnotesize{}median market capitalization} & {\footnotesize{}total debt } & {\footnotesize{}leverage } & {\footnotesize{}\#}\tabularnewline
\hline
{\footnotesize{}Consumer discretionary} & {\footnotesize{}12.41} & {\footnotesize{}3.15 } & {\footnotesize{}77.85 } & {\footnotesize{}70}\tabularnewline
{\footnotesize{}Consumer staples} & {\footnotesize{}28.29 } & {\footnotesize{}6.64 } & {\footnotesize{}80.52 } & {\footnotesize{}32}\tabularnewline
{\footnotesize{}Energy} & {\footnotesize{}14.39 } & {\footnotesize{}6.97 } & {\footnotesize{}58.11 } & {\footnotesize{}32}\tabularnewline
{\footnotesize{}Financials} & {\footnotesize{}21.34 } & {\footnotesize{}7.58 } & {\footnotesize{}67.77 } & {\footnotesize{}56}\tabularnewline
{\footnotesize{}Health care} & {\footnotesize{}25.76 } & {\footnotesize{}6.36 } & {\footnotesize{}60.10 } & {\footnotesize{}53}\tabularnewline
{\footnotesize{}Industrials} & {\footnotesize{}15.46 } & {\footnotesize{}3.41 } & {\footnotesize{}75.59 } & {\footnotesize{}57}\tabularnewline
{\footnotesize{}Information technology} & {\footnotesize{}18.32 } & {\footnotesize{}2.63 } & {\footnotesize{}54.13} & {\footnotesize{}54}\tabularnewline
{\footnotesize{}Materials} & {\footnotesize{}14.90 } & {\footnotesize{}5.57 } & {\footnotesize{}109.65 } & {\footnotesize{}23}\tabularnewline
{\footnotesize{}Real Estate} & {\footnotesize{}18.73 } & {\footnotesize{}6.49 } & {\footnotesize{}116.64 } & {\footnotesize{}24}\tabularnewline
{\footnotesize{}Telecommunication services} & {\footnotesize{}16.57 } & {\footnotesize{}20.23 } & {\footnotesize{}143.85 } & {\footnotesize{}5}\tabularnewline
{\footnotesize{}Utilities} & {\footnotesize{}19.41 } & {\footnotesize{}13.88 } & {\footnotesize{}119.72 } & {\footnotesize{}26}\tabularnewline
\hline
\end{tabular}
\par\end{centering}{\footnotesize \par}

\begin{centering}
\vspace*{0.15cm}
\par\end{centering}

{\footnotesize{}This table reports the median market capitalization,
total debt and Leverage for the firms belonging to the S\&P $500$
index. The values refer to the last available annual report (fiscal
year $2015$). Market capitalization and total debt are in billions
while leverage is in percentage points. The last column, ``\#'',
reports the number of firms that belongs to each sector according
to the GICS classification scheme. Source: Datastream.}{\footnotesize \par}

\vspace{-0.45in}
\end{table}
\vspace{-0.20in}
\par\end{center}

\begin{center}
\begin{table}[H]
{\footnotesize{}\caption{\label{Table 6} Pairwise density forecast comparison using daily
S\&P $500$ equity returns}
}{\footnotesize \par}

\begin{centering}
\setlength\tabcolsep{23.2pt}{\footnotesize{}}%
\begin{tabular}{lllll}
\hline
{\footnotesize{}Model} & {\footnotesize{}t-EGARCH} & {\footnotesize{}Beta-t-EGARCH } & {\footnotesize{}SPEGARCH} & {\footnotesize{}SV}\tabularnewline
\hline
{\footnotesize{}$h=1$} &  &  & {\footnotesize{} } & \tabularnewline
{\footnotesize{}Uniform } & {\footnotesize{}82 (38)} & {\footnotesize{}89 (48)} & {\footnotesize{}74 (35) } & {\footnotesize{}89 (48)}\tabularnewline
{\footnotesize{}Center } & {\footnotesize{}77 (25)} & {\footnotesize{}79 (23)} & {\footnotesize{}69 (21) } & {\footnotesize{}85 (30)}\tabularnewline
{\footnotesize{}Tails } & {\footnotesize{}82 (40)} & {\footnotesize{}89 (51)} & {\footnotesize{}75 (36) } & {\footnotesize{}89 (44)}\tabularnewline
{\footnotesize{}Tail-r } & {\footnotesize{}46 (4)} & {\footnotesize{}57 (11)} & {\footnotesize{}39 (5) } & {\footnotesize{}61 (11)}\tabularnewline
{\footnotesize{}Tail-l } & {\footnotesize{}91 (58)} & {\footnotesize{}96 (58)} & {\footnotesize{}86 (52) } & {\footnotesize{}94 (54)}\tabularnewline
{\footnotesize{}$h=5$} &  &  & {\footnotesize{} } & \tabularnewline
{\footnotesize{}Uniform } & {\footnotesize{}88 (63)} & {\footnotesize{}70 (29)} & {\footnotesize{}84 (60) } & {\footnotesize{}76 (39)}\tabularnewline
{\footnotesize{}Center } & {\footnotesize{}78 (36)} & {\footnotesize{}62 (15)} & {\footnotesize{}77 (40)} & {\footnotesize{}72 (24)}\tabularnewline
{\footnotesize{}Tails } & {\footnotesize{}90 (67)} & {\footnotesize{}73 (34)} & {\footnotesize{}84 (64) } & {\footnotesize{}77 (39)}\tabularnewline
{\footnotesize{}Tail-r } & {\footnotesize{}71 (27)} & {\footnotesize{}59 (11)} & {\footnotesize{}72 (33) } & {\footnotesize{}66 (22)}\tabularnewline
{\footnotesize{}Tail-l } & {\footnotesize{}94 (68)} & {\footnotesize{}79 (26)} & {\footnotesize{}90 (61) } & {\footnotesize{}82 (30)}\tabularnewline
{\footnotesize{}$h=20$} &  &  & {\footnotesize{} } & \tabularnewline
{\footnotesize{}Uniform } & {\footnotesize{}99 (96)} & {\footnotesize{}51 (18)} & {\footnotesize{}94 (91) } & {\footnotesize{}45 (19)}\tabularnewline
{\footnotesize{}Center } & {\footnotesize{}99 (96)} & {\footnotesize{}33 (8)} & {\footnotesize{}95 (89) } & {\footnotesize{}38 (9)}\tabularnewline
{\footnotesize{}Tails } & {\footnotesize{}99 (95)} & {\footnotesize{}54 (24)} & {\footnotesize{}93 (90) } & {\footnotesize{}47 (23)}\tabularnewline
{\footnotesize{}Tail-r } & {\footnotesize{}97 (89)} & {\footnotesize{}44 (10)} & {\footnotesize{}92 (84) } & {\footnotesize{}41 (13)}\tabularnewline
{\footnotesize{}Tail-l } & {\footnotesize{}100 (96)} & {\footnotesize{}53 (13)} & {\footnotesize{}95 (89) } & {\footnotesize{}53 (16)}\tabularnewline
\hline
\end{tabular}
\par\end{centering}{\footnotesize \par}

\begin{centering}
\vspace*{0.15cm}
\par\end{centering}

{\footnotesize{}This table reports the percentages for each model-type,
where the version that considers the leverage effect generates more
accurate density forecasts than the version that does not consider
the leverage effect. The numbers in parentheses indicate rejection
of the null-hypothesis of equal predictive ability according to the
Diebold and Mariano (1995) test at $5\%$ level. The out-of-sample
period consists of $1768$ observations from $24$ December $2007$
until $31$ December, $2014$.}{\footnotesize \par}

\vspace{1.15in}
\end{table}
\vspace{-0.20in}
\par\end{center}

\begin{center}
\begin{table}[H]
{\footnotesize{}\caption{\label{Table 7} Density forecast comparison using daily S\&P $500$
equity returns, $w\left(z\right)=1$, i.e. uniform}
}{\footnotesize \par}

\begin{centering}
\setlength\tabcolsep{19.8pt}%
\begin{tabular}{lllll}
\hline
{\footnotesize{}Model} & {\footnotesize{}t-EGARCH} & {\footnotesize{}Beta-t-EGARCH} & {\footnotesize{}SPEGARCH} & {\footnotesize{}SV}\tabularnewline
\hline
{\footnotesize{}$h=1$} &  &  &  & \tabularnewline
{\footnotesize{}t-EGARCH} &  & {\footnotesize{}1 (0)} & {\footnotesize{}81 (41)} & {\footnotesize{}54 (25)}\tabularnewline
{\footnotesize{}Beta-t-EGARCH} & {\footnotesize{}99 (93) } &  & {\footnotesize{}99 (93)} & {\footnotesize{}98 (85)}\tabularnewline
{\footnotesize{}SPEGARCH} & {\footnotesize{}19 (2)} & {\footnotesize{}1 (0)} &  & {\footnotesize{}33 (12)}\tabularnewline
{\footnotesize{}SV} & {\footnotesize{}46 (23)} & {\footnotesize{}2 (0)} & {\footnotesize{}67 (35)} & \tabularnewline
{\footnotesize{}$h=5$} &  &  &  & \tabularnewline
{\footnotesize{}t-EGARCH} &  & {\footnotesize{}0 (0)} & {\footnotesize{}66 (41)} & {\footnotesize{}43 (27)}\tabularnewline
{\footnotesize{}Beta-t-EGARCH} & {\footnotesize{}100 (98)} &  & {\footnotesize{}99 (96)} & {\footnotesize{}99 (94)}\tabularnewline
{\footnotesize{}SPEGARCH} & {\footnotesize{}34 (13)} & {\footnotesize{}1 (0)} &  & {\footnotesize{}34 (23)}\tabularnewline
{\footnotesize{}SV} & {\footnotesize{}57 (40)} & {\footnotesize{}1 (0)} & {\footnotesize{}66 (45)} & \tabularnewline
{\footnotesize{}$h=20$} &  &  &  & \tabularnewline
{\footnotesize{}t-EGARCH} &  & {\footnotesize{}1 (0)} & {\footnotesize{}45 (30)} & {\footnotesize{}24 (13)}\tabularnewline
{\footnotesize{}Beta-t-EGARCH} & {\footnotesize{}99 (98)} &  & {\footnotesize{}98 (94)} & {\footnotesize{}100 (100)}\tabularnewline
{\footnotesize{}SPEGARCH} & {\footnotesize{}55 (39)} & {\footnotesize{}2 (0)} &  & {\footnotesize{}27 (18)}\tabularnewline
{\footnotesize{}SV} & {\footnotesize{}76 (67)} & {\footnotesize{}0 (0)} & {\footnotesize{}73 (62)} & \tabularnewline
\hline
\end{tabular}
\par\end{centering}

\begin{centering}
\vspace*{0.15cm}
\par\end{centering}

{\footnotesize{}Each row in this table reports the percentages, where
each model generates more accurate density forecasts relative to the
other models reported in each column. The numbers in parentheses indicate
rejection of the null-hypothesis of equal predictive ability according
to the Diebold and Mariano (1995) test at $5\%$ level. The out-of-sample
period consists of $1768$ observations from $24$ December $2007$
until $31$ December, $2014$.}{\footnotesize \par}

\vspace{-0.45in}
\end{table}
\vspace{-0.20in}
\par\end{center}

\begin{center}
\begin{table}[H]
{\footnotesize{}\caption{\label{Table 8} Density forecast comparison using daily S\&P $500$
equity returns, $w\left(z\right)=1-\Phi\left(z\right)$, i.e. left
tail}
}{\footnotesize \par}

\begin{centering}
\setlength\tabcolsep{19.9pt}{\footnotesize{}}%
\begin{tabular}{lllll}
\hline
{\footnotesize{}Model} & {\footnotesize{}t-EGARCH} & {\footnotesize{}Beta-t-EGARCH} & {\footnotesize{}SPEGARCH} & {\footnotesize{}SV}\tabularnewline
\hline
{\footnotesize{}$h=1$} &  &  &  & \tabularnewline
{\footnotesize{}t-EGARCH} &  & {\footnotesize{}4 (0)} & {\footnotesize{}82 (36)} & {\footnotesize{}59 (16)}\tabularnewline
{\footnotesize{}Beta-t-EGARCH} & {\footnotesize{}96 (49)} &  & {\footnotesize{}98 (63)} & {\footnotesize{}93 (55)}\tabularnewline
{\footnotesize{}SPEGARCH} & {\footnotesize{}18 (1)} & {\footnotesize{}2 (0)} &  & {\footnotesize{}36 (7)}\tabularnewline
{\footnotesize{}SV} & {\footnotesize{}41 (6)} & {\footnotesize{}7 (0)} & {\footnotesize{}64 (17)} & \tabularnewline
{\footnotesize{}$h=5$} &  &  & {\footnotesize{} } & \tabularnewline
{\footnotesize{}t-EGARCH} &  & {\footnotesize{}0 (0)} & {\footnotesize{}65 (32) } & {\footnotesize{}41 (20)}\tabularnewline
{\footnotesize{}Beta-t-EGARCH} & {\footnotesize{}100 (84)} &  & {\footnotesize{}100 (100)} & {\footnotesize{}98 (79)}\tabularnewline
{\footnotesize{}SPEGARCH} & {\footnotesize{}35 (10)} & {\footnotesize{}0 (0)} &  & {\footnotesize{}36 (17)}\tabularnewline
{\footnotesize{}SV} & {\footnotesize{}59 (30)} & {\footnotesize{}2 (0)} & {\footnotesize{}64 (39) } & \tabularnewline
{\footnotesize{}$h=20$} &  &  &  & \tabularnewline
{\footnotesize{}t-EGARCH} &  & {\footnotesize{}0 (0)} & {\footnotesize{}42 (25) } & {\footnotesize{}19 (6)}\tabularnewline
{\footnotesize{}Beta-t-EGARCH} & {\footnotesize{}100 (97)} &  & {\footnotesize{}100 (100)} & {\footnotesize{}100 (95)}\tabularnewline
{\footnotesize{}SPEGARCH} & {\footnotesize{}58 (38)} & {\footnotesize{}0 (0)} &  & {\footnotesize{}23 (12)}\tabularnewline
{\footnotesize{}SV} & {\footnotesize{}81 (67)} & {\footnotesize{}0 (0)} & {\footnotesize{}77 (61) } & \tabularnewline
\hline
\end{tabular}
\par\end{centering}{\footnotesize \par}

\begin{centering}
\vspace*{0.15cm}
\par\end{centering}

{\footnotesize{}Each row in this table reports the percentages, where
each model generates more accurate density forecasts relative to the
other models reported in each column. The numbers in parentheses indicate
rejection of the null-hypothesis of equal predictive ability according
to the Diebold and Mariano (1995) test at $5\%$ level. The out-of-sample
period consists of $1768$ observations from $24$ December $2007$
until $31$ December, $2014$.}{\footnotesize \par}

\vspace{-0.45in}
\end{table}
\vspace{-0.20in}
\par\end{center}

\begin{center}
\begin{table}[H]
{\footnotesize{}\caption{\label{Table 9} Density forecast comparison using daily S\&P $500$
equity returns by sectors, $w\left(z\right)=1$, i.e. uniform}
}{\footnotesize \par}

\begin{centering}
\setlength\tabcolsep{15.6pt}{\footnotesize{}}%
\begin{tabular}{lllll}
\hline
{\footnotesize{}Model} & {\footnotesize{}t-EGARCH} & {\footnotesize{}Beta-t-EGARCH} & {\footnotesize{}SPEGARCH} & {\footnotesize{}SV}\tabularnewline
\hline
{\footnotesize{}$h=1$} &  &  & {\footnotesize{} } & \tabularnewline
{\footnotesize{}Consumer discretionary} & {\footnotesize{}76 (39)} & {\footnotesize{}89 (39)} & {\footnotesize{}80 (39) } & {\footnotesize{}89 (56)}\tabularnewline
{\footnotesize{}Consumer staples} & {\footnotesize{}84 (44)} & {\footnotesize{}91 (62)} & {\footnotesize{}66 (31) } & {\footnotesize{}84 (31)}\tabularnewline
{\footnotesize{}Energy} & {\footnotesize{}100 (53)} & {\footnotesize{}81 (28)} & {\footnotesize{}84 (47) } & {\footnotesize{}78 (31)}\tabularnewline
{\footnotesize{}Financials} & {\footnotesize{}77 (32)} & {\footnotesize{}80 (38)} & {\footnotesize{}71 (23) } & {\footnotesize{}88 (45)}\tabularnewline
{\footnotesize{}Health care} & {\footnotesize{}91 (47)} & {\footnotesize{}92 (70)} & {\footnotesize{}68 (32) } & {\footnotesize{}94 (53)}\tabularnewline
{\footnotesize{}Industrials} & {\footnotesize{}91 (44)} & {\footnotesize{}98 (68)} & {\footnotesize{}88 (56) } & {\footnotesize{}98 (75)}\tabularnewline
{\footnotesize{}Information technology} & {\footnotesize{}81 (28)} & {\footnotesize{}100 (61)} & {\footnotesize{}70 (22) } & {\footnotesize{}94 (46)}\tabularnewline
{\footnotesize{}Materials} & {\footnotesize{}83 (48)} & {\footnotesize{}91 (57)} & {\footnotesize{}78 (43) } & {\footnotesize{}91 (65)}\tabularnewline
{\footnotesize{}Real Estate} & {\footnotesize{}67 (21)} & {\footnotesize{}71 (12)} & {\footnotesize{}62 (38) } & {\footnotesize{}67 (17)}\tabularnewline
{\footnotesize{}Telecommunication services} & {\footnotesize{}40 (20)} & {\footnotesize{}60 (20)} & {\footnotesize{}20 (0) } & {\footnotesize{}80 (40)}\tabularnewline
{\footnotesize{}Utilities} & {\footnotesize{}65 (23)} & {\footnotesize{}92 (23)} & {\footnotesize{}73 (19) } & {\footnotesize{}88 (31)}\tabularnewline
{\footnotesize{}$h=5$} &  &  & {\footnotesize{} } & \tabularnewline
{\footnotesize{}Consumer discretionary} & {\footnotesize{}84 (60)} & {\footnotesize{}67 (21)} & {\footnotesize{}83 (69) } & {\footnotesize{}81 (46)}\tabularnewline
{\footnotesize{}Consumer staples} & {\footnotesize{}78 (44)} & {\footnotesize{}75 (38)} & {\footnotesize{}78 (44) } & {\footnotesize{}78 (47)}\tabularnewline
{\footnotesize{}Energy} & {\footnotesize{}84 (44)} & {\footnotesize{}38 (12)} & {\footnotesize{}84 (41) } & {\footnotesize{}41 (9)}\tabularnewline
{\footnotesize{}Financials} & {\footnotesize{}95 (79)} & {\footnotesize{}70 (18)} & {\footnotesize{}93 (82) } & {\footnotesize{}79 (38)}\tabularnewline
{\footnotesize{}Health care} & {\footnotesize{}92 (74)} & {\footnotesize{}89 (72)} & {\footnotesize{}83 (64) } & {\footnotesize{}89 (49)}\tabularnewline
{\footnotesize{}Industrials} & {\footnotesize{}95 (75)} & {\footnotesize{}82 (32)} & {\footnotesize{}86 (67) } & {\footnotesize{}89 (54)}\tabularnewline
{\footnotesize{}Information technology} & {\footnotesize{}80 (37)} & {\footnotesize{}83 (43)} & {\footnotesize{}74 (48) } & {\footnotesize{}80 (37)}\tabularnewline
{\footnotesize{}Materials} & {\footnotesize{}83 (61)} & {\footnotesize{}74 (17)} & {\footnotesize{}78 (39) } & {\footnotesize{}74 (43)}\tabularnewline
{\footnotesize{}Real Estate} & {\footnotesize{}88 (79)} & {\footnotesize{}29 (4)} & {\footnotesize{}79 (71) } & {\footnotesize{}38 (4)}\tabularnewline
{\footnotesize{}Telecommunication services} & {\footnotesize{}80 (80)} & {\footnotesize{}80 (20)} & {\footnotesize{}80 (20) } & {\footnotesize{}100 (60)}\tabularnewline
{\footnotesize{}Utilities} & {\footnotesize{}100 (73)} & {\footnotesize{}54 (4)} & {\footnotesize{}96 (58) } & {\footnotesize{}65 (27)}\tabularnewline
{\footnotesize{}$h=20$} &  &  & {\footnotesize{} } & \tabularnewline
{\footnotesize{}Consumer discretionary} & {\footnotesize{}99 (96)} & {\footnotesize{}40 (13)} & {\footnotesize{}94 (93) } & {\footnotesize{}47 (23)}\tabularnewline
{\footnotesize{}Consumer staples} & {\footnotesize{}97 (94)} & {\footnotesize{}56 (19)} & {\footnotesize{}88 (78) } & {\footnotesize{}59 (22)}\tabularnewline
{\footnotesize{}Energy} & {\footnotesize{}100 (97)} & {\footnotesize{}38 (3)} & {\footnotesize{}97 (97) } & {\footnotesize{}22 (3)}\tabularnewline
{\footnotesize{}Financials} & {\footnotesize{}100 (98)} & {\footnotesize{}55 (21)} & {\footnotesize{}98 (98) } & {\footnotesize{}61 (36)}\tabularnewline
{\footnotesize{}Health care} & {\footnotesize{}98 (94)} & {\footnotesize{}66 (38)} & {\footnotesize{}85 (83) } & {\footnotesize{}47 (9)}\tabularnewline
{\footnotesize{}Industrials} & {\footnotesize{}98 (98)} & {\footnotesize{}53 (21)} & {\footnotesize{}95 (91) } & {\footnotesize{}47 (21)}\tabularnewline
{\footnotesize{}Information technology} & {\footnotesize{}100 (89)} & {\footnotesize{}65 (17)} & {\footnotesize{}89 (85) } & {\footnotesize{}37 (17)}\tabularnewline
{\footnotesize{}Materials} & {\footnotesize{}100 (96)} & {\footnotesize{}26 (4)} & {\footnotesize{}100 (96) } & {\footnotesize{}30 (17)}\tabularnewline
{\footnotesize{}Real Estate} & {\footnotesize{}100 (100)} & {\footnotesize{}25 (4)} & {\footnotesize{}96 (96) } & {\footnotesize{}21 (12)}\tabularnewline
{\footnotesize{}Telecommunication services} & {\footnotesize{}100 (80)} & {\footnotesize{}60 (20)} & {\footnotesize{}100 (80) } & {\footnotesize{}60 (20)}\tabularnewline
{\footnotesize{}Utilities} & {\footnotesize{}100 (100)} & {\footnotesize{}58 (27)} & {\footnotesize{}100 (100) } & {\footnotesize{}62 (15)}\tabularnewline
\hline
\end{tabular}
\par\end{centering}{\footnotesize \par}

\begin{centering}
\vspace*{0.15cm}
\par\end{centering}

{\footnotesize{}This table reports the percentages for each model-type,
where the version that considers the leverage effect generates more
accurate density forecasts than the version that does not consider
the leverage effect. The numbers in parentheses indicate rejection
of the null-hypothesis of equal predictive ability according to the
Diebold and Mariano (1995) test at $5\%$ level. The out-of-sample
period consists of $1768$ observations from $24$ December $2007$
until $31$ December, $2014$.}{\footnotesize \par}

\vspace{-0.45in}
\end{table}
\vspace{-0.20in}
\par\end{center}

\begin{center}
\begin{table}[H]
{\footnotesize{}\caption{\label{Table 10} Density forecast comparison using daily S\&P $500$
equity returns by sectors, $w\left(z\right)=1-\Phi\left(z\right)$,
i.e. left tail}
}{\footnotesize \par}

\begin{centering}
\setlength\tabcolsep{15.6pt}{\footnotesize{}}%
\begin{tabular}{lllll}
\hline
{\footnotesize{}Model} & {\footnotesize{}t-EGARCH} & {\footnotesize{}Beta-t-EGARCH} & {\footnotesize{}SPEGARCH} & {\footnotesize{}SV}\tabularnewline
\hline
{\footnotesize{}$h=1$} &  &  & {\footnotesize{} } & \tabularnewline
{\footnotesize{}Consumer discretionary} & {\footnotesize{}86 (43)} & {\footnotesize{}96 (39)} & {\footnotesize{}90 (50) } & {\footnotesize{}91 (51)}\tabularnewline
{\footnotesize{}Consumer staples} & {\footnotesize{}91 (59)} & {\footnotesize{}97 (62)} & {\footnotesize{}78 (44) } & {\footnotesize{}88 (41)}\tabularnewline
{\footnotesize{}Energy} & {\footnotesize{}100 (62)} & {\footnotesize{}91 (53)} & {\footnotesize{}91 (59) } & {\footnotesize{}81 (53)}\tabularnewline
{\footnotesize{}Financials} & {\footnotesize{}88 (52)} & {\footnotesize{}91 (62)} & {\footnotesize{}84 (57) } & {\footnotesize{}98 (54)}\tabularnewline
{\footnotesize{}Health care} & {\footnotesize{}94 (68)} & {\footnotesize{}98 (66)} & {\footnotesize{}81 (43) } & {\footnotesize{}96 (58)}\tabularnewline
{\footnotesize{}Industrials} & {\footnotesize{}95 (61)} & {\footnotesize{}98 (67)} & {\footnotesize{}89 (63) } & {\footnotesize{}96 (63)}\tabularnewline
{\footnotesize{}Information technology} & {\footnotesize{}93 (56)} & {\footnotesize{}98 (67)} & {\footnotesize{}89 (48) } & {\footnotesize{}96 (50)}\tabularnewline
{\footnotesize{}Materials} & {\footnotesize{}87 (65)} & {\footnotesize{}96 (57)} & {\footnotesize{}83 (61) } & {\footnotesize{}96 (52)}\tabularnewline
{\footnotesize{}Real Estate} & {\footnotesize{}88 (71)} & {\footnotesize{}96 (62)} & {\footnotesize{}79 (54) } & {\footnotesize{}96 (46)}\tabularnewline
{\footnotesize{}Telecommunication services} & {\footnotesize{}60 (20)} & {\footnotesize{}80 (40)} & {\footnotesize{}60 (0) } & {\footnotesize{}100 (40)}\tabularnewline
{\footnotesize{}Utilities} & {\footnotesize{}100 (65)} & {\footnotesize{}100 (54)} & {\footnotesize{}100 (54) } & {\footnotesize{}96 (65)}\tabularnewline
{\footnotesize{}$h=5$} &  &  & {\footnotesize{} } & \tabularnewline
{\footnotesize{}Consumer discretionary} & {\footnotesize{}94 (64)} & {\footnotesize{}76 (16)} & {\footnotesize{}90 (71) } & {\footnotesize{}83 (24)}\tabularnewline
{\footnotesize{}Consumer staples} & {\footnotesize{}91 (59)} & {\footnotesize{}78 (38)} & {\footnotesize{}84 (47) } & {\footnotesize{}78 (38)}\tabularnewline
{\footnotesize{}Energy} & {\footnotesize{}97 (69)} & {\footnotesize{}62 (12)} & {\footnotesize{}91 (53) } & {\footnotesize{}66 (9)}\tabularnewline
{\footnotesize{}Financials} & {\footnotesize{}96 (77)} & {\footnotesize{}80 (16)} & {\footnotesize{}98 (82) } & {\footnotesize{}80 (29)}\tabularnewline
{\footnotesize{}Health care} & {\footnotesize{}100 (81)} & {\footnotesize{}92 (57)} & {\footnotesize{}91 (62) } & {\footnotesize{}92 (43)}\tabularnewline
{\footnotesize{}Industrials} & {\footnotesize{}96 (75)} & {\footnotesize{}89 (37)} & {\footnotesize{}95 (68) } & {\footnotesize{}93 (44)}\tabularnewline
{\footnotesize{}Information technology} & {\footnotesize{}91 (39)} & {\footnotesize{}85 (35)} & {\footnotesize{}81 (41) } & {\footnotesize{}87 (37)}\tabularnewline
{\footnotesize{}Materials} & {\footnotesize{}87 (61)} & {\footnotesize{}78 (9)} & {\footnotesize{}87 (48) } & {\footnotesize{}87 (22)}\tabularnewline
{\footnotesize{}Real Estate} & {\footnotesize{}88 (71)} & {\footnotesize{}50 (4)} & {\footnotesize{}83 (58) } & {\footnotesize{}50 (4)}\tabularnewline
{\footnotesize{}Telecommunication services} & {\footnotesize{}80 (60)} & {\footnotesize{}80 (0)} & {\footnotesize{}80 (0) } & {\footnotesize{}80 (20)}\tabularnewline
{\footnotesize{}Utilities} & {\footnotesize{}100 (85)} & {\footnotesize{}77 (8)} & {\footnotesize{}100 (62) } & {\footnotesize{}81 (27)}\tabularnewline
{\footnotesize{}$h=20$} &  &  & {\footnotesize{} } & \tabularnewline
{\footnotesize{}Consumer discretionary} & {\footnotesize{}100 (99)} & {\footnotesize{}49 (13)} & {\footnotesize{}97 (90) } & {\footnotesize{}51 (16)}\tabularnewline
{\footnotesize{}Consumer staples} & {\footnotesize{}97 (94)} & {\footnotesize{}56 (22)} & {\footnotesize{}91 (81) } & {\footnotesize{}53 (16)}\tabularnewline
{\footnotesize{}Energy} & {\footnotesize{}100 (97)} & {\footnotesize{}44 (0)} & {\footnotesize{}97 (97) } & {\footnotesize{}31 (3)}\tabularnewline
{\footnotesize{}Financials} & {\footnotesize{}100 (98)} & {\footnotesize{}55 (14)} & {\footnotesize{}98 (96) } & {\footnotesize{}73 (29)}\tabularnewline
{\footnotesize{}Health care} & {\footnotesize{}98 (94)} & {\footnotesize{}74 (25)} & {\footnotesize{}85 (77) } & {\footnotesize{}57 (13)}\tabularnewline
{\footnotesize{}Industrials} & {\footnotesize{}100 (98)} & {\footnotesize{}53 (16)} & {\footnotesize{}95 (89) } & {\footnotesize{}49 (19)}\tabularnewline
{\footnotesize{}Information technology} & {\footnotesize{}100 (87)} & {\footnotesize{}59 (7)} & {\footnotesize{}93 (83) } & {\footnotesize{}46 (13)}\tabularnewline
{\footnotesize{}Materials} & {\footnotesize{}100 (96)} & {\footnotesize{}30 (4)} & {\footnotesize{}100 (91) } & {\footnotesize{}43 (13)}\tabularnewline
{\footnotesize{}Real Estate} & {\footnotesize{}100 (100)} & {\footnotesize{}25 (0)} & {\footnotesize{}96 (92) } & {\footnotesize{}33 (4)}\tabularnewline
{\footnotesize{}Telecommunication services} & {\footnotesize{}100 (80)} & {\footnotesize{}60 (20)} & {\footnotesize{}100 (80) } & {\footnotesize{}80 (0)}\tabularnewline
{\footnotesize{}Utilities} & {\footnotesize{}100 (100)} & {\footnotesize{}62 (19)} & {\footnotesize{}100 (100) } & {\footnotesize{}69 (19)}\tabularnewline
\hline
\end{tabular}
\par\end{centering}{\footnotesize \par}

\begin{centering}
\vspace*{0.15cm}
\par\end{centering}

{\footnotesize{}This table reports the percentages for each model-type,
where the version that considers the leverage effect generates more
accurate density forecasts than the version that does not consider
the leverage effect. The numbers in parentheses indicate rejection
of the null-hypothesis of equal predictive ability according to the
Diebold and Mariano (1995) test at $5\%$ level. The out-of-sample
period consists of $1768$ observations from $24$ December $2007$
until $31$ December, $2014$.}{\footnotesize \par}

\vspace{-0.45in}
\end{table}
\vspace{-0.20in}
\par\end{center}

\begin{center}
\begin{table}[H]
{\footnotesize{}\caption{\label{Table 11} Density forecast comparison of individual stocks
from daily S\&P $500$ equity returns, $h=1$ and $w\left(z\right)=1$}
}{\footnotesize \par}

\begin{centering}
\setlength\tabcolsep{1.8pt}{\footnotesize{}}%
\begin{tabular}{llllllll}
\hline
{\footnotesize{}$\begin{array}{c}
\textrm{Model}\\
\\
\end{array}$} & {\footnotesize{}$\begin{array}{c}
\textrm{t-}\\
\textrm{EGARCH}
\end{array}$} & {\footnotesize{}$\begin{array}{c}
\textrm{Beta-t-}\\
\textrm{EGARCH-NL}
\end{array}$} & {\footnotesize{}$\begin{array}{c}
\textrm{Beta-t-}\\
\textrm{EGARCH}
\end{array}$} & {\footnotesize{}$\begin{array}{c}
\textrm{SPEGARCH}\\
\textrm{-NL}
\end{array}$} & {\footnotesize{}$\begin{array}{c}
\textrm{SPEGARCH}\\
\\
\end{array}$} & {\footnotesize{}$\begin{array}{c}
\textrm{SV}\\
\textrm{-NL}
\end{array}$} & {\footnotesize{}$\begin{array}{c}
\textrm{SV}\\
\\
\end{array}$}\tabularnewline
\hline
{\footnotesize{}Alcoa} & {\footnotesize{}1.001} & {\footnotesize{}0.993$^{\left(a\right)}$ } & {\footnotesize{}0.993$^{\left(a\right)}$ } & {\footnotesize{}1.004$^{\left(a\right)}$ } & {\footnotesize{}1.003$^{\left(a\right)}$ } & {\footnotesize{}0.997$^{\left(b\right)}$ } & {\footnotesize{}0.998}\tabularnewline
{\footnotesize{}American Express} & {\footnotesize{}1.001} & {\footnotesize{}0.991$^{\left(a\right)}$ } & {\footnotesize{}0.992$^{\left(a\right)}$ } & {\footnotesize{}1.002$^{\left(b\right)}$ } & {\footnotesize{}1.002} & {\footnotesize{}0.992$^{\left(a\right)}$ } & {\footnotesize{}0.996$^{\left(b\right)}$ }\tabularnewline
{\footnotesize{}Boeing} & {\footnotesize{}0.997$^{\left(a\right)}$ } & {\footnotesize{}0.995$^{\left(a\right)}$ } & {\footnotesize{}0.993$^{\left(a\right)}$ } & {\footnotesize{}1.000} & {\footnotesize{}0.996$^{\left(a\right)}$ } & {\footnotesize{}0.998$^{\left(c\right)}$ } & {\footnotesize{}1.001}\tabularnewline
{\footnotesize{}Caterpillar} & {\footnotesize{}0.998$^{\left(c\right)}$ } & {\footnotesize{}0.993$^{\left(a\right)}$ } & {\footnotesize{}0.991$^{\left(a\right)}$ } & {\footnotesize{}1.004$^{\left(a\right)}$ } & {\footnotesize{}1.000} & {\footnotesize{}0.998$^{\left(c\right)}$ } & {\footnotesize{}1.001}\tabularnewline
{\footnotesize{}Chevron} & {\footnotesize{}0.997$^{\left(a\right)}$ } & {\footnotesize{}0.992$^{\left(a\right)}$ } & {\footnotesize{}0.993$^{\left(a\right)}$ } & {\footnotesize{}0.999} & {\footnotesize{}0.995$^{\left(b\right)}$ } & {\footnotesize{}0.998$^{\left(c\right)}$ } & {\footnotesize{}1.000}\tabularnewline
{\footnotesize{}Walt Disney} & {\footnotesize{}0.996$^{\left(a\right)}$ } & {\footnotesize{}0.992$^{\left(a\right)}$ } & {\footnotesize{}0.989$^{\left(a\right)}$ } & {\footnotesize{}0.999} & {\footnotesize{}0.995$^{\left(a\right)}$ } & {\footnotesize{}0.998$^{\left(c\right)}$ } & {\footnotesize{}1.000}\tabularnewline
{\footnotesize{}General Electric} & {\footnotesize{}0.998$^{\left(b\right)}$ } & {\footnotesize{}0.990$^{\left(a\right)}$ } & {\footnotesize{}0.987$^{\left(a\right)}$ } & {\footnotesize{}1.000} & {\footnotesize{}0.998$^{\left(c\right)}$ } & {\footnotesize{}0.992$^{\left(a\right)}$ } & {\footnotesize{}0.996$^{\left(b\right)}$ }\tabularnewline
{\footnotesize{}IBM} & {\footnotesize{}0.998$^{\left(b\right)}$ } & {\footnotesize{}0.995$^{\left(a\right)}$ } & {\footnotesize{}0.992$^{\left(a\right)}$ } & {\footnotesize{}1.001} & {\footnotesize{}0.998$^{\left(c\right)}$ } & {\footnotesize{}0.998$^{\left(c\right)}$ } & {\footnotesize{}1.000}\tabularnewline
{\footnotesize{}Intel} & {\footnotesize{}1.003$^{\left(a\right)}$ } & {\footnotesize{}0.993$^{\left(a\right)}$ } & {\footnotesize{}0.993$^{\left(a\right)}$ } & {\footnotesize{}1.001$^{\left(b\right)}$ } & {\footnotesize{}1.002$^{\left(b\right)}$ } & {\footnotesize{}1.001} & {\footnotesize{}1.002$^{\left(b\right)}$ }\tabularnewline
{\footnotesize{}Johnson \& Johnson} & {\footnotesize{}0.996$^{\left(a\right)}$ } & {\footnotesize{}0.999} & {\footnotesize{}0.993$^{\left(a\right)}$ } & {\footnotesize{}0.999} & {\footnotesize{}0.996$^{\left(a\right)}$ } & {\footnotesize{}0.995$^{\left(a\right)}$ } & {\footnotesize{}1.001}\tabularnewline
{\footnotesize{}JPMorgan} & {\footnotesize{}1.005$^{\left(b\right)}$ } & {\footnotesize{}0.993$^{\left(a\right)}$ } & {\footnotesize{}0.991$^{\left(a\right)}$ } & {\footnotesize{}1.005$^{\left(a\right)}$ } & {\footnotesize{}1.003$^{\left(c\right)}$ } & {\footnotesize{}0.991$^{\left(a\right)}$ } & {\footnotesize{}0.995$^{\left(a\right)}$ }\tabularnewline
{\footnotesize{}Coca-Cola} & {\footnotesize{}0.996$^{\left(a\right)}$ } & {\footnotesize{}0.997$^{\left(b\right)}$ } & {\footnotesize{}0.993$^{\left(a\right)}$ } & {\footnotesize{}1.002$^{\left(b\right)}$ } & {\footnotesize{}0.999} & {\footnotesize{}0.997$^{\left(b\right)}$ } & {\footnotesize{}1.002$^{\left(b\right)}$ }\tabularnewline
{\footnotesize{}McDonald\textquoteright s} & {\footnotesize{}0.999 } & {\footnotesize{}0.995$^{\left(a\right)}$ } & {\footnotesize{}0.994$^{\left(a\right)}$ } & {\footnotesize{}1.000} & {\footnotesize{}0.999} & {\footnotesize{}1.000} & {\footnotesize{}1.001}\tabularnewline
{\footnotesize{}Merck} & {\footnotesize{}0.998$^{\left(a\right)}$ } & {\footnotesize{}0.994$^{\left(a\right)}$ } & {\footnotesize{}0.990$^{\left(a\right)}$ } & {\footnotesize{}1.008$^{\left(a\right)}$ } & {\footnotesize{}1.004$^{\left(b\right)}$ } & {\footnotesize{}0.999} & {\footnotesize{}1.000}\tabularnewline
{\footnotesize{}Microsoft} & {\footnotesize{}1.000} & {\footnotesize{}0.994$^{\left(a\right)}$ } & {\footnotesize{}0.993$^{\left(a\right)}$ } & {\footnotesize{}1.005$^{\left(a\right)}$ } & {\footnotesize{}1.006$^{\left(a\right)}$ } & {\footnotesize{}1.001} & {\footnotesize{}1.002}\tabularnewline
{\footnotesize{}Pfizer} & {\footnotesize{}0.998$^{\left(b\right)}$ } & {\footnotesize{}0.995$^{\left(a\right)}$ } & {\footnotesize{}0.994$^{\left(a\right)}$ } & {\footnotesize{}1.000} & {\footnotesize{}1.036} & {\footnotesize{}1.000} & {\footnotesize{}1.001}\tabularnewline
{\footnotesize{}Procter \& Gamble} & {\footnotesize{}0.997$^{\left(a\right)}$ } & {\footnotesize{}0.993$^{\left(a\right)}$ } & {\footnotesize{}0.991$^{\left(a\right)}$ } & {\footnotesize{}1.003$^{\left(a\right)}$ } & {\footnotesize{}1.039} & {\footnotesize{}1.000} & {\footnotesize{}1.002$^{\left(b\right)}$ }\tabularnewline
{\footnotesize{}AT\&T} & {\footnotesize{}1.001 } & {\footnotesize{}0.994$^{\left(a\right)}$ } & {\footnotesize{}0.993$^{\left(a\right)}$ } & {\footnotesize{}1.003$^{\left(a\right)}$ } & {\footnotesize{}1.004$^{\left(a\right)}$ } & {\footnotesize{}0.996$^{\left(a\right)}$ } & {\footnotesize{}0.998}\tabularnewline
{\footnotesize{}Walmart} & {\footnotesize{}1.001} & {\footnotesize{}0.998$^{\left(c\right)}$ } & {\footnotesize{}0.998} & {\footnotesize{}1.000} & {\footnotesize{}1.000} & {\footnotesize{}1.003$^{\left(b\right)}$ } & {\footnotesize{}1.003$^{\left(a\right)}$ }\tabularnewline
{\footnotesize{}ExxonMobil} & {\footnotesize{}0.997$^{\left(b\right)}$ } & {\footnotesize{}0.998$^{\left(b\right)}$ } & {\footnotesize{}0.997$^{\left(a\right)}$ } & {\footnotesize{}1.002$^{\left(b\right)}$ } & {\footnotesize{}0.999} & {\footnotesize{}0.999} & {\footnotesize{}1.004$^{\left(a\right)}$ }\tabularnewline
\hline
\end{tabular}
\par\end{centering}{\footnotesize \par}

\begin{centering}
\vspace*{0.15cm}
\par\end{centering}

{\footnotesize{}This table reports the average wCRPS for different
models relative to t-EGARCH-NL using twenty equities from the S\&P
$500$ index. The apexes $a$, $b$, and $c$ indicate rejection of
the null-hypothesis of equal predictive ability according to the Diebold
and Mariano (1995) test at $1\%$, $5\%$ and $10\%$, respectively.
The out-of-sample period consists of $1768$ observations from $24$
December $2007$ until $31$ December, $2014$.}{\footnotesize \par}

\vspace{-0.45in}
\end{table}
\vspace{-0.20in}
\par\end{center}

\begin{center}
\begin{table}[H]
{\footnotesize{}\caption{\label{Table 12} Density forecast comparison of individual stocks
from daily S\&P $500$ equity returns, $h=5$ and $w\left(z\right)=1$}
}{\footnotesize \par}

\begin{centering}
\setlength\tabcolsep{1.8pt}{\footnotesize{}}%
\begin{tabular}{llllllll}
\hline
{\footnotesize{}$\begin{array}{c}
\textrm{Model}\\
\\
\end{array}$} & {\footnotesize{}$\begin{array}{c}
\textrm{t-}\\
\textrm{EGARCH}
\end{array}$} & {\footnotesize{}$\begin{array}{c}
\textrm{Beta-t-}\\
\textrm{EGARCH-NL}
\end{array}$} & {\footnotesize{}$\begin{array}{c}
\textrm{Beta-t-}\\
\textrm{EGARCH}
\end{array}$} & {\footnotesize{}$\begin{array}{c}
\textrm{SPEGARCH}\\
\textrm{-NL}
\end{array}$} & {\footnotesize{}$\begin{array}{c}
\textrm{SPEGARCH}\\
\\
\end{array}$} & {\footnotesize{}$\begin{array}{c}
\textrm{SV}\\
\textrm{-NL}
\end{array}$} & {\footnotesize{}$\begin{array}{c}
\textrm{SV}\\
\\
\end{array}$}\tabularnewline
\hline
{\footnotesize{}Alcoa} & {\footnotesize{}1.002$^{\left(b\right)}$ } & {\footnotesize{}0.986$^{\left(a\right)}$ } & {\footnotesize{}0.987$^{\left(a\right)}$ } & {\footnotesize{}1.003$^{\left(a\right)}$ } & {\footnotesize{}1.002$^{\left(b\right)}$ } & {\footnotesize{}0.994$^{\left(a\right)}$ } & {\footnotesize{}0.993$^{\left(a\right)}$ }\tabularnewline
{\footnotesize{}American Express} & {\footnotesize{}0.995$^{\left(c\right)}$ } & {\footnotesize{}0.979$^{\left(a\right)}$ } & {\footnotesize{}0.983$^{\left(a\right)}$ } & {\footnotesize{}1.004$^{\left(a\right)}$ } & {\footnotesize{}0.994$^{\left(b\right)}$ } & {\footnotesize{}0.984$^{\left(a\right)}$ } & {\footnotesize{}0.984$^{\left(a\right)}$ }\tabularnewline
{\footnotesize{}Boeing} & {\footnotesize{}0.994$^{\left(a\right)}$ } & {\footnotesize{}0.989$^{\left(a\right)}$ } & {\footnotesize{}0.989$^{\left(a\right)}$ } & {\footnotesize{}1.000} & {\footnotesize{}0.993$^{\left(a\right)}$ } & {\footnotesize{}0.995$^{\left(a\right)}$ } & {\footnotesize{}0.998}\tabularnewline
{\footnotesize{}Caterpillar} & {\footnotesize{}0.997$^{\left(b\right)}$ } & {\footnotesize{}0.990$^{\left(a\right)}$ } & {\footnotesize{}0.988$^{\left(a\right)}$ } & {\footnotesize{}1.000} & {\footnotesize{}0.996$^{\left(b\right)}$ } & {\footnotesize{}0.999} & {\footnotesize{}1.002}\tabularnewline
{\footnotesize{}Chevron} & {\footnotesize{}0.998$^{\left(b\right)}$ } & {\footnotesize{}0.988$^{\left(a\right)}$ } & {\footnotesize{}0.992$^{\left(a\right)}$ } & {\footnotesize{}0.998} & {\footnotesize{}0.996$^{\left(b\right)}$ } & {\footnotesize{}1.000} & {\footnotesize{}1.000}\tabularnewline
{\footnotesize{}Walt Disney} & {\footnotesize{}0.991$^{\left(a\right)}$ } & {\footnotesize{}0.984$^{\left(a\right)}$ } & {\footnotesize{}0.982$^{\left(a\right)}$ } & {\footnotesize{}0.995$^{\left(a\right)}$ } & {\footnotesize{}0.992$^{\left(a\right)}$ } & {\footnotesize{}0.993$^{\left(a\right)}$ } & {\footnotesize{}0.997$^{\left(c\right)}$ }\tabularnewline
{\footnotesize{}General Electric} & {\footnotesize{}0.994$^{\left(a\right)}$ } & {\footnotesize{}0.978$^{\left(a\right)}$ } & {\footnotesize{}0.976$^{\left(a\right)}$ } & {\footnotesize{}0.998$^{\left(b\right)}$ } & {\footnotesize{}0.994$^{\left(a\right)}$ } & {\footnotesize{}0.982$^{\left(a\right)}$ } & {\footnotesize{}0.986$^{\left(a\right)}$ }\tabularnewline
{\footnotesize{}IBM} & {\footnotesize{}0.996$^{\left(a\right)}$ } & {\footnotesize{}0.989$^{\left(a\right)}$ } & {\footnotesize{}0.987$^{\left(a\right)}$ } & {\footnotesize{}1.005$^{\left(a\right)}$ } & {\footnotesize{}0.997$^{\left(b\right)}$ } & {\footnotesize{}0.994$^{\left(a\right)}$ } & {\footnotesize{}0.997$^{\left(b\right)}$ }\tabularnewline
{\footnotesize{}Intel} & {\footnotesize{}1.002$^{\left(a\right)}$ } & {\footnotesize{}0.989$^{\left(a\right)}$ } & {\footnotesize{}0.989$^{\left(a\right)}$ } & {\footnotesize{}0.999} & {\footnotesize{}1.001} & {\footnotesize{}0.999} & {\footnotesize{}0.999}\tabularnewline
{\footnotesize{}Johnson \& Johnson} & {\footnotesize{}0.995$^{\left(a\right)}$ } & {\footnotesize{}0.988$^{\left(a\right)}$ } & {\footnotesize{}0.984$^{\left(a\right)}$ } & {\footnotesize{}1.000} & {\footnotesize{}0.994$^{\left(a\right)}$ } & {\footnotesize{}0.986$^{\left(a\right)}$ } & {\footnotesize{}0.991$^{\left(a\right)}$ }\tabularnewline
{\footnotesize{}JPMorgan} & {\footnotesize{}1.006$^{\left(a\right)}$ } & {\footnotesize{}0.978$^{\left(a\right)}$ } & {\footnotesize{}0.979$^{\left(a\right)}$ } & {\footnotesize{}1.009$^{\left(a\right)}$ } & {\footnotesize{}1.001} & {\footnotesize{}0.983$^{\left(a\right)}$ } & {\footnotesize{}0.984$^{\left(a\right)}$ }\tabularnewline
{\footnotesize{}Coca-Cola} & {\footnotesize{}0.993$^{\left(a\right)}$ } & {\footnotesize{}0.985$^{\left(a\right)}$ } & {\footnotesize{}0.983$^{\left(a\right)}$ } & {\footnotesize{}1.008$^{\left(a\right)}$ } & {\footnotesize{}1.003$^{\left(b\right)}$ } & {\footnotesize{}0.990$^{\left(a\right)}$ } & {\footnotesize{}0.993$^{\left(a\right)}$ }\tabularnewline
{\footnotesize{}McDonald\textquoteright s} & {\footnotesize{}1.000} & {\footnotesize{}0.993$^{\left(a\right)}$ } & {\footnotesize{}0.993$^{\left(a\right)}$ } & {\footnotesize{}0.999} & {\footnotesize{}1.000} & {\footnotesize{}1.002} & {\footnotesize{}1.001}\tabularnewline
{\footnotesize{}Merck} & {\footnotesize{}0.995$^{\left(a\right)}$ } & {\footnotesize{}0.989$^{\left(a\right)}$ } & {\footnotesize{}0.982$^{\left(a\right)}$ } & {\footnotesize{}1.002$^{\left(c\right)}$ } & {\footnotesize{}0.995$^{\left(a\right)}$ } & {\footnotesize{}0.997$^{\left(b\right)}$ } & {\footnotesize{}0.998$^{\left(c\right)}$ }\tabularnewline
{\footnotesize{}Microsoft} & {\footnotesize{}1.001} & {\footnotesize{}0.988$^{\left(a\right)}$ } & {\footnotesize{}0.988$^{\left(a\right)}$ } & {\footnotesize{}1.003$^{\left(a\right)}$ } & {\footnotesize{}1.004$^{\left(b\right)}$ } & {\footnotesize{}0.997$^{\left(c\right)}$ } & {\footnotesize{}0.998}\tabularnewline
{\footnotesize{}Pfizer} & {\footnotesize{}0.998$^{\left(b\right)}$ } & {\footnotesize{}0.989$^{\left(a\right)}$ } & {\footnotesize{}0.990$^{\left(a\right)}$ } & {\footnotesize{}1.004$^{\left(a\right)}$ } & {\footnotesize{}1.039} & {\footnotesize{}0.998} & {\footnotesize{}1.000}\tabularnewline
{\footnotesize{}Procter \& Gamble} & {\footnotesize{}0.993$^{\left(a\right)}$ } & {\footnotesize{}0.983$^{\left(a\right)}$ } & {\footnotesize{}0.980$^{\left(a\right)}$ } & {\footnotesize{}1.002$^{\left(b\right)}$ } & {\footnotesize{}1.031} & {\footnotesize{}0.995$^{\left(a\right)}$ } & {\footnotesize{}0.998}\tabularnewline
{\footnotesize{}AT\&T} & {\footnotesize{}0.997$^{\left(b\right)}$ } & {\footnotesize{}0.988$^{\left(a\right)}$ } & {\footnotesize{}0.987$^{\left(a\right)}$ } & {\footnotesize{}1.002$^{\left(b\right)}$ } & {\footnotesize{}1.001} & {\footnotesize{}0.990$^{\left(a\right)}$ } & {\footnotesize{}0.992$^{\left(a\right)}$ }\tabularnewline
{\footnotesize{}Walmart} & {\footnotesize{}1.000} & {\footnotesize{}0.997$^{\left(c\right)}$ } & {\footnotesize{}0.997$^{\left(c\right)}$ } & {\footnotesize{}1.002$^{\left(b\right)}$ } & {\footnotesize{}1.002$^{\left(c\right)}$ } & {\footnotesize{}1.008$^{\left(a\right)}$ } & {\footnotesize{}1.008$^{\left(a\right)}$ }\tabularnewline
{\footnotesize{}ExxonMobil} & {\footnotesize{}0.999} & {\footnotesize{}0.995$^{\left(a\right)}$ } & {\footnotesize{}0.996$^{\left(b\right)}$ } & {\footnotesize{}1.000} & {\footnotesize{}1.000} & {\footnotesize{}1.000} & {\footnotesize{}1.001}\tabularnewline
\hline
\end{tabular}
\par\end{centering}{\footnotesize \par}

\begin{centering}
\vspace*{0.15cm}
\par\end{centering}

{\footnotesize{}This table reports the average wCRPS for different
models relative to t-EGARCH-NL using twenty equities from the S\&P
$500$ index. The apexes $a$, $b$, and $c$ indicate rejection of
the null-hypothesis of equal predictive ability according to the Diebold
and Mariano (1995) test at $1\%$, $5\%$ and $10\%$, respectively.
The out-of-sample period consists of $1768$ observations from $24$
December $2007$ until $31$ December, $2014$.}{\footnotesize \par}

\vspace{-0.45in}
\end{table}
\vspace{-0.20in}
\par\end{center}

\begin{center}
\begin{table}[H]
{\footnotesize{}\caption{\label{Table 13} Density forecast comparison of individual stocks
from daily S\&P $500$ equity returns, $h=20$ and $w\left(z\right)=1$}
}{\footnotesize \par}

\begin{centering}
\setlength\tabcolsep{1.8pt}{\footnotesize{}}%
\begin{tabular}{llllllll}
\hline
{\footnotesize{}$\begin{array}{c}
\textrm{Model}\\
\\
\end{array}$} & {\footnotesize{}$\begin{array}{c}
\textrm{t-}\\
\textrm{EGARCH}
\end{array}$} & {\footnotesize{}$\begin{array}{c}
\textrm{Beta-t-}\\
\textrm{EGARCH-NL}
\end{array}$} & {\footnotesize{}$\begin{array}{c}
\textrm{Beta-t-}\\
\textrm{EGARCH}
\end{array}$} & {\footnotesize{}$\begin{array}{c}
\textrm{SPEGARCH}\\
\textrm{-NL}
\end{array}$} & {\footnotesize{}$\begin{array}{c}
\textrm{SPEGARCH}\\
\\
\end{array}$} & {\footnotesize{}$\begin{array}{c}
\textrm{SV}\\
\textrm{-NL}
\end{array}$} & {\footnotesize{}$\begin{array}{c}
\textrm{SV}\\
\\
\end{array}$}\tabularnewline
\hline
{\footnotesize{}Alcoa} & {\footnotesize{}0.994$^{\left(a\right)}$ } & {\footnotesize{}0.953$^{\left(a\right)}$ } & {\footnotesize{}0.953$^{\left(a\right)}$ } & {\footnotesize{}1.000} & {\footnotesize{}0.988$^{\left(a\right)}$ } & {\footnotesize{}0.970$^{\left(a\right)}$ } & {\footnotesize{}0.969$^{\left(a\right)}$ }\tabularnewline
{\footnotesize{}American Express} & {\footnotesize{}0.959$^{\left(a\right)}$ } & {\footnotesize{}0.929$^{\left(a\right)}$ } & {\footnotesize{}0.934$^{\left(a\right)}$ } & {\footnotesize{}1.001} & {\footnotesize{}0.959$^{\left(a\right)}$ } & {\footnotesize{}0.942$^{\left(a\right)}$ } & {\footnotesize{}0.941$^{\left(a\right)}$ }\tabularnewline
{\footnotesize{}Boeing} & {\footnotesize{}0.981$^{\left(a\right)}$ } & {\footnotesize{}0.962$^{\left(a\right)}$ } & {\footnotesize{}0.967$^{\left(a\right)}$ } & {\footnotesize{}0.997$^{\left(a\right)}$ } & {\footnotesize{}0.978$^{\left(a\right)}$ } & {\footnotesize{}0.975$^{\left(a\right)}$ } & {\footnotesize{}0.975$^{\left(a\right)}$ }\tabularnewline
{\footnotesize{}Caterpillar} & {\footnotesize{}0.983$^{\left(a\right)}$ } & {\footnotesize{}0.969$^{\left(a\right)}$ } & {\footnotesize{}0.970$^{\left(a\right)}$ } & {\footnotesize{}0.991$^{\left(a\right)}$ } & {\footnotesize{}0.974$^{\left(a\right)}$ } & {\footnotesize{}0.985$^{\left(a\right)}$ } & {\footnotesize{}0.984$^{\left(a\right)}$ }\tabularnewline
{\footnotesize{}Chevron} & {\footnotesize{}0.987$^{\left(a\right)}$ } & {\footnotesize{}0.971$^{\left(a\right)}$ } & {\footnotesize{}0.976$^{\left(a\right)}$ } & {\footnotesize{}0.999} & {\footnotesize{}0.986$^{\left(a\right)}$ } & {\footnotesize{}0.984$^{\left(a\right)}$ } & {\footnotesize{}0.981$^{\left(a\right)}$ }\tabularnewline
{\footnotesize{}Walt Disney} & {\footnotesize{}0.977$^{\left(a\right)}$ } & {\footnotesize{}0.960$^{\left(a\right)}$ } & {\footnotesize{}0.960$^{\left(a\right)}$ } & {\footnotesize{}0.994$^{\left(a\right)}$ } & {\footnotesize{}0.982$^{\left(a\right)}$ } & {\footnotesize{}0.975$^{\left(a\right)}$ } & {\footnotesize{}0.974$^{\left(a\right)}$ }\tabularnewline
{\footnotesize{}General Electric} & {\footnotesize{}0.984$^{\left(a\right)}$ } & {\footnotesize{}0.936$^{\left(a\right)}$ } & {\footnotesize{}0.934$^{\left(a\right)}$ } & {\footnotesize{}0.993$^{\left(a\right)}$ } & {\footnotesize{}0.982$^{\left(a\right)}$ } & {\footnotesize{}0.952$^{\left(a\right)}$ } & {\footnotesize{}0.956$^{\left(a\right)}$ }\tabularnewline
{\footnotesize{}IBM} & {\footnotesize{}0.982$^{\left(a\right)}$ } & {\footnotesize{}0.963$^{\left(a\right)}$ } & {\footnotesize{}0.962$^{\left(a\right)}$ } & {\footnotesize{}1.007$^{\left(a\right)}$ } & {\footnotesize{}0.986$^{\left(a\right)}$ } & {\footnotesize{}0.975$^{\left(a\right)}$ } & {\footnotesize{}0.974$^{\left(a\right)}$ }\tabularnewline
{\footnotesize{}Intel} & {\footnotesize{}0.998$^{\left(c\right)}$ } & {\footnotesize{}0.973$^{\left(a\right)}$ } & {\footnotesize{}0.973$^{\left(a\right)}$ } & {\footnotesize{}0.998$^{\left(a\right)}$ } & {\footnotesize{}0.992$^{\left(a\right)}$ } & {\footnotesize{}0.986$^{\left(a\right)}$ } & {\footnotesize{}0.986$^{\left(a\right)}$ }\tabularnewline
{\footnotesize{}Johnson \& Johnson} & {\footnotesize{}0.991$^{\left(a\right)}$ } & {\footnotesize{}0.963$^{\left(a\right)}$ } & {\footnotesize{}0.961$^{\left(a\right)}$ } & {\footnotesize{}0.997$^{\left(a\right)}$ } & {\footnotesize{}0.987$^{\left(a\right)}$ } & {\footnotesize{}0.963$^{\left(a\right)}$ } & {\footnotesize{}0.963$^{\left(a\right)}$ }\tabularnewline
{\footnotesize{}JPMorgan} & {\footnotesize{}0.989$^{\left(a\right)}$ } & {\footnotesize{}0.920$^{\left(a\right)}$ } & {\footnotesize{}0.919$^{\left(a\right)}$ } & {\footnotesize{}1.005$^{\left(a\right)}$ } & {\footnotesize{}0.978$^{\left(a\right)}$ } & {\footnotesize{}0.928$^{\left(a\right)}$ } & {\footnotesize{}0.932$^{\left(a\right)}$ }\tabularnewline
{\footnotesize{}Coca-Cola} & {\footnotesize{}0.983$^{\left(a\right)}$ } & {\footnotesize{}0.950$^{\left(a\right)}$ } & {\footnotesize{}0.949$^{\left(a\right)}$ } & {\footnotesize{}1.006$^{\left(a\right)}$ } & {\footnotesize{}0.993$^{\left(a\right)}$ } & {\footnotesize{}0.961$^{\left(a\right)}$ } & {\footnotesize{}0.962$^{\left(a\right)}$ }\tabularnewline
{\footnotesize{}McDonald\textquoteright s} & {\footnotesize{}0.995$^{\left(a\right)}$ } & {\footnotesize{}0.979$^{\left(a\right)}$ } & {\footnotesize{}0.981$^{\left(a\right)}$ } & {\footnotesize{}1.000} & {\footnotesize{}0.997$^{\left(b\right)}$ } & {\footnotesize{}0.993$^{\left(b\right)}$ } & {\footnotesize{}0.992$^{\left(a\right)}$ }\tabularnewline
{\footnotesize{}Merck} & {\footnotesize{}0.989$^{\left(a\right)}$ } & {\footnotesize{}0.977$^{\left(a\right)}$ } & {\footnotesize{}0.969$^{\left(a\right)}$ } & {\footnotesize{}0.992$^{\left(a\right)}$ } & {\footnotesize{}0.976$^{\left(a\right)}$ } & {\footnotesize{}0.982$^{\left(a\right)}$ } & {\footnotesize{}0.983$^{\left(a\right)}$ }\tabularnewline
{\footnotesize{}Microsoft} & {\footnotesize{}0.995$^{\left(a\right)}$ } & {\footnotesize{}0.961$^{\left(a\right)}$ } & {\footnotesize{}0.961$^{\left(a\right)}$ } & {\footnotesize{}0.988$^{\left(a\right)}$ } & {\footnotesize{}0.987$^{\left(a\right)}$ } & {\footnotesize{}0.976$^{\left(a\right)}$ } & {\footnotesize{}0.976$^{\left(a\right)}$ }\tabularnewline
{\footnotesize{}Pfizer} & {\footnotesize{}0.990$^{\left(a\right)}$ } & {\footnotesize{}0.969$^{\left(a\right)}$ } & {\footnotesize{}0.970$^{\left(a\right)}$ } & {\footnotesize{}1.007$^{\left(a\right)}$ } & {\footnotesize{}1.037} & {\footnotesize{}0.982$^{\left(a\right)}$ } & {\footnotesize{}0.983$^{\left(a\right)}$ }\tabularnewline
{\footnotesize{}Procter \& Gamble} & {\footnotesize{}0.985$^{\left(a\right)}$ } & {\footnotesize{}0.960$^{\left(a\right)}$ } & {\footnotesize{}0.955$^{\left(a\right)}$ } & {\footnotesize{}0.987$^{\left(a\right)}$ } & {\footnotesize{}0.994} & {\footnotesize{}0.974$^{\left(a\right)}$ } & {\footnotesize{}0.975$^{\left(a\right)}$ }\tabularnewline
{\footnotesize{}AT\&T} & {\footnotesize{}0.990$^{\left(a\right)}$ } & {\footnotesize{}0.965$^{\left(a\right)}$ } & {\footnotesize{}0.964$^{\left(a\right)}$ } & {\footnotesize{}0.999} & {\footnotesize{}0.991$^{\left(a\right)}$ } & {\footnotesize{}0.974$^{\left(a\right)}$ } & {\footnotesize{}0.975$^{\left(a\right)}$ }\tabularnewline
{\footnotesize{}Walmart} & {\footnotesize{}0.995$^{\left(a\right)}$ } & {\footnotesize{}0.985$^{\left(a\right)}$ } & {\footnotesize{}0.985$^{\left(a\right)}$ } & {\footnotesize{}1.004$^{\left(a\right)}$ } & {\footnotesize{}0.999} & {\footnotesize{}1.000} & {\footnotesize{}1.000}\tabularnewline
{\footnotesize{}ExxonMobil} & {\footnotesize{}0.990$^{\left(a\right)}$ } & {\footnotesize{}0.975$^{\left(a\right)}$ } & {\footnotesize{}0.977$^{\left(a\right)}$ } & {\footnotesize{}0.996$^{\left(a\right)}$ } & {\footnotesize{}0.990$^{\left(a\right)}$ } & {\footnotesize{}0.983$^{\left(a\right)}$ } & {\footnotesize{}0.982$^{\left(a\right)}$ }\tabularnewline
\hline
\end{tabular}
\par\end{centering}{\footnotesize \par}

\begin{centering}
\vspace*{0.15cm}
\par\end{centering}

{\footnotesize{}This table reports the average wCRPS for different
models relative to t-EGARCH-NL using twenty equities from the S\&P
$500$ index. The apexes $a$, $b$, and $c$ indicate rejection of
the null-hypothesis of equal predictive ability according to the Diebold
and Mariano (1995) test at $1\%$, $5\%$ and $10\%$, respectively.
The out-of-sample period consists of $1768$ observations from $24$
December $2007$ until $31$ December, $2014$.}{\footnotesize \par}

\vspace{-0.45in}
\end{table}
\vspace{-0.20in}
\par\end{center}

\begin{center}
\begin{figure}[H]
\caption{\label{Figure 1} Out-of-sample cumulative wCRPS using daily Dow Jones
returns }

\begin{centering}
{\footnotesize{}}%
\begin{tabular}{cc}
{\footnotesize{}(a)} & {\footnotesize{}(b)}\tabularnewline
{\footnotesize{}\includegraphics[scale=0.52]{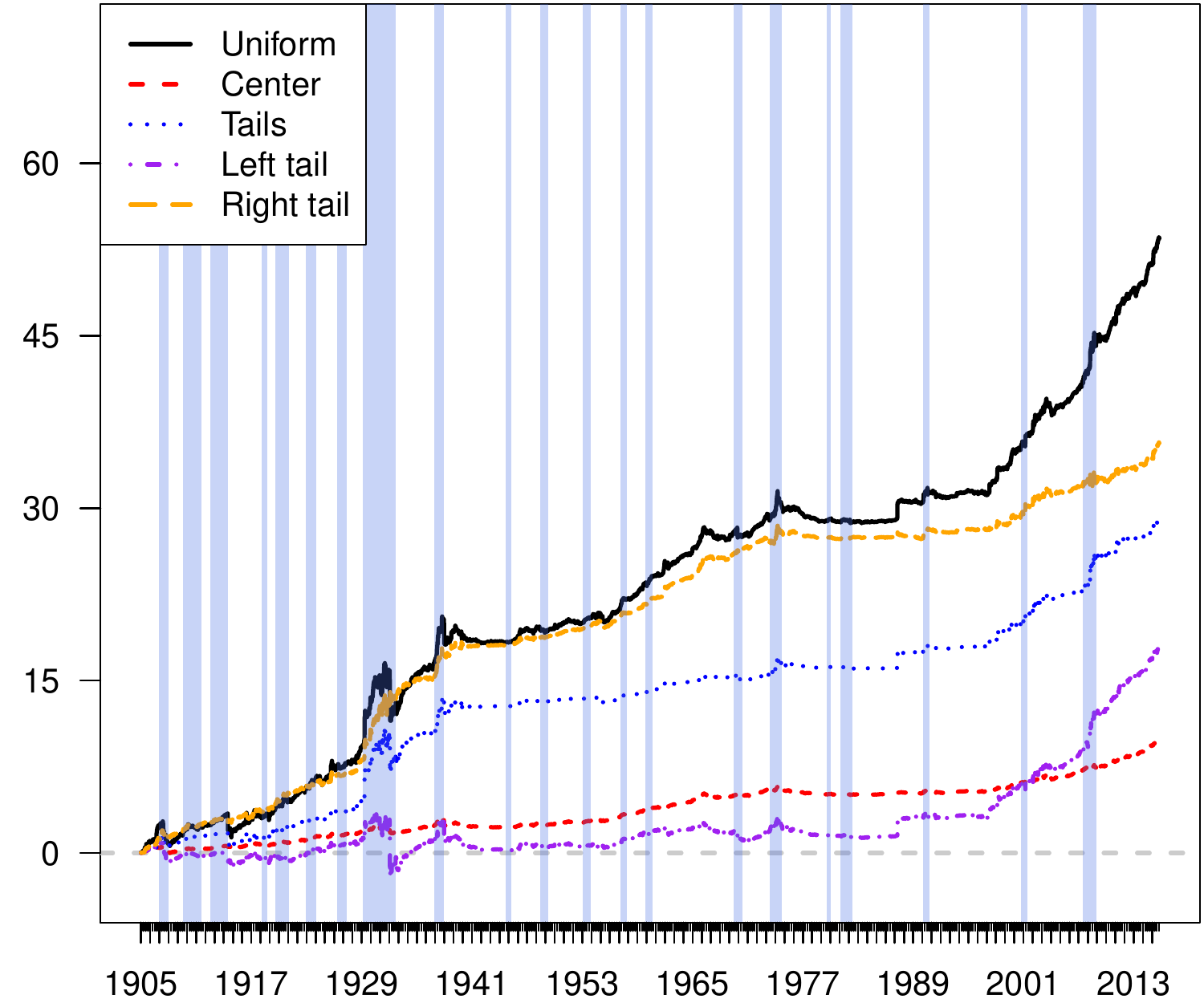}} & {\footnotesize{}\includegraphics[scale=0.52]{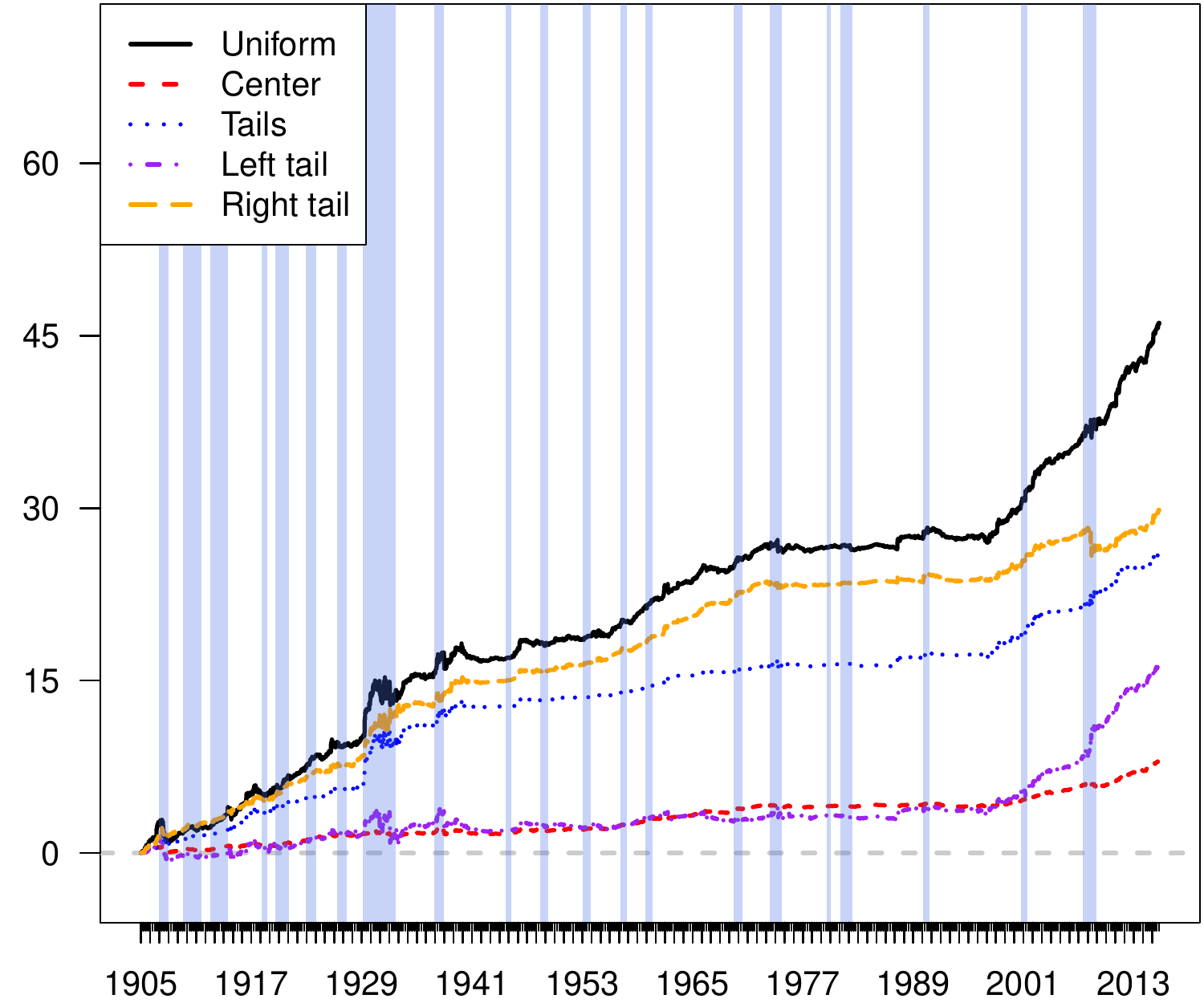}}\tabularnewline
{\footnotesize{}(c)} & {\footnotesize{}(d)}\tabularnewline
{\footnotesize{}\includegraphics[scale=0.52]{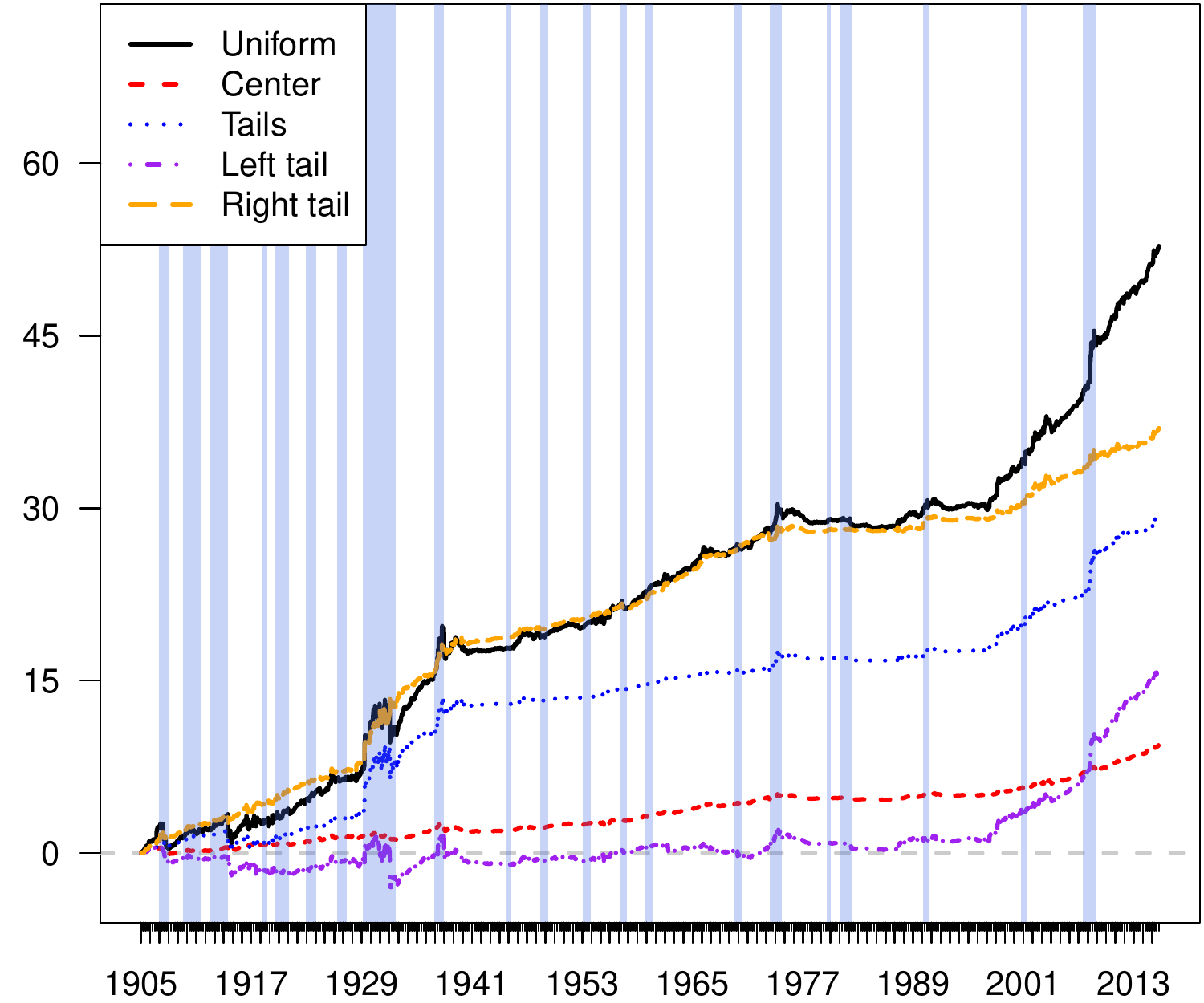}} & {\footnotesize{}\includegraphics[scale=0.52]{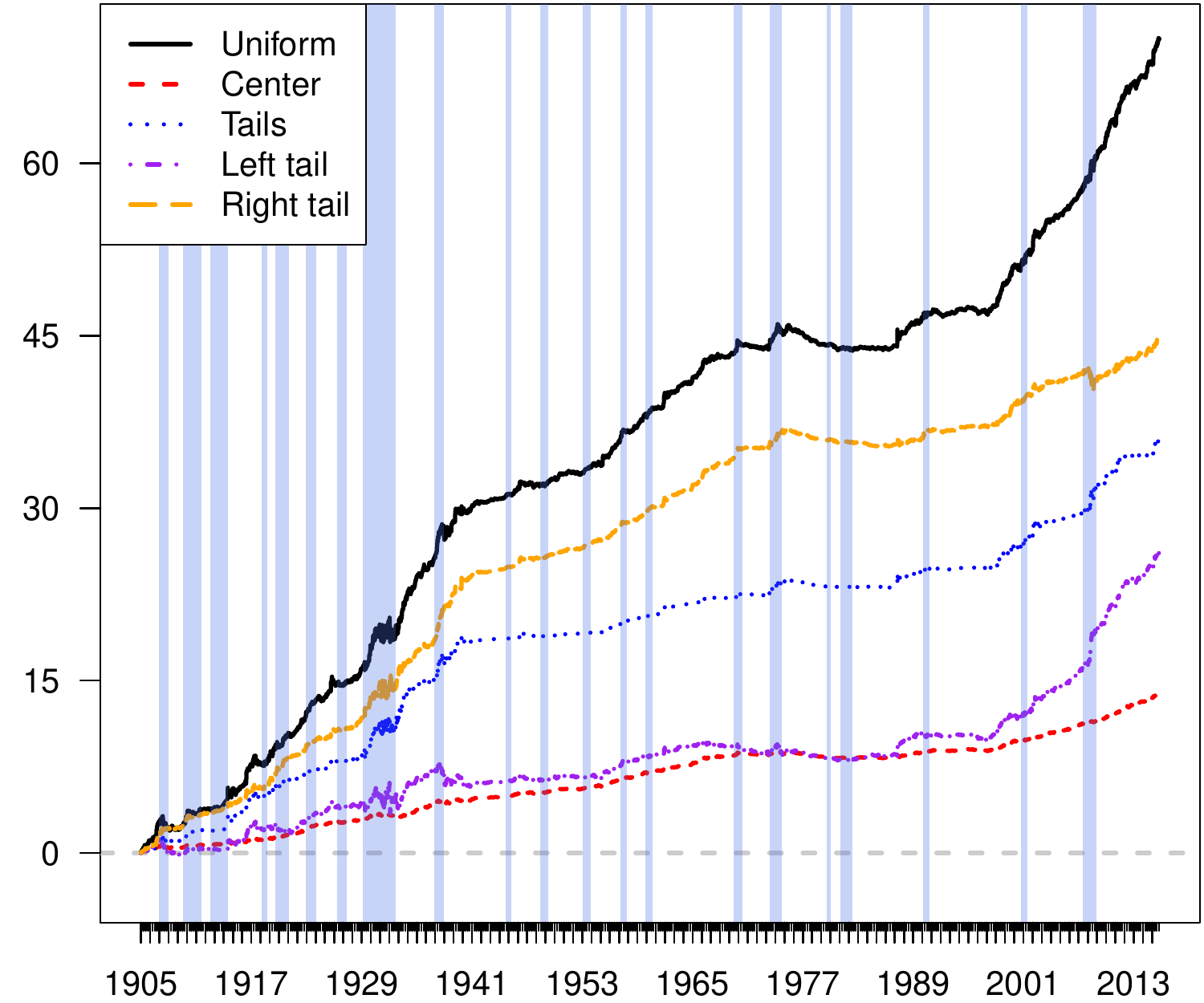}}\tabularnewline
\end{tabular}
\par\end{centering}{\footnotesize \par}

\vspace*{0.15cm}

{\footnotesize{}Panel (a): t-EGARCH-NL relative to EGARCH using different
weights, $w\left(z\right)$, (see Table \ref{Table 2}). Panel (b):
Beta-t-EGARCH-NL relative to Beta-t-EGARCH using different weights,
$w\left(z\right)$, (see Table \ref{Table 2}). Panel (c): SPEGARCH-NL
relative to SPEGARCH using different weights, $w\left(z\right)$,
(see Table \ref{Table 2}). Panel (d): SV-NL compared to SV using
different weights, $w\left(z\right)$, (see Table \ref{Table 2}).
The blue vertical lines indicate business cycle peaks, i.e. the point
at which an economic expansion transitions to a recession, based on
National Bureau of Economic Research (NBER) business cycle dating.}{\footnotesize \par}

\vspace{-0.45in}
\end{figure}
\vspace{-0.20in}
\par\end{center}

\appendix

\section{Appendices\label{sec:Appendices}}

\subsection{Supplementary results\label{sec: Supplementary results}}

\begin{center}
\begin{table}[H]
{\footnotesize{}\caption{\label{Table 14} Density forecast comparison using weekly Dow Jones
returns}
}{\footnotesize \par}

\begin{centering}
\setlength\tabcolsep{2.4pt}{\footnotesize{}}%
\begin{tabular}{llllllll}
\hline
{\footnotesize{}Model} & {\footnotesize{}t-EGARCH } & {\footnotesize{}Beta-t-EGARCH-NL} & {\footnotesize{}Beta-t-EGARCH } & {\footnotesize{}SPEGARCH-NL} & {\footnotesize{}SPEGARCH} & {\footnotesize{}SV-NL} & {\footnotesize{}SV}\tabularnewline
\hline
{\footnotesize{}$h=1$} &  &  &  &  &  &  & \tabularnewline
{\footnotesize{}Uniform } & {\footnotesize{}0.997$^{\left(a\right)}$ } & {\footnotesize{}0.998$^{\left(a\right)}$ } & {\footnotesize{}0.996$^{\left(a\right)}$ } & {\footnotesize{}1.000} & {\footnotesize{}0.997$^{\left(b\right)}$ } & {\footnotesize{}1.000} & {\footnotesize{}0.997$^{\left(a\right)}$ }\tabularnewline
{\footnotesize{}Center } & {\footnotesize{}0.999$^{\left(b\right)}$ } & {\footnotesize{}1.000} & {\footnotesize{}0.999$^{\left(a\right)}$ } & {\footnotesize{}0.999} & {\footnotesize{}0.998 } & {\footnotesize{}1.001} & {\footnotesize{}0.999$^{\left(b\right)}$ }\tabularnewline
{\footnotesize{}Tails } & {\footnotesize{}0.996$^{\left(a\right)}$ } & {\footnotesize{}0.996$^{\left(a\right)}$ } & {\footnotesize{}0.994$^{\left(a\right)}$ } & {\footnotesize{}1.001} & {\footnotesize{}0.997$^{\left(a\right)}$ } & {\footnotesize{}1.000} & {\footnotesize{}0.996$^{\left(a\right)}$ }\tabularnewline
{\footnotesize{}Tail-r } & {\footnotesize{}0.996$^{\left(a\right)}$ } & {\footnotesize{}0.998$^{\left(b\right)}$ } & {\footnotesize{}0.995$^{\left(a\right)}$ } & {\footnotesize{}1.001} & {\footnotesize{}0.997$^{\left(b\right)}$ } & {\footnotesize{}1.001$^{\left(c\right)}$ } & {\footnotesize{}0.997$^{\left(a\right)}$ }\tabularnewline
{\footnotesize{}Tail-l } & {\footnotesize{}0.998 } & {\footnotesize{}0.997$^{\left(b\right)}$ } & {\footnotesize{}0.997$^{\left(b\right)}$ } & {\footnotesize{}1.000} & {\footnotesize{}0.998 } & {\footnotesize{}0.999} & {\footnotesize{}0.997$^{\left(b\right)}$ }\tabularnewline
{\footnotesize{}$h=4$} &  &  &  &  &  &  & \tabularnewline
{\footnotesize{}Uniform } & {\footnotesize{}1.000} & {\footnotesize{}0.991$^{\left(a\right)}$ } & {\footnotesize{}0.993$^{\left(a\right)}$ } & {\footnotesize{}1.003$^{\left(a\right)}$ } & {\footnotesize{}1.000} & {\footnotesize{}0.995$^{\left(a\right)}$ } & {\footnotesize{}0.996$^{\left(a\right)}$ }\tabularnewline
{\footnotesize{}Center } & {\footnotesize{}1.001 } & {\footnotesize{}0.998$^{\left(a\right)}$ } & {\footnotesize{}0.999$^{\left(b\right)}$ } & {\footnotesize{}1.000} & {\footnotesize{}0.999} & {\footnotesize{}0.999$^{\left(b\right)}$ } & {\footnotesize{}1.000 }\tabularnewline
{\footnotesize{}Tails } & {\footnotesize{}0.999 } & {\footnotesize{}0.986$^{\left(a\right)}$ } & {\footnotesize{}0.988$^{\left(a\right)}$ } & {\footnotesize{}1.005$^{\left(a\right)}$ } & {\footnotesize{}1.001} & {\footnotesize{}0.992$^{\left(a\right)}$ } & {\footnotesize{}0.993$^{\left(a\right)}$ }\tabularnewline
{\footnotesize{}Tail-r } & {\footnotesize{}1.000} & {\footnotesize{}0.991$^{\left(a\right)}$ } & {\footnotesize{}0.992$^{\left(a\right)}$ } & {\footnotesize{}1.003$^{\left(a\right)}$ } & {\footnotesize{}1.001} & {\footnotesize{}0.996$^{\left(a\right)}$ } & {\footnotesize{}0.997$^{\left(b\right)}$ }\tabularnewline
{\footnotesize{}Tail-l } & {\footnotesize{}1.000 } & {\footnotesize{}0.992$^{\left(a\right)}$ } & {\footnotesize{}0.993$^{\left(a\right)}$ } & {\footnotesize{}1.003$^{\left(a\right)}$ } & {\footnotesize{}1.000} & {\footnotesize{}0.994$^{\left(a\right)}$ } & {\footnotesize{}0.996$^{\left(a\right)}$ }\tabularnewline
{\footnotesize{}$h=12$} &  &  &  &  &  &  & \tabularnewline
{\footnotesize{}Uniform } & {\footnotesize{}0.994$^{\left(a\right)}$ } & {\footnotesize{}0.972$^{\left(a\right)}$ } & {\footnotesize{}0.975$^{\left(a\right)}$ } & {\footnotesize{}1.004$^{\left(a\right)}$ } & {\footnotesize{}0.996$^{\left(a\right)}$ } & {\footnotesize{}0.978$^{\left(a\right)}$ } & {\footnotesize{}0.980$^{\left(a\right)}$ }\tabularnewline
{\footnotesize{}Center } & {\footnotesize{}0.995$^{\left(a\right)}$ } & {\footnotesize{}0.987$^{\left(a\right)}$ } & {\footnotesize{}0.988$^{\left(a\right)}$ } & {\footnotesize{}1.001} & {\footnotesize{}0.995$^{\left(a\right)}$ } & {\footnotesize{}0.989$^{\left(a\right)}$ } & {\footnotesize{}0.990$^{\left(a\right)}$ }\tabularnewline
{\footnotesize{}Tails } & {\footnotesize{}0.993$^{\left(a\right)}$ } & {\footnotesize{}0.961$^{\left(a\right)}$ } & {\footnotesize{}0.965$^{\left(a\right)}$ } & {\footnotesize{}1.007$^{\left(a\right)}$ } & {\footnotesize{}0.996$^{\left(a\right)}$ } & {\footnotesize{}0.969$^{\left(a\right)}$ } & {\footnotesize{}0.972$^{\left(a\right)}$ }\tabularnewline
{\footnotesize{}Tail-r } & {\footnotesize{}0.994$^{\left(a\right)}$ } & {\footnotesize{}0.973$^{\left(a\right)}$ } & {\footnotesize{}0.975$^{\left(a\right)}$ } & {\footnotesize{}1.003$^{\left(a\right)}$ } & {\footnotesize{}0.996$^{\left(a\right)}$ } & {\footnotesize{}0.979$^{\left(a\right)}$ } & {\footnotesize{}0.980$^{\left(a\right)}$ }\tabularnewline
{\footnotesize{}Tail-l } & {\footnotesize{}0.994$^{\left(a\right)}$ } & {\footnotesize{}0.971$^{\left(a\right)}$ } & {\footnotesize{}0.974$^{\left(a\right)}$ } & {\footnotesize{}1.005$^{\left(a\right)}$ } & {\footnotesize{}0.996$^{\left(a\right)}$ } & {\footnotesize{}0.977$^{\left(a\right)}$ } & {\footnotesize{}0.979$^{\left(a\right)}$ }\tabularnewline
\hline
\end{tabular}
\par\end{centering}{\footnotesize \par}

\begin{centering}
\vspace*{0.15cm}
\par\end{centering}

{\footnotesize{}This table reports the average wCRPS for different
models relative to t-EGARCH-NL. The apexes $a$, $b$, and $c$ indicate
rejection of the null-hypothesis of equal predictive ability according
to the Diebold and Mariano (1995) test at $1\%$, $5\%$ and $10\%$,
respectively. The out-of-sample period consists of $4936$ observations
from $16$ September $1921$ until$15$ April of $2016$.}{\footnotesize \par}

\vspace{-0.45in}
\end{table}
\vspace{-0.10in}
\par\end{center}

\begin{center}
\begin{table}[H]
{\footnotesize{}\caption{\label{Table 15} Density forecast comparison using daily S\&P $500$
equity returns, $w\left(z\right)=\phi\left(z\right)$, i.e. center}
}{\footnotesize \par}

\begin{centering}
\setlength\tabcolsep{19.9pt}{\footnotesize{}}%
\begin{tabular}{lllll}
\hline
{\footnotesize{}Model} & {\footnotesize{}t-EGARCH} & {\footnotesize{}Beta-t-EGARCH} & {\footnotesize{}SPEGARCH} & {\footnotesize{}SV}\tabularnewline
\hline
{\footnotesize{}$h=1$} &  &  &  & {\footnotesize{} }\tabularnewline
{\footnotesize{}t-EGARCH} &  & {\footnotesize{}1 (0)} & {\footnotesize{}71 (22)} & {\footnotesize{}59 (17) }\tabularnewline
{\footnotesize{}Beta-t-EGARCH} & {\footnotesize{}99 (75)} &  & {\footnotesize{}96 (60)} & {\footnotesize{}98 (74) }\tabularnewline
{\footnotesize{}SPEGARCH} & {\footnotesize{}29 (1)} & {\footnotesize{}4 (0)} &  & {\footnotesize{}41 (7) }\tabularnewline
{\footnotesize{}SV} & {\footnotesize{}41 (10)} & {\footnotesize{}2 (0)} & {\footnotesize{}59 (17)} & {\footnotesize{} }\tabularnewline
{\footnotesize{}$h=5$} &  &  &  & {\footnotesize{} }\tabularnewline
{\footnotesize{}t-EGARCH} &  & {\footnotesize{}2 (0)} & {\footnotesize{}66 (22)} & {\footnotesize{}54 (28) }\tabularnewline
{\footnotesize{}Beta-t-EGARCH} & {\footnotesize{}98 (85)} &  & {\footnotesize{}95 (70)} & {\footnotesize{}100 (88) }\tabularnewline
{\footnotesize{}SPEGARCH} & {\footnotesize{}34 (6)} & {\footnotesize{}5 (0)} &  & {\footnotesize{}42 (17) }\tabularnewline
{\footnotesize{}SV} & {\footnotesize{}46 (16)} & {\footnotesize{}0 (0)} & {\footnotesize{}58 (21)} & {\footnotesize{} }\tabularnewline
{\footnotesize{}$h=20$} &  &  &  & {\footnotesize{} }\tabularnewline
{\footnotesize{}t-EGARCH} &  & {\footnotesize{}2 (0)} & {\footnotesize{}43 (24)} & {\footnotesize{}23 (10) }\tabularnewline
{\footnotesize{}Beta-t-EGARCH} & {\footnotesize{}98 (93)} &  & {\footnotesize{}95 (84)} & {\footnotesize{}100 (97) }\tabularnewline
{\footnotesize{}SPEGARCH} & {\footnotesize{}57 (35)} & {\footnotesize{}5 (0)} &  & {\footnotesize{}27 (15) }\tabularnewline
{\footnotesize{}SV} & {\footnotesize{}77 (63)} & {\footnotesize{}0 (0)} & {\footnotesize{}73 (56)} & {\footnotesize{} }\tabularnewline
\hline
\end{tabular}
\par\end{centering}{\footnotesize \par}

\begin{centering}
\vspace*{0.15cm}
\par\end{centering}

{\footnotesize{}Each row in this table reports the percentages,  where
each model generates more accurate density forecasts relative to the
other models reported in each column. The numbers in parentheses indicate
rejection of the null-hypothesis of equal predictive ability according
to the Diebold and Mariano (1995) test at $5\%$ level. The out-of-sample
period consists of $1768$ observations from $24$ December $2007$
until $31$ December, $2014$.}{\footnotesize \par}

\vspace{-0.45in}
\end{table}
\vspace{-0.20in}
\par\end{center}

\begin{center}
\begin{table}[H]
{\footnotesize{}\caption{\label{Table 16} Density forecast comparison using daily S\&P $500$
equity returns, $w\left(z\right)=1-\phi\left(z\right)/\phi\left(0\right)$,
i.e. tails}
}{\footnotesize \par}

\begin{centering}
\setlength\tabcolsep{19.9pt}{\footnotesize{}}%
\begin{tabular}{lllll}
\hline
{\footnotesize{}Model} & {\footnotesize{}t-EGARCH} & {\footnotesize{}Beta-t-EGARCH} & {\footnotesize{}SPEGARCH} & {\footnotesize{}SV}\tabularnewline
\hline
{\footnotesize{}$h=1$} &  &  &  & {\footnotesize{} }\tabularnewline
{\footnotesize{}t-EGARCH} &  & {\footnotesize{}1 (0)} & {\footnotesize{}82 (48)} & {\footnotesize{}53 (26) }\tabularnewline
{\footnotesize{}Beta-t-EGARCH} & {\footnotesize{}99 (92)} &  & {\footnotesize{}100 (95)} & {\footnotesize{}97 (83) }\tabularnewline
{\footnotesize{}SPEGARCH} & {\footnotesize{}18 (2)} & {\footnotesize{}0 (0)} &  & {\footnotesize{}34 (13) }\tabularnewline
{\footnotesize{}SV} & {\footnotesize{}47 (25)} & {\footnotesize{}3 (0)} & {\footnotesize{}66 (37)} & {\footnotesize{} }\tabularnewline
{\footnotesize{}$h=5$} &  &  &  & {\footnotesize{} }\tabularnewline
{\footnotesize{}t-EGARCH} &  & {\footnotesize{}0 (0)} & {\footnotesize{}66 (46)} & {\footnotesize{}41 (27) }\tabularnewline
{\footnotesize{}Beta-t-EGARCH} & {\footnotesize{}100 (97)} &  & {\footnotesize{}100 (96)} & {\footnotesize{}99 (93) }\tabularnewline
{\footnotesize{}SPEGARCH} & {\footnotesize{}34 (17)} & {\footnotesize{}0 (0)} &  & {\footnotesize{}34 (22) }\tabularnewline
{\footnotesize{}SV} & {\footnotesize{}59 (43)} & {\footnotesize{}1 (0)} & {\footnotesize{}66 (50)} & {\footnotesize{} }\tabularnewline
{\footnotesize{}$h=20$} &  &  &  & \tabularnewline
{\footnotesize{}t-EGARCH} &  & {\footnotesize{}1 (0)} & {\footnotesize{}47 (34)} & {\footnotesize{}23 (14) }\tabularnewline
{\footnotesize{}Beta-t-EGARCH} & {\footnotesize{}99 (98)} &  & {\footnotesize{}99 (95)} & {\footnotesize{}100 (99) }\tabularnewline
{\footnotesize{}SPEGARCH} & {\footnotesize{}53 (41)} & {\footnotesize{}1 (0)} &  & {\footnotesize{}27 (19) }\tabularnewline
{\footnotesize{}SV} & {\footnotesize{}77 (66)} & {\footnotesize{}0 (0)} & {\footnotesize{}73 (62)} & {\footnotesize{} }\tabularnewline
\hline
\end{tabular}
\par\end{centering}{\footnotesize \par}

\begin{centering}
\vspace*{0.15cm}
\par\end{centering}

{\footnotesize{}Each row in this table reports the percentages, where
each model generates more accurate density forecasts relative to the
other models reported in each column. The numbers in parentheses indicate
rejection of the null-hypothesis of equal predictive ability according
to the Diebold and Mariano (1995) test at $5\%$ level. The out-of-sample
period consists of $1768$ observations from $24$ December $2007$
until $31$ December, $2014$.}{\footnotesize \par}

\vspace{-0.45in}
\end{table}
\vspace{-0.20in}
\par\end{center}

\begin{center}
\begin{table}[H]
{\footnotesize{}\caption{\label{Table 17} Density forecast comparison using daily S\&P $500$
equity returns, $w\left(z\right)=\Phi\left(z\right)$, i.e. right
tail}
}{\footnotesize \par}

\begin{centering}
\setlength\tabcolsep{19.9pt}{\footnotesize{}}%
\begin{tabular}{lllll}
\hline
{\footnotesize{}Model} & {\footnotesize{}t-EGARCH} & {\footnotesize{}Beta-t-EGARCH} & {\footnotesize{}SPEGARCH} & {\footnotesize{}SV}\tabularnewline
\hline
{\footnotesize{}$h=1$} &  &  &  & {\footnotesize{} }\tabularnewline
{\footnotesize{}t-EGARCH} &  & {\footnotesize{}1 (0)} & {\footnotesize{}69 (32)} & {\footnotesize{}47 (18) }\tabularnewline
{\footnotesize{}Beta-t-EGARCH} & {\footnotesize{}99 (86)} &  & {\footnotesize{}99 (85)} & {\footnotesize{}97 (76) }\tabularnewline
{\footnotesize{}SPEGARCH} & {\footnotesize{}31 (4)} & {\footnotesize{}1 (0)} &  & {\footnotesize{}35 (9) }\tabularnewline
{\footnotesize{}SV} & {\footnotesize{}53 (19)} & {\footnotesize{}3 (0)} & {\footnotesize{}65 (28)} & {\footnotesize{} }\tabularnewline
{\footnotesize{}$h=5$} &  &  &  & {\footnotesize{} }\tabularnewline
{\footnotesize{}t-EGARCH} &  & {\footnotesize{}1 (0)} & {\footnotesize{}64 (36)} & {\footnotesize{}46 (24) }\tabularnewline
{\footnotesize{}Beta-t-EGARCH} & {\footnotesize{}99 (90)} &  & {\footnotesize{}99 (87)} & {\footnotesize{}99 (87) }\tabularnewline
{\footnotesize{}SPEGARCH} & {\footnotesize{}36 (13)} & {\footnotesize{}1 (0)} &  & {\footnotesize{}38 (18) }\tabularnewline
{\footnotesize{}SV} & {\footnotesize{}54 (29)} & {\footnotesize{}1 (0)} & {\footnotesize{}62 (35)} & {\footnotesize{} }\tabularnewline
{\footnotesize{}$h=20$} &  &  &  & {\footnotesize{} }\tabularnewline
{\footnotesize{}t-EGARCH} &  & {\footnotesize{}2 (0)} & {\footnotesize{}49 (31)} & {\footnotesize{}27 (15) }\tabularnewline
{\footnotesize{}Beta-t-EGARCH} & {\footnotesize{}98 (91)} &  & {\footnotesize{}97 (87)} & {\footnotesize{}100 (99) }\tabularnewline
{\footnotesize{}SPEGARCH} & {\footnotesize{}51 (33)} & {\footnotesize{}3 (0)} &  & {\footnotesize{}31 (17) }\tabularnewline
{\footnotesize{}SV} & {\footnotesize{}73 (50)} & {\footnotesize{}0 (0)} & {\footnotesize{}69 (48)} & {\footnotesize{} }\tabularnewline
\hline
\end{tabular}
\par\end{centering}{\footnotesize \par}

\begin{centering}
\vspace*{0.15cm}
\par\end{centering}

{\footnotesize{}Each row in this table reports the percentages, where
each model generates more accurate density forecasts relative to the
other models reported in each column. The numbers in parentheses indicate
rejection of the null-hypothesis of equal predictive ability according
to the Diebold and Mariano (1995) test at $5\%$ level. The out-of-sample
period consists of $1768$ observations from $24$ December $2007$
until $31$ December, $2014$.}{\footnotesize \par}

\vspace{-0.45in}
\end{table}
\vspace{-0.20in}
\par\end{center}

\begin{center}
\begin{table}[H]
{\footnotesize{}\caption{\label{Table 18} Density forecast comparison using daily S\&P $500$
equity returns by sectors, $w\left(z\right)=\phi\left(z\right)$,
i.e. center}
}{\footnotesize \par}

\begin{centering}
\setlength\tabcolsep{15.5pt}{\footnotesize{}}%
\begin{tabular}{lllll}
\hline
{\footnotesize{}Model} & {\footnotesize{}t-EGARCH} & {\footnotesize{}Beta-t-EGARCH} & {\footnotesize{}SPEGARCH} & {\footnotesize{}SV}\tabularnewline
\hline
{\footnotesize{}$h=1$} &  &  & {\footnotesize{} } & \tabularnewline
{\footnotesize{}Consumer discretionary} & {\footnotesize{}83 (26)} & {\footnotesize{}70 (20)} & {\footnotesize{}71 (20) } & {\footnotesize{}84 (34)}\tabularnewline
{\footnotesize{}Consumer staples} & {\footnotesize{}75 (25)} & {\footnotesize{}81 (25)} & {\footnotesize{}53 (12) } & {\footnotesize{}81 (22)}\tabularnewline
{\footnotesize{}Energy} & {\footnotesize{}78 (25)} & {\footnotesize{}78 (12)} & {\footnotesize{}69 (16) } & {\footnotesize{}69 (25)}\tabularnewline
{\footnotesize{}Financials} & {\footnotesize{}79 (43)} & {\footnotesize{}82 (30)} & {\footnotesize{}75 (32) } & {\footnotesize{}89 (32)}\tabularnewline
{\footnotesize{}Health care} & {\footnotesize{}79 (28)} & {\footnotesize{}85 (30)} & {\footnotesize{}62 (23) } & {\footnotesize{}91 (32)}\tabularnewline
{\footnotesize{}Industrials} & {\footnotesize{}81 (28)} & {\footnotesize{}84 (37)} & {\footnotesize{}84 (26) } & {\footnotesize{}89 (47)}\tabularnewline
{\footnotesize{}Information technology} & {\footnotesize{}80 (15)} & {\footnotesize{}81 (20)} & {\footnotesize{}63 (19) } & {\footnotesize{}93 (33)}\tabularnewline
{\footnotesize{}Materials} & {\footnotesize{}87 (30)} & {\footnotesize{}87 (22)} & {\footnotesize{}70 (30) } & {\footnotesize{}96 (43)}\tabularnewline
{\footnotesize{}Real Estate} & {\footnotesize{}62 (25)} & {\footnotesize{}67 (12)} & {\footnotesize{}71 (17) } & {\footnotesize{}75 (0)}\tabularnewline
{\footnotesize{}Telecommunication services} & {\footnotesize{}60 (0)} & {\footnotesize{}100 (20)} & {\footnotesize{}60 (0) } & {\footnotesize{}100 (20)}\tabularnewline
{\footnotesize{}Utilities} & {\footnotesize{}54 (0)} & {\footnotesize{}69 (4)} & {\footnotesize{}65 (4) } & {\footnotesize{}69 (0)}\tabularnewline
{\footnotesize{}$h=5$} &  &  & {\footnotesize{} } & \tabularnewline
{\footnotesize{}Consumer discretionary} & {\footnotesize{}76 (39)} & {\footnotesize{}53 (10)} & {\footnotesize{}79 (51) } & {\footnotesize{}73 (26)}\tabularnewline
{\footnotesize{}Consumer staples} & {\footnotesize{}72 (34)} & {\footnotesize{}72 (16)} & {\footnotesize{}81 (41) } & {\footnotesize{}78 (19)}\tabularnewline
{\footnotesize{}Energy} & {\footnotesize{}72 (16)} & {\footnotesize{}41 (6)} & {\footnotesize{}56 (9) } & {\footnotesize{}50 (3)}\tabularnewline
{\footnotesize{}Financials} & {\footnotesize{}91 (55)} & {\footnotesize{}62 (7)} & {\footnotesize{}93 (68) } & {\footnotesize{}66 (27)}\tabularnewline
{\footnotesize{}Health care} & {\footnotesize{}91 (55)} & {\footnotesize{}83 (47)} & {\footnotesize{}83 (55) } & {\footnotesize{}81 (30)}\tabularnewline
{\footnotesize{}Industrials} & {\footnotesize{}81 (33)} & {\footnotesize{}68 (23)} & {\footnotesize{}75 (32) } & {\footnotesize{}81 (39)}\tabularnewline
{\footnotesize{}Information technology} & {\footnotesize{}59 (15)} & {\footnotesize{}74 (11)} & {\footnotesize{}61 (24) } & {\footnotesize{}81 (28)}\tabularnewline
{\footnotesize{}Materials} & {\footnotesize{}65 (26)} & {\footnotesize{}65 (4)} & {\footnotesize{}61 (26) } & {\footnotesize{}78 (22)}\tabularnewline
{\footnotesize{}Real Estate} & {\footnotesize{}79 (8)} & {\footnotesize{}38 (4)} & {\footnotesize{}79 (29) } & {\footnotesize{}38 (4)}\tabularnewline
{\footnotesize{}Telecommunication services} & {\footnotesize{}80 (40)} & {\footnotesize{}60 (0)} & {\footnotesize{}80 (0) } & {\footnotesize{}100 (0)}\tabularnewline
{\footnotesize{}Utilities} & {\footnotesize{}96 (54)} & {\footnotesize{}38 (0)} & {\footnotesize{}88 (31) } & {\footnotesize{}65 (19)}\tabularnewline
{\footnotesize{}$h=20$} &  &  &  & \tabularnewline
{\footnotesize{}Consumer discretionary} & {\footnotesize{}99 (94)} & {\footnotesize{}21 (3)} & {\footnotesize{}96 (90) } & {\footnotesize{}37 (13)}\tabularnewline
{\footnotesize{}Consumer staples} & {\footnotesize{}100 (97)} & {\footnotesize{}38 (16)} & {\footnotesize{}91 (75) } & {\footnotesize{}56 (6)}\tabularnewline
{\footnotesize{}Energy} & {\footnotesize{}100 (97)} & {\footnotesize{}31 (0)} & {\footnotesize{}97 (97) } & {\footnotesize{}34 (0)}\tabularnewline
{\footnotesize{}Financials} & {\footnotesize{}100 (100)} & {\footnotesize{}27 (4)} & {\footnotesize{}100 (100)} & {\footnotesize{}50 (20)}\tabularnewline
{\footnotesize{}Health care} & {\footnotesize{}96 (92)} & {\footnotesize{}60 (21)} & {\footnotesize{}87 (85) } & {\footnotesize{}38 (8)}\tabularnewline
{\footnotesize{}Industrials} & {\footnotesize{}98 (96)} & {\footnotesize{}33 (9)} & {\footnotesize{}95 (91) } & {\footnotesize{}37 (11)}\tabularnewline
{\footnotesize{}Information technology} & {\footnotesize{}98 (93)} & {\footnotesize{}35 (6)} & {\footnotesize{}89 (76) } & {\footnotesize{}35 (4)}\tabularnewline
{\footnotesize{}Materials} & {\footnotesize{}100 (100)} & {\footnotesize{}17 (4)} & {\footnotesize{}100 (96) } & {\footnotesize{}26 (4)}\tabularnewline
{\footnotesize{}Real Estate} & {\footnotesize{}100 (100)} & {\footnotesize{}8 (0)} & {\footnotesize{}100 (92) } & {\footnotesize{}4 (0)}\tabularnewline
{\footnotesize{}Telecommunication services} & {\footnotesize{}80 (80)} & {\footnotesize{}40 (20)} & {\footnotesize{}100 (80) } & {\footnotesize{}20 (0)}\tabularnewline
{\footnotesize{}Utilities} & {\footnotesize{}100 (100)} & {\footnotesize{}42 (12)} & {\footnotesize{}100 (100) } & {\footnotesize{}50 (8)}\tabularnewline
\hline
\end{tabular}
\par\end{centering}{\footnotesize \par}

\begin{centering}
\vspace*{0.15cm}
\par\end{centering}

{\footnotesize{}This table reports the percentages for each model-type,
where the version that considers the leverage effect generates more
accurate density forecasts than the version that does not consider
the leverage effect. The numbers in parentheses indicate rejection
of the null-hypothesis of equal predictive ability according to the
Diebold and Mariano (1995) test at $5\%$ level. The out-of-sample
period consists of $1768$ observations from $24$ December $2007$
until $31$ December, $2014$.}{\footnotesize \par}

\vspace{-0.45in}
\end{table}
\vspace{-0.10in}
\par\end{center}

\begin{center}
\begin{table}[H]
{\footnotesize{}\caption{\label{Table 19} Density forecast comparison using daily S\&P $500$
equity returns by sectors, $w\left(z\right)=1-\phi\left(z\right)/\phi\left(0\right)$,
i.e. tails}
}{\footnotesize \par}

\begin{centering}
\setlength\tabcolsep{15.9pt}{\footnotesize{}}%
\begin{tabular}{lllll}
\hline
{\footnotesize{}Model} & {\footnotesize{}t-EGARCH} & {\footnotesize{}Beta-t-EGARCH} & {\footnotesize{}SPEGARCH} & {\footnotesize{}SV}\tabularnewline
\hline
{\footnotesize{}$h=1$} &  &  &  & \tabularnewline
{\footnotesize{}Consumer discretionary} & {\footnotesize{}76 (37)} & {\footnotesize{}91 (43)} & {\footnotesize{}83 (37) } & {\footnotesize{}91 (47)}\tabularnewline
{\footnotesize{}Consumer staples} & {\footnotesize{}88 (50)} & {\footnotesize{}91 (66)} & {\footnotesize{}75 (38) } & {\footnotesize{}81 (38)}\tabularnewline
{\footnotesize{}Energy} & {\footnotesize{}100 (47)} & {\footnotesize{}78 (25)} & {\footnotesize{}84 (50) } & {\footnotesize{}84 (25)}\tabularnewline
{\footnotesize{}Financials} & {\footnotesize{}70 (32)} & {\footnotesize{}79 (36)} & {\footnotesize{}66 (25) } & {\footnotesize{}88 (36)}\tabularnewline
{\footnotesize{}Health care} & {\footnotesize{}92 (55)} & {\footnotesize{}92 (75)} & {\footnotesize{}72 (32) } & {\footnotesize{}96 (47)}\tabularnewline
{\footnotesize{}Industrials} & {\footnotesize{}91 (49)} & {\footnotesize{}98 (68)} & {\footnotesize{}89 (58) } & {\footnotesize{}95 (72)}\tabularnewline
{\footnotesize{}Information technology} & {\footnotesize{}76 (30)} & {\footnotesize{}100 (67)} & {\footnotesize{}69 (24) } & {\footnotesize{}93 (41)}\tabularnewline
{\footnotesize{}Materials} & {\footnotesize{}83 (43)} & {\footnotesize{}91 (52)} & {\footnotesize{}74 (30) } & {\footnotesize{}91 (43)}\tabularnewline
{\footnotesize{}Real Estate} & {\footnotesize{}67 (21)} & {\footnotesize{}71 (17)} & {\footnotesize{}62 (33) } & {\footnotesize{}67 (17)}\tabularnewline
{\footnotesize{}Telecommunication services} & {\footnotesize{}40 (20)} & {\footnotesize{}60 (0)} & {\footnotesize{}20 (0) } & {\footnotesize{}80 (60)}\tabularnewline
{\footnotesize{}Utilities} & {\footnotesize{}88 (35)} & {\footnotesize{}88 (38)} & {\footnotesize{}81 (31) } & {\footnotesize{}88 (46)}\tabularnewline
{\footnotesize{}$h=5$} &  &  &  & \tabularnewline
{\footnotesize{}Consumer discretionary} & {\footnotesize{}89 (63)} & {\footnotesize{}74 (23)} & {\footnotesize{}86 (70) } & {\footnotesize{}83 (44)}\tabularnewline
{\footnotesize{}Consumer staples} & {\footnotesize{}81 (44)} & {\footnotesize{}78 (47)} & {\footnotesize{}78 (50) } & {\footnotesize{}78 (50)}\tabularnewline
{\footnotesize{}Energy} & {\footnotesize{}91 (47)} & {\footnotesize{}41 (12)} & {\footnotesize{}84 (41) } & {\footnotesize{}41 (6)}\tabularnewline
{\footnotesize{}Financials} & {\footnotesize{}93 (77)} & {\footnotesize{}66 (20)} & {\footnotesize{}93 (82) } & {\footnotesize{}77 (39)}\tabularnewline
{\footnotesize{}Health care} & {\footnotesize{}96 (79)} & {\footnotesize{}94 (81)} & {\footnotesize{}85 (66) } & {\footnotesize{}91 (53)}\tabularnewline
{\footnotesize{}Industrials} & {\footnotesize{}95 (81)} & {\footnotesize{}88 (37)} & {\footnotesize{}89 (74) } & {\footnotesize{}91 (53)}\tabularnewline
{\footnotesize{}Information technology} & {\footnotesize{}78 (44)} & {\footnotesize{}87 (46)} & {\footnotesize{}72 (56) } & {\footnotesize{}85 (39)}\tabularnewline
{\footnotesize{}Materials} & {\footnotesize{}83 (74)} & {\footnotesize{}74 (22)} & {\footnotesize{}78 (52) } & {\footnotesize{}70 (39)}\tabularnewline
{\footnotesize{}Real Estate} & {\footnotesize{}88 (83)} & {\footnotesize{}38 (4)} & {\footnotesize{}79 (71) } & {\footnotesize{}42 (8)}\tabularnewline
{\footnotesize{}Telecommunication services} & {\footnotesize{}100 (80)} & {\footnotesize{}60 (40)} & {\footnotesize{}80 (20) } & {\footnotesize{}80 (60)}\tabularnewline
{\footnotesize{}Utilities} & {\footnotesize{}100 (73)} & {\footnotesize{}54 (12)} & {\footnotesize{}96 (62) } & {\footnotesize{}65 (23)}\tabularnewline
{\footnotesize{}$h=20$} &  &  &  & \tabularnewline
{\footnotesize{}Consumer discretionary} & {\footnotesize{}99 (94)} & {\footnotesize{}44 (14)} & {\footnotesize{}94 (93) } & {\footnotesize{}50 (23)}\tabularnewline
{\footnotesize{}Consumer staples} & {\footnotesize{}94 (94)} & {\footnotesize{}59 (22)} & {\footnotesize{}88 (81) } & {\footnotesize{}50 (34)}\tabularnewline
{\footnotesize{}Energy} & {\footnotesize{}97 (97)} & {\footnotesize{}38 (6)} & {\footnotesize{}97 (97) } & {\footnotesize{}22 (3)}\tabularnewline
{\footnotesize{}Financials} & {\footnotesize{}100 (98)} & {\footnotesize{}61 (25)} & {\footnotesize{}98 (96) } & {\footnotesize{}64 (45)}\tabularnewline
{\footnotesize{}Health care} & {\footnotesize{}98 (94)} & {\footnotesize{}72 (47)} & {\footnotesize{}85 (83) } & {\footnotesize{}49 (11)}\tabularnewline
{\footnotesize{}Industrials} & {\footnotesize{}98 (98)} & {\footnotesize{}56 (25)} & {\footnotesize{}93 (89) } & {\footnotesize{}51 (21)}\tabularnewline
{\footnotesize{}Information technology} & {\footnotesize{}100 (89)} & {\footnotesize{}69 (28)} & {\footnotesize{}91 (83) } & {\footnotesize{}39 (15)}\tabularnewline
{\footnotesize{}Materials} & {\footnotesize{}100 (91)} & {\footnotesize{}35 (13)} & {\footnotesize{}96 (91) } & {\footnotesize{}39 (17)}\tabularnewline
{\footnotesize{}Real Estate} & {\footnotesize{}100 (100)} & {\footnotesize{}29 (8)} & {\footnotesize{}96 (96) } & {\footnotesize{}29 (17)}\tabularnewline
{\footnotesize{}Telecommunication services} & {\footnotesize{}100 (100)} & {\footnotesize{}60 (60)} & {\footnotesize{}100 (80) } & {\footnotesize{}60 (60)}\tabularnewline
{\footnotesize{}Utilities} & {\footnotesize{}100 (100)} & {\footnotesize{}50 (31)} & {\footnotesize{}100 (100)} & {\footnotesize{}62 (31)}\tabularnewline
\hline
\end{tabular}
\par\end{centering}{\footnotesize \par}

\begin{centering}
\vspace*{0.15cm}
\par\end{centering}

{\footnotesize{}This table reports the percentages for each model-type,
where the version that considers the leverage effect generates more
accurate density forecasts than the version that does not consider
the leverage effect. The numbers in parentheses indicate rejection
of the null-hypothesis of equal predictive ability according to the
Diebold and Mariano (1995) test at $5\%$ level. The out-of-sample
period consists of $1768$ observations from $24$ December $2007$
until $31$ December, $2014$.}{\footnotesize \par}

\vspace{-0.45in}
\end{table}
\vspace{-0.10in}
\par\end{center}

\begin{center}
\begin{table}[H]
{\footnotesize{}\caption{\label{Table 20} Density forecast comparison using daily S\&P $500$
equity returns by sectors, $w\left(z\right)=\Phi\left(z\right)$,
i.e. right tail}
}{\footnotesize \par}

\begin{centering}
\setlength\tabcolsep{15.5pt}{\footnotesize{}}%
\begin{tabular}{lllll}
\hline
{\footnotesize{}Model} & {\footnotesize{}t-EGARCH} & {\footnotesize{}Beta-t-EGARCH} & {\footnotesize{}SPEGARCH} & {\footnotesize{}SV}\tabularnewline
\hline
{\footnotesize{}$h=1$} &  &  & {\footnotesize{} } & \tabularnewline
{\footnotesize{}Consumer discretionary} & {\footnotesize{}47 (6)} & {\footnotesize{}61 (6)} & {\footnotesize{}44 (6) } & {\footnotesize{}66 (10)}\tabularnewline
{\footnotesize{}Consumer staples} & {\footnotesize{}56 (3)} & {\footnotesize{}72 (12)} & {\footnotesize{}47 (0) } & {\footnotesize{}56 (3)}\tabularnewline
{\footnotesize{}Energy} & {\footnotesize{}56 (3)} & {\footnotesize{}47 (6)} & {\footnotesize{}44 (6) } & {\footnotesize{}50 (12)}\tabularnewline
{\footnotesize{}Financials} & {\footnotesize{}32 (2)} & {\footnotesize{}36 (0)} & {\footnotesize{}18 (4) } & {\footnotesize{}52 (2)}\tabularnewline
{\footnotesize{}Health care} & {\footnotesize{}60 (6)} & {\footnotesize{}70 (25)} & {\footnotesize{}51 (2) } & {\footnotesize{}66 (6)}\tabularnewline
{\footnotesize{}Industrials} & {\footnotesize{}56 (7)} & {\footnotesize{}77 (21)} & {\footnotesize{}63 (11) } & {\footnotesize{}75 (32)}\tabularnewline
{\footnotesize{}Information technology} & {\footnotesize{}44 (0)} & {\footnotesize{}70 (15)} & {\footnotesize{}35 (11) } & {\footnotesize{}65 (20)}\tabularnewline
{\footnotesize{}Materials} & {\footnotesize{}57 (4)} & {\footnotesize{}65 (13)} & {\footnotesize{}39 (4) } & {\footnotesize{}70 (4)}\tabularnewline
{\footnotesize{}Real Estate} & {\footnotesize{}12 (4)} & {\footnotesize{}12 (4)} & {\footnotesize{}4 (4) } & {\footnotesize{}25 (4)}\tabularnewline
{\footnotesize{}Telecommunication services} & {\footnotesize{}20 (0)} & {\footnotesize{}40 (0)} & {\footnotesize{}0 (0) } & {\footnotesize{}80 (20)}\tabularnewline
{\footnotesize{}Utilities} & {\footnotesize{}23 (0)} & {\footnotesize{}31 (0)} & {\footnotesize{}23 (0) } & {\footnotesize{}54 (0)}\tabularnewline
{\footnotesize{}$h=5$} &  &  & {\footnotesize{} } & \tabularnewline
{\footnotesize{}Consumer discretionary} & {\footnotesize{}67 (30)} & {\footnotesize{}63 (9)} & {\footnotesize{}76 (39) } & {\footnotesize{}70 (30)}\tabularnewline
{\footnotesize{}Consumer staples} & {\footnotesize{}59 (25)} & {\footnotesize{}72 (25)} & {\footnotesize{}66 (22) } & {\footnotesize{}81 (38)}\tabularnewline
{\footnotesize{}Energy} & {\footnotesize{}47 (6)} & {\footnotesize{}28 (3)} & {\footnotesize{}47 (6) } & {\footnotesize{}28 (3)}\tabularnewline
{\footnotesize{}Financials} & {\footnotesize{}84 (46)} & {\footnotesize{}48 (0)} & {\footnotesize{}88 (62) } & {\footnotesize{}70 (16)}\tabularnewline
{\footnotesize{}Health care} & {\footnotesize{}79 (34)} & {\footnotesize{}87 (36)} & {\footnotesize{}77 (40) } & {\footnotesize{}75 (26)}\tabularnewline
{\footnotesize{}Industrials} & {\footnotesize{}75 (23)} & {\footnotesize{}61 (5)} & {\footnotesize{}72 (32) } & {\footnotesize{}79 (30)}\tabularnewline
{\footnotesize{}Information technology} & {\footnotesize{}57 (15)} & {\footnotesize{}81 (13)} & {\footnotesize{}61 (31) } & {\footnotesize{}63 (24)}\tabularnewline
{\footnotesize{}Materials} & {\footnotesize{}78 (26)} & {\footnotesize{}65 (13)} & {\footnotesize{}61 (22) } & {\footnotesize{}65 (17)}\tabularnewline
{\footnotesize{}Real Estate} & {\footnotesize{}79 (42)} & {\footnotesize{}17 (0)} & {\footnotesize{}75 (33) } & {\footnotesize{}42 (8)}\tabularnewline
{\footnotesize{}Telecommunication services} & {\footnotesize{}80 (20)} & {\footnotesize{}40 (20)} & {\footnotesize{}80 (20) } & {\footnotesize{}100 (20)}\tabularnewline
{\footnotesize{}Utilities} & {\footnotesize{}85 (19)} & {\footnotesize{}31 (0)} & {\footnotesize{}81 (12) } & {\footnotesize{}42 (12)}\tabularnewline
{\footnotesize{}$h=20$} &  &  & {\footnotesize{} } & \tabularnewline
{\footnotesize{}Consumer discretionary} & {\footnotesize{}94 (84)} & {\footnotesize{}30 (7)} & {\footnotesize{}93 (90) } & {\footnotesize{}46 (20)}\tabularnewline
{\footnotesize{}Consumer staples} & {\footnotesize{}100 (94)} & {\footnotesize{}44 (16)} & {\footnotesize{}81 (75) } & {\footnotesize{}56 (19)}\tabularnewline
{\footnotesize{}Energy} & {\footnotesize{}100 (97)} & {\footnotesize{}25 (3)} & {\footnotesize{}97 (94) } & {\footnotesize{}16 (0)}\tabularnewline
{\footnotesize{}Financials} & {\footnotesize{}100 (96)} & {\footnotesize{}54 (12)} & {\footnotesize{}98 (96) } & {\footnotesize{}57 (23)}\tabularnewline
{\footnotesize{}Health care} & {\footnotesize{}96 (85)} & {\footnotesize{}66 (15)} & {\footnotesize{}85 (75) } & {\footnotesize{}36 (2)}\tabularnewline
{\footnotesize{}Industrials} & {\footnotesize{}98 (93)} & {\footnotesize{}46 (11)} & {\footnotesize{}93 (86) } & {\footnotesize{}44 (18)}\tabularnewline
{\footnotesize{}Information technology} & {\footnotesize{}93 (69)} & {\footnotesize{}57 (7)} & {\footnotesize{}87 (63) } & {\footnotesize{}35 (11)}\tabularnewline
{\footnotesize{}Materials} & {\footnotesize{}100 (87)} & {\footnotesize{}30 (4)} & {\footnotesize{}100 (87) } & {\footnotesize{}35 (13)}\tabularnewline
{\footnotesize{}Real Estate} & {\footnotesize{}100 (100)} & {\footnotesize{}17 (0)} & {\footnotesize{}96 (88) } & {\footnotesize{}25 (8)}\tabularnewline
{\footnotesize{}Telecommunication services} & {\footnotesize{}100 (80)} & {\footnotesize{}60 (20)} & {\footnotesize{}100 (80) } & {\footnotesize{}40 (20)}\tabularnewline
{\footnotesize{}Utilities} & {\footnotesize{}100 (100)} & {\footnotesize{}46 (19)} & {\footnotesize{}100 (100) } & {\footnotesize{}42 (4)}\tabularnewline
\hline
\end{tabular}
\par\end{centering}{\footnotesize \par}

\begin{centering}
\vspace*{0.15cm}
\par\end{centering}

{\footnotesize{}This table reports the percentages for each model-type,
where the version that considers the leverage effect generates more
accurate density forecasts than the version that does not consider
the leverage effect. The numbers in parentheses indicate rejection
of the null-hypothesis of equal predictive ability according to the
Diebold and Mariano (1995) test at $5\%$ level. The out-of-sample
period consists of $1768$ observations from $24$ December $2007$
until $31$ December, $2014$.}{\footnotesize \par}

\vspace{-0.45in}
\end{table}
\vspace{-0.10in}
\par\end{center}

\begin{center}
\begin{figure}[H]
\caption{\label{Figure 2} Out-of-sample cumulative wCRPS using weekly Dow
Jones returns }

\begin{centering}
{\footnotesize{}}%
\begin{tabular}{cc}
{\footnotesize{}(a)} & {\footnotesize{}(b)}\tabularnewline
{\footnotesize{}\includegraphics[scale=0.52]{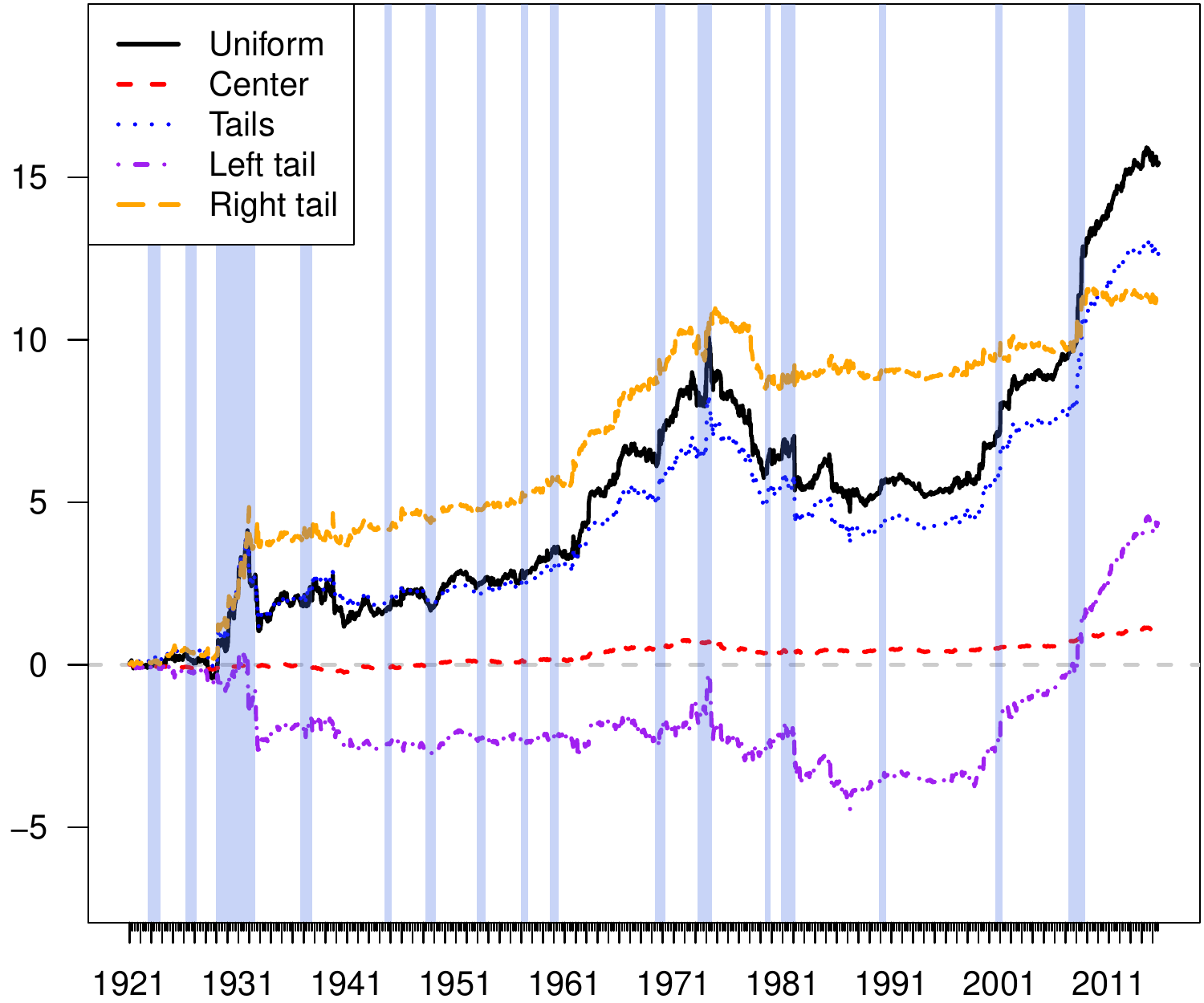}} & {\footnotesize{}\includegraphics[scale=0.52]{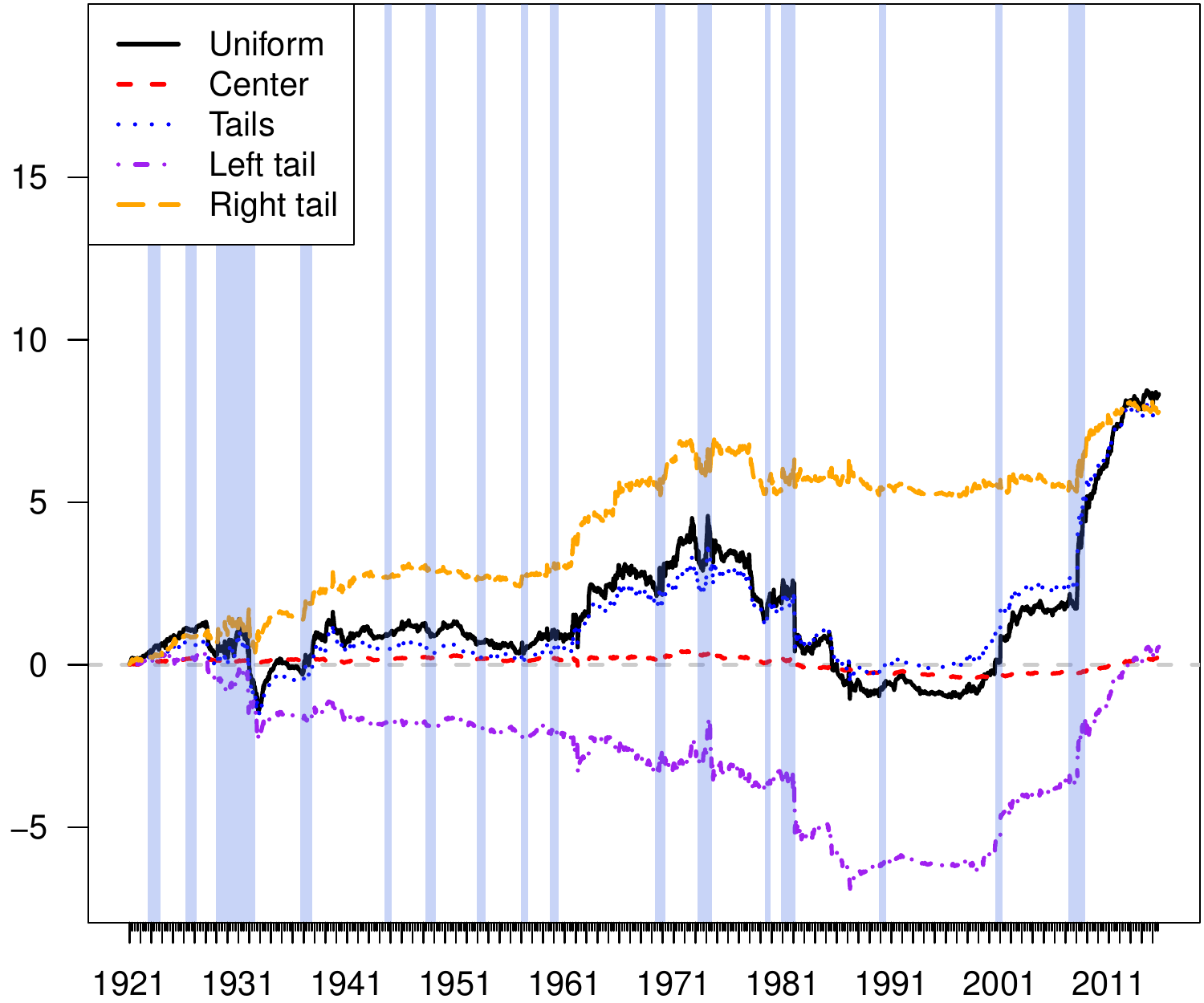}}\tabularnewline
{\footnotesize{}(c)} & {\footnotesize{}(d)}\tabularnewline
{\footnotesize{}\includegraphics[scale=0.52]{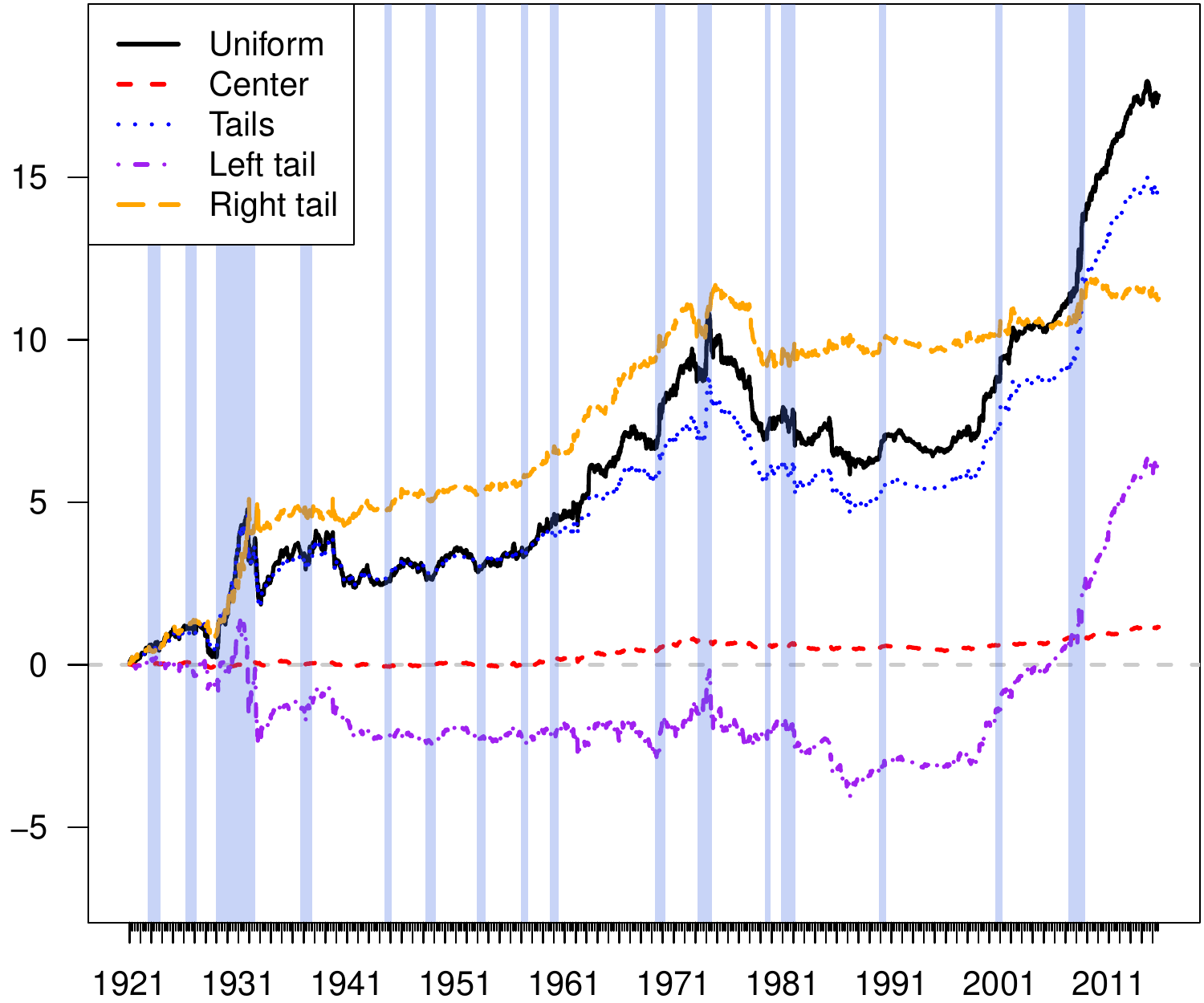}} & {\footnotesize{}\includegraphics[scale=0.52]{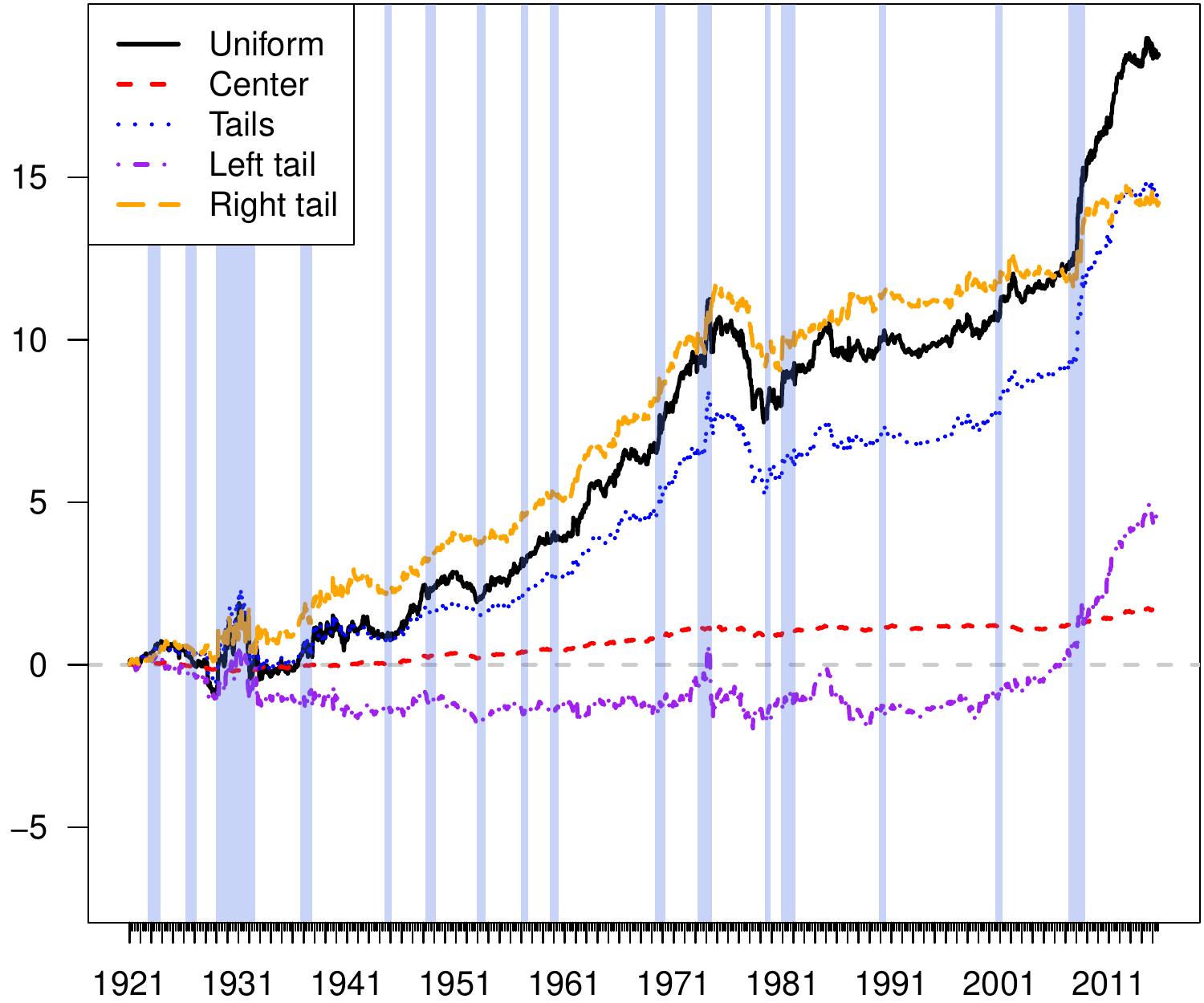}}\tabularnewline
\end{tabular}
\par\end{centering}{\footnotesize \par}

\vspace*{0.15cm}

{\footnotesize{}Panel (a): t-EGARCH-NL relative to EGARCH using different
weights, $w\left(z\right)$, (see Table \ref{Table 2}). Panel (b):
Beta-t-EGARCH-NL relative to Beta-t-EGARCH using different weights,
$w\left(z\right)$, (see Table \ref{Table 2}). Panel (c): SPEGARCH-NL
relative to SPEGARCH using different weights, $w\left(z\right)$,
(see Table \ref{Table 2}). Panel (d): SV-NL compared to SV using
different weights, $w\left(z\right)$, (see Table \ref{Table 2}).
The blue vertical lines indicate business cycle peaks, i.e. the point
at which an economic expansion transitions to a recession, based on
National Bureau of Economic Research (NBER) business cycle dating.}{\footnotesize \par}

\vspace{-0.45in}
\end{figure}
\vspace{-0.10in}
\par\end{center}
\end{document}